\documentclass[12pt,preprint,eps]{aastex}

\slugcomment{}

\shorttitle{Lobel et al.}
\shortauthors{The 2000-2001 outburst of $\rho$ Cas}

\begin{document}

\title{High-Resolution Spectroscopy of the Yellow Hypergiant $\rho$~Cassiopeiae from 1993 Through the Outburst of 2000-2001}

\author{A. Lobel\footnote{Guest investigator of the UK Astronomy Data Centre.}, A. K. Dupree, 
R. P. Stefanik, G. Torres, \affil{Harvard-Smithsonian Center for Astrophysics, 60 Garden Street, 
Cambridge MA 02138} 
G. Israelian,\affil{Instituto de Astrofisica de Canarias, E-38200 La Laguna, Tenerife, Spain} 
N. Morrison,\affil{Ritter Astrophysical Research Center, University of Toledo, Toledo, OH 43606} 
C. de Jager, H. Nieuwenhuijzen,\affil{SRON Laboratory for Space Research, Sorbonnelaan 2, 3584 CA Utrecht, the Netherlands } 
I. Ilyin,\affil{Astronomy Division, PO Box 3000, 90014 University of Oulu, Finland} 
and F. Musaev\affil{Special Astrophysical Observatory, Nizhnij Arkhyz 369167, Russia}}

\begin{abstract}
We present an overview of the spectral variability of the peculiar F-type hypergiant 
$\rho$ Cas, obtained from our long-term monitoring campaigns over the past 8.5 years 
with four spectrographs in the northern hemisphere. 
Between 2000 June and September an exceptional variability phase occurred 
when the $V$-brightness dimmed by about a full magnitude.  The star recovered 
from this deep minimum by 2001 April. It is the third outburst of $\rho$ Cas 
on record in the last century. We observe TiO absorption bands in 
high-resolution near-IR spectra obtained with the Utrecht Echelle Spectrograph
during the summer of 2000. TiO formation in the outer atmosphere occurred 
before the deep brightness minimum. Atmospheric models reveal that the 
effective temperature decreases by at least 3000 K, and the TiO shell is 
driven supersonically with $\dot{M}$$\simeq$5.4$\times$$10^{-2}$ $\rm M_{\odot}\,yr^{-1}$. 
Strong episodic mass loss and TiO have also been observed during the 
outbursts of 1945-47 and 1985-86. 
 
A detailed analysis of the exceptional outburst spectra is provided, 
by comparing with high-resolution optical spectra of the early M-type 
supergiants $\mu$ Cep ($\rm Ia$) and Betelgeuse ($\rm Iab$). 
During the outburst, central emission appears above the local continuum level 
in the split Na $D$ lines. A prominent optical emission line spectrum 
appears in variability phases of fast wind expansion. The radial velocity 
curves of H$\alpha$, and of photospheric metal absorption lines signal a 
very extended, and velocity stratified dynamic atmosphere. The outburst 
spectra indicate the formation of a low-temperature, optically thick 
circumstellar gas shell of 3$\times$$10^{-2}$ $\rm M_{\odot}$ during ~200 d, 
caused by dynamic instability of the upper atmosphere of this pulsating 
massive supergiant near the Eddington luminosity limit. 
We observe that the mass-loss rate during the outburst is of 
the same order of magnitude as has been proposed for the outbursts of $\eta$ Carinae.
We present calculations that correctly predict the outburst time-scale,
whereby the shell ejection is driven by the release 
of hydrogen ionization-recombination energy.  
\end{abstract}


\section{Introduction}
The recurrent eruptions of $\rho$ Cas (HD 224014) recorded over the past century are 
the hallmark of the exceptional atmospheric physics manifested by the yellow hypergiants.
These cool luminous stars are thought to be post-red supergiants, 
rapidly evolving toward the blue supergiant phase (de Jager 1998).
They are rare enigmatic objects, and continuous high-resolution spectroscopic investigations are 
limited to a small sample of bright stars, often showing dissimilar spectra, but with very peculiar
spectral properties. Yellow hypergiants are the candidates `par excellence' among 
the cool luminous stars to investigate the physical causes for the luminosity limit of evolved stars
(i.e. de Jager et al. 2001; Lobel 2001a; Stothers \& Chin 2001).
The post-red supergiant phase assumed for $\rho$ Cas mainly relies, among other cool hypergiants
such as HR~8752 and IRC+10420, on a substantial atmospheric overabundance observed in N and Na 
(Takeda \& Takeda-Hidai 1994; El Eid \& Champagne 1995). There is evidence that $\rho$ Cas is 
surrounded by a tenuous circumstellar gas shell with the observation of forbidden [Ca~{\sc ii}] 
optical emission lines (Lobel 1997)\footnote{Partly available at {\tt http://www.shaker.nl/Boekencatalogus.asp}.}. 
Recent high-resolution near-IR imaging with HST-WFPC2 does not show evidence of 
distant shells or a reflection nebula in $\rho$ Cas (Schuster \& Humphreys 2001), although     
complex circumstellar environments have been imaged in the cool 
hypergiants IRC+10420 and VY CMa (Humphreys et al. 1997; Smith et al. 2001).  
The distance to $\rho$ Cas is large ($d$=3.1$\pm$0.5 kpc), and presently based on
interstellar absorption and absolute visual magnitude estimates (e.g. Zsoldos \& Percy 1991).
Since the distance to IRC+10420 is estimated $\sim$5 kpc, the lack of detection 
in $\rho$ Cas may result from the imaging technique. Its brightness causes bleeding near the star, 
although a range of integration times was used. The images do not show any extended structure 
within 0.1$\arcsec$ of the star, and down to the WFPC2's read noise 2$\arcsec$ away. 

A spectroscopic study during 1993-95 showed that $\rho$ Cas is a slowly 
pulsating supergiant with changes in the optical spectrum corresponding to 
variations of $T_{\rm eff}$ less than 750 K ($T_{\rm eff}$=7250K $-$ 6500 K; 
Lobel et al. 1998). Its optical spectrum is peculiar because prominent permitted emission 
lines from neutral atoms sometimes appear, which are rarely seen in stellar spectra
(e.g. in 1993 December; Lobel 1997).   
An important aspect of hypergiant spectra is the unusual large broadening
observed in photospheric absorption lines. This broadening is currently attributed to 
large-scale atmospheric movements which assume highly supersonic velocities,
while the stellar rotational broadening is small.        
Another remarkable aspect of $\rho$ Cas' spectrum is the 
permanent, and cyclic, core doubling in low-energy resonance and subordinate 
photospheric absorption lines, such as Fe~{\sc i}, Fe~{\sc ii}, and Ba~{\sc ii}. 
The phenomenon is presently thought to result from a static central emission 
reversal which is optically thick with an excitation temperature of 3050 K, 
formed at a large distance from the photosphere (Lobel 1997). The intensity of both 
adjacent absorption components periodically alters with the Doppler shifts of the 
pulsating photosphere. High-resolution spectroscopic studies of stellar atmospheric dynamics
based on split absorption lines are important because they are also observed in RR Lyrae 
variables, Cepheids, and FU Ori objects (Lobel 2001b). 
The line doubling appears to be physically linked with the 
changing spectral type of the hypergiant, because it was also observed in HR 8752 in 1960-70,
when this binary star had a smaller $T_{\rm eff}$, and appeared to be a spectroscopic twin 
of $\rho$ Cas (Israelian, Lobel, \& Schmidt 1999). There are presently no indications for 
a companion star in $\rho$ Cas which, in addition to its visual brightness, makes it an important target for
long-term spectroscopic monitoring to investigate the exceptional stellar atmospheric dynamics.    

Of the bright cool hypergiants we investigated spectroscopically over the past decade,
the spectrum of $\rho$ Cas appears similar to RW Cep in late 1998, 
displaying comparable spectral properties mentioned above. A number of fainter yellow Ia-supergiants
as HD 12399 and HD 179821, are spectroscopically even more compatible.   
RW Cep is classified as a semi-regular SRd variable, although the SRd class
covers a variety of variable star types (Percy \& Kolin 2000). $\rho$ Cas is 
too luminous to be related to the majority of low-mass Asymptotic Giant Branch stars in this class, 
although its quiescent small light amplitudes and spectral properties suggest kinship, 
which could result from its advanced evolutionary stage.  
  
Surprisingly, the spectrum of $\rho$ Cas lacks core emission in 
the broad near-UV Mg~{\sc ii} and Ca~{\sc ii} resonance 
absorption lines (\textsection\, 7.3), although it is so prominently observed in the less luminous F- and G-type 
(Ib) supergiants (i.e. Lobel \& Dupree 2000a). Variable and temporal emission line wings observed in H$\alpha$ 
may signal weak quasi-chromospheric activity (de Jager, Lobel, \& Israelian 1997) 
during phases of enhanced mass-loss with $\dot{M}$ $\geq$ $10^{-5}$ $\rm M_{\odot}\,yr^{-1}$.
Observational evidence for a permanent kinetic temperature minimum in its  
extended outer atmosphere ($\sim$10 \% of $R_{*}$=400$\pm$100~$\rm R_{\odot}$) is, however, presently not available.    

Jura \& Kleinmann (1990) presented a model for the 
weak silicate emission feature observed at 9.7 $\mu$m with $IRAS$ in 1983. They argued  
that dust formed at large distance from the star between 1973 and 1983, which
could be the consequence of an outburst event observed in 1945-47. During the event 
many zero-volt excitation energy lines appeared, not previously observed in the 
absorption spectrum. These atomic lines, in fact normally observed in M-type supergiants,
were strongly blue-shifted, signaling the ejection of a cool circumstellar gas shell (Beardsley 1961). 
In the following two decades after this event, a number of noteworthy papers were published
in the astrophysical literature, discussing $\rho$ Cas' peculiar spectroscopic changes in the 
years during and after the outburst (Popper 1947; Gaposchkin-Payne, Gaposchkin, \& Mayall 1947; 
Keenan 1947; Thackeray 1948; Tai \& Thackeray 1948; Beardsley 1953; Bidelmann \& McKellar 1957). 
Between 1945 and 1946 the star rapidly dimmed and developed optical and near-IR TiO bands. 
Within a couple of years, the hypergiant brightened up by 
nearly a full magnitude, and a mid G-type spectrum was recovered around 1950. 
Beardsley (1961) discusses a possible brightness minimum during 1893, 
which could be the first reported outburst of $\rho$ Cas. More recently,
Boyarchuk, Boyarchuk, \& Petrov (1988) also reported the detection of TiO bands in the 
optical spectrum during the moderate outburst of 1985-87 (\textsection\, 10). 

This paper presents a detailed spectroscopic study of the recent outburst in $\rho$ Cas 
in 2000-01. Our observations presently cover over 8.5 years, beginning in 1993, of continuous 
monitoring with high spectral resolution, including 
the recent outburst event. Throughout the paper we adopt the term `outburst' or `eruption' 
to indicate the variability phases of mid 2000 when newly formed TiO bands are observed 
in the spectrum. The terminology of `shell or mass ejections', which has been used in the 
literature for these dramatic spectral changes is not adopted here, because we presently 
also report high-resolution observations of migrating discrete absorption features 
in the far violet extended wings of atomic lines during other variability phases. 
Section 2 discusses the spectroscopic observations and 
data reduction procedures. 
Section 3 compares, for the first time, long-term radial velocity observations
with photometric monitoring, and we can therefore assign the peculiar spectral evolutions 
to different variability phases (\textsection\, 4 \& 5). We observe, for example, that a prominent emission 
line spectrum develops during phases of fast expansion of the photosphere. 
A discussion of the spectral changes during the 2000-01 outburst is provided in \textsection\, 7.
We analyze prominent TiO bands newly detected in the spectra, before and during the outburst minimum. 
From these measurements we compute in \textsection\, 8 the gas mass-loss rate during the eruption, 
for a semi-empirical model of the supersonic expanding photosphere. 
Based on these observations we propose in \textsection\, 11 a thermal mechanism that drives 
the eruptions of $\rho$ Cas. New equations are presented that correctly predict the time-scale for 
the outbursts. We further demonstrate numerically that these 
atmospheric explosions are driven by the release of ionization-recombination energy 
of hydrogen during the expansion, which causes $T_{\rm eff}$ to decrease from $\sim$7250 K 
to below 4000 K (\textsection\, 6). A discussion of stellar outburst physics is provided in \textsection\, 12.
The conclusions of this investigation are summarized in \textsection\, 13.

\section{Observations and Data Reductions}

\subsection{Optical Line Profile Monitoring}

Table 1 lists our high-resolution spectroscopic observations of $\rho$ Cas between 1993 January 29 
and 2002 July 7. A total of 78 spectra were observed on different nights with the 
Utrecht Echelle Spectrograph (UES) at the Nasmyth focus of the 4.2 m William Herschel 
Telescope at the Observatorio del Roque de los Muchachos (ING-WHT, Canary Islands), 
the Sofin spectrograph at the Cassegrain focus of the 2.6 m Nordic Optical Telescope (NOT, Canary Islands), 
the echelle spectrographs of the 1 m Zeiss-1000 telescope of the Special Astrophysical Observatory of the 
Russian Academy of Science (SAO/AN, Russia), and the fiberfed echelle spectrograph of the 1 m Ritter Observatory telescope (Toledo, OH). 
One echelle spectrum was recently obtained with the MuSiCoS echelle spectrograph of the 2 m Bernard Lyot Telescope 
at Pic du Midi Observatory (PMO, France) (S. Bagnulo 2002, priv.\ comm.), and an additional two
coud\'{e} spectra were observed with the 2.6 m Shajn telescope of the Crimean Astrophysical Observatory (CrAO, Ukraine)
(A. Rosenbush 2002, priv.\ comm.). The absorption line monitoring program includes H$\alpha$, Na~$D$, 
the single and unblended Fe~{\sc i} lines around 5570~\AA\, of the Revised Multiplet Table (Moore 1945) (RMT) 686, 
the split Fe~{\sc i} lines around 5500 \AA\, (RMT 15), and the Ba~{\sc ii} lines at $\lambda$5853 and $\lambda$6141 (RMT 2).

The wavelength coverage and nominal wavelength dispersions of the spectra are listed in columns [7]
and [9] of Table 1, respectively. The majority of these echelle spectra are co-added from separate exposures 
during the same night, with exposure times provided in column [10]. 
The typical continuum signal-to-noise ratios (S/N) therefore exceed 100 with WHT-UES (mostly obtained from the 
service time programs in good weather conditions), while some of our dedicated observation runs with 
NOT-Sofin yielded exceptional spectra with S/N$>$300, dependent on the quality of the seeing conditions. Two very 
high-resolution ($R$=170,000, but lower S/N) spectra were observed at SAO/AN in 1997 February and March, 
while three similar spectra were obtained in 1998 October and November with NOT-Sofin. These spectra 
cover only very limited wavelength portions, with broad echelle gaps, because they focus on the detailed 
profiles of the selected absorption lines. Five of the WHT-UES spectra do not include H$\alpha$ because the data 
were obtained with the 18-order echelle E79 grating to improve the background measurements, or with a shorter 
central wavelength setting to investigate the near-UV spectrum (Israelian et al. 1999)
or the Ca~{\sc ii} resonance lines. In 1998 October a number of NOT-Sofin spectra were observed with 
different spectral resolutions on consecutive nights, to investigate the influence of the wavelength 
dispersion on the detailed line shapes, and to also extend the free wavelength range to beyond 
1~$\mu$m. Unlike HR~8752, we do not detect He~{\sc i} $\lambda$10830 in $\rho$ Cas. 
The line is also absent in the near-IR spectra of 1999, and in the most recent 
observations of 2002 February. 

The WHT-UES spectra were reduced with the standard {\sc IRAF}\footnote{IRAF is distributed 
by the National Optical Astronomy Observatories, which is operated by the Association of Universities
for Research in Astronomy, Inc., under contract with the National
Science Foundation.} procedures.  These include the average bias subtractions, order definition
based on flat fields, removal of cosmic ray hits, and average flat division.     
A polynomial fit to the inter-orders is used to remove the background.
The wavelength calibration is performed with a Th-Ar hollow-cathode lamp spectrum, and the pixel-to-wavelength 
mapping is iterated by increasing the number of calibration emission lines 
in the echelle order solutions to accuracies typically better than 0.02~\AA.
 
The NOT-Sofin spectra were also calibrated with a Th-Ar source (or lamp) spectrum obtained
after each stellar exposure, to eliminate any temporal changes in the spectrograph 
in a combined wavelength solution with the use of the {\sc 4A} software package (Ilyin 2000)\footnote{I. Ilyin's  
{\it Acquisition, Archiving, and Analysis Software Package User's Manual} 
is available at {\tt http://www.not.iac.es/newwww/html/sofin.html}.}. 
It also includes a 2D-spline fit to remove scattered light, and weighted spectral order 
extraction which excludes the cosmic spikes from the fit.  
The sum of numerous master flat fields with overlapping orders is used to correct 
the pixel-to-pixel noise for the high S/N echelle images.
The exposure times for the high-resolution observations ($R$$\geq$80,000) have been 
increased to exceed 100, while values above 50 are obtained for the lower resolution observations.
The SAO/AN spectra were reduced with the {\sc DECH} code (Galazutdinov 1992),
which performs the standard echelle reduction steps. Most of these 
spectra have a S/N$\geq$100 per resolution element. Comparison Th-Ar 
exposures are obtained for the wavelength calibrations, although some of the 
dispersion solutions were also based on sharp telluric lines.   
The S/N of the Ritter spectra is between 100 and 200.
Further details about the Ritter observations
and data reduction procedures are given in Morrison et al. (1997). 

The mean accuracy of our wavelength calibrations for the spectra of Table 1 
is better than $\pm$1~$\rm km\,s^{-1}$. Note that these echelle spectra are not calibrated in 
the absolute flux scale, and that the placement of local continuum levels is 
either visually estimated, or results from a detailed spectral synthesis (\textsection\, 7).  

\subsection{Radial Velocity Monitoring}

Separate from the line profile monitoring, 88 radial velocity observations were obtained between 1990 July 10 and 2002 February 7
at the Oak Ridge Observatory (ORO of the Smithsonian Astrophysical Observatory) 
with the 1.5-m Wyeth reflector and an echelle spectrograph.
A single echelle order was recorded using a photon-counting intensified Reticon 
detector at a central wavelength of 5187~\AA, with a spectral coverage 
of 45~\AA. The strongest features present in this window are the lines 
of the \ion{Mg}{1}~b triplet.  The resolving power is $R\approx
35,\!000$, and the nominal S/N ranges from about 25 to 100 per resolution 
element of 8.5~km~s$^{-1}$. Wavelength calibration was performed in the 
standard way with Th-Ar lamp exposures taken before and after each stellar
exposure. The rms residual, from the 30$-$35 calibration lines used in the
dispersion solution, was typically around 0.01~\AA. 
The stability of the zero-point of our velocity system was monitored
by means of exposures of the dusk and dawn sky, and small systematic
run-to-run corrections were applied with the methods described by
Latham (1992). Internal errors for each of our velocities were
computed using the {\sc IRAF} task {\sc XCSAO} (Kurtz \& Mink 1998), 
and range mostly from 1.5~km~s$^{-1}$ to 2.0~km~s$^{-1}$. 

In the next two sections we investigate the radial velocity changes 
obtained from individual line profiles in the high-resolution spectra. 
A comparison with the radial velocity monitoring at ORO 
is provided in \textsection\, 5.  

\section{Light Curve vs. Radial Velocity Curve}

In Figure 1 we compare the radial velocity curve with the $V$-brightness
curve observed over the last 8.5 years. The solid dots are 
photo-electric measurements offered in Percy, Kolin, \& Henry (2000).
Their high-precision $V$, $R$, and $I$ observations are complemented with 
visual brightness estimates from the AAVSO ({\em solid boxes}).
Visual brightness estimates during the outburst between JD 2451100 and JD 2452400
are offered by the AFOEV (French Association of
Variable Star Observers)\footnote{\tt http://cdsweb.u-strasbg.fr/afoev/.}.
The light curve shows how the star dims by $1^{\rm m}.2$ $-$ $1^{\rm m}.4$,
from an unusual high brightness maximum of $V$$\sim$$4^{\rm m}.0$, to a 
deep minimum between 2000 September and December. In the half 
year after the deep minimum the star rapidly brightened up 
to average $V$-magnitude values observed before the outburst, however 
recently assuming another very bright visual maximum in late December 2001.
The lower panel of Figure 1 shows the radial velocity curve ($V_{\rm rad}$) of the photosphere
that we measure from the bisector at half intensity minimum of the 
unblended Fe~{\sc i} $\lambda$5572 absorption line ({\em solid dots}).
These values have been added after JD 2450100 (1995 December) to the 
radial velocity curve in Figure 8 of Lobel et al. (1998). The dotted 
line is computed from a linear interpolation of the Fe~{\sc i}
line {\it profiles} between consecutive observations, marked by the vertical lines.
Note that the detailed changes of the line shape also contain information about the 
intensity changes between the observations, but which would not 
be represented with a simple spline fit to the radial velocity data. 
Although usually small, the difference is appreciable during phases when the 
line rapidly becomes much deeper and more asymmetric, while shifting blueward, 
as we observe during the outburst. 
The numbers correspond to the spectrum numbers of Table 1.     
The monitoring suggests that during the quiescent variability phases before JD 2451000 
the radial velocity variations with a (rapid) decrease of absolute velocity 
are followed by a velocity increase that lasts considerably longer. 
After this time the unusually large velocity variations observed during the 
pre-outburst and the outburst cycle become rather symmetric.

For comparison the $V_{\rm rad}$-curve is overplotted with the light
curve ({\em dotted line in upper panel of Fig. 1}). The 
star becomes brightest for phases soon after the photosphere begins to expand.
The stellar center-of-mass velocity of $-$47~$\rm km\,s^{-1}$ was determined from the
line bisector at half maximum of the forbidden [Ca~{\sc ii}] $\lambda$7323 
emission line, which was prominent in 1993 December (Lobel 1997, p. 53). 
We observe four phases between JD 2449300 and JD 2451000
with rapid brightness decreases that lag the fast (absolute) radial velocity decrements by $\sim$3.5 months. 
The radial velocity minima, marking maximum photospheric 
expansion, do not coincide with the $V$-brightness minima.
The phase lag is clearly observed during the pre-outburst variability
cycle of JD 2451000$-$2451500, and during the actual outburst.
During the outburst, the $V_{\rm rad}$-values decrease from 
$\sim$$-$30~$\rm km\,s^{-1}$ to $-$65~$\rm km\,s^{-1}$ in $\sim$200 d,
which is followed $\sim$100 d later by a steep decline of the $V$-magnitude over the same period of time.    
We observe a strong correlation between the velocity and brightness amplitudes 
during the past 8.5 years. These observations show that 
the flow dynamics of the photospheric layers where the Fe~{\sc i} $\lambda$5572 
line forms are strongly related to the overall thermal conditions of the atmosphere 
($T_{\rm eff}$ and $\log{g}$) in which the stellar $V$-band continuum is produced, 
near the maximum of the spectral energy distribution of this cool star. It signals 
that the global photospheric dynamics due to pulsation causes the 
thermal conditions of the atmosphere to vary with time.     

\section{Line Profile Evolution}

Figure 2 shows the detailed line profile changes of H$\alpha$, 
Fe~{\sc i} $\lambda$5572, and Fe~{\sc i} $\lambda$5506 
({\em panels left to right}). The line profiles are plotted in 
heliocentric velocity scale and have been normalized to 
the local continuum level, indicated by the right-hand
horizontal line for each observation night. 
The stellar continuum level is estimated with a polynomial 
fit through wavelength regions of maximum intensity in the 
vicinity of the line, which remain sufficiently `flat' with the spectral changes,
and indicate the absence of weak absorption features.
The Fe~{\sc i} $\lambda$5572 line has been continuum normalized through 
the flat intensity maxima beyond the line wings, while H$\alpha$ 
requires a local continuum region around $-$320~$\rm km\,s^{-1}$  
because its violet wing sometimes blends with Ti~{\sc ii} $\lambda$6559.
The split Fe~{\sc i} $\lambda$5506 line requires even farther continuum 
definition points because the violet line wing is more extended, 
while the red line wing is weakly blended. The same continuum fitting  
points have systematically been applied for a consistent measurement 
of equivalent width changes in H$\alpha$ and Fe~{\sc i} $\lambda$5572 (\textsection\, 6),
and for the radial velocity measurements from the line bisectors.            
The vertical dotted line is drawn at the stellar rest velocity of 
$-$47$\pm$2~$\rm km\,s^{-1}$.

For an improved visualization of the detailed line profile evolutions we plot 
dynamic line spectra in Figure 3. 
The observed line profiles are 
linearly interpolated between consecutive observation nights ({\em left-hand
tickmarks}).  While the velocity amplitudes and overall line profile
evolutions in Fe~{\sc i} $\lambda$5572 and Fe~{\sc i} $\lambda$5506 
are similar, the H$\alpha$ line shape and velocity changes are notably different.  
The difference signals different regions for the mean formation 
of H$\alpha$ and the Fe~{\sc i} lines. The absorption portions of the 
Fe~{\sc i} lines are formed around photospheric levels of $-$2$\leq$$\log{\tau_{\rm Ross}}$$\leq$0, 
and these low-excitation energy lines can develop extended violet 
wings, produced in a cool optically thick expanding stellar wind at smaller optical depths.
The body of the Fe~{\sc i} lines therefore tracks the photospheric movements
over time, while H$\alpha$ forms higher, over a much more extended 
region of the upper atmosphere. The extended H$\alpha$ envelope of the 
hypergiant produces broad emission line wings at either or both sides
of the central absorption core due to photon scattering in our line of sight.    
On the other hand, the dynamic H$\alpha$ spectrum also reveals large 
Doppler shifts between 1993 and 1997. This strongly suggests
that the H$\alpha$ absorption core is rather optically thick, and
results from radiation transfer through a moving line formation region, 
in front of the stellar disk, where photons are removed from or scattered 
out of our line of sight. 

The outburst of 2000-01 is characterized by the strong 
blue-shifts of the Fe~{\sc i} lines in the upper portions of Figs. 2 \& 3. 
These neutral lines become much deeper and broader than usual 
in the spectra of summer 2000. Note that the split Fe~{\sc i} $\lambda$5506 line has 
a central emission reversal which remains around the stellar 
rest velocity, while a far extended violet wing unfolds
during the outburst. Our monitoring also reveals that many of the split lines 
in the optical spectrum of $\rho$ Cas develop discrete and weak absorption features 
at high velocity in the violet line wing. These features migrate toward the central emission 
core over a long period of time (several years). It indicates the regular formation 
in the supersonic stellar wind of circumstellar gas `shells' that gradually decelerate 
and dissolve. A more detailed discussion of the split line features  
will be given elsewhere. A very remarkable aspect of the dynamic spectra 
is that the H$\alpha$ core does not display the strong blue-shifts
observed in the photospheric lines during the outburst. Instead,
the H$\alpha$ absorption core becomes very weak during the deep brightness minimum. 
A comparison of the shape changes with H$\alpha$ in Betelgeuse and $\mu$ Cep is 
discussed in \textsection\, 7.2.                

For an improved comparison of the detailed evolutions of the three line profiles 
with the stellar brightness changes, we convert the dynamic spectra
into a movie sequence\footnote{\tt http://cfa-www.harvard.edu/$\sim$alobel.}. 
The day-to-day changes of the line shapes are shown together with the $V$-magnitude curve ({\em lower panel}). 
The movie reveals how the H$\alpha$ absorption core strongly red-shifts
between 1995 and early 1999 (around JD 2451200), following the trend of the Fe~{\sc i} lines.
After this time, the star rapidly brightens up to the very bright maximum of JD 2451600 (early 2000), leading up to the outburst event.
During this pre-outburst phase unusual strong emission develops in the blue wing of H$\alpha$
(the white emission `spot' in Figure 3 is at maximum normalized intensity of 25\% above the continuum level), 
while the red-shifted absorption core becomes very broad and extended, indicating the collapse of the 
H$\alpha$-envelope in our line of sight. However, during this pre-outburst cycle the Fe~{\sc i} lines 
also strongly red-shift, indicating a similar collapse or compression of the deeper photosphere. 
The combined spectral and photometric monitoring indicates
that the outbursts of the hypergiant are preceded by phases of unusual fast contraction 
of the upper and lower atmosphere, during which the star quickly brightens up and strongly increases 
its $T_{\rm eff}$ (\textsection\, 6).

We note that a comparable strong emission increase of the blue H$\alpha$ wing has more 
recently been observed during a new very bright maximum around JD 2452270 (2001 December). 
The very extended red absorption core of H$\alpha$ observed during this period, followed by
fast blueshifts of Fe~{\sc i} in early 2002, could suggest that a new outburst of $\rho$ Cas
is imminent. This appears supported by the fact that in 2002 July 07 an unusual H$\alpha$
profile was observed with a very extended red absorption wing. The profile is shown      
in the left-hand panel of Figure 2 ({\em top}). Most remarkably, the core of the line is split, 
and its large broadening has not been observed before. The recent evolution of the 
Fe~{\sc i} $\lambda$5572 is also strikingly similar to the changes observed before the 2000-01
outburst. If the star brightens up to a new very large brightness maximum over the 
next half year a new, and possibly more dramatic, outburst event could be expected.
        
\section{Comparison of Radial Velocity Curves}

The radial velocity curves of Fe~{\sc i} $\lambda$5572 
({\em filled circles}) and H$\alpha$ ({\em triangles}) are compared in Figure 4. 
These velocity values are measured from the absorption line bisectors, and listed in Table 2. 
The box symbols show the ORO radial velocity values, also listed in Table 3. 
These velocities are obtained from a cross-correlation technique applied to
ORO spectra observed in the 5166 \AA $-$ 5211 \AA\, wavelength range.
The template for $\rho$~Cas is selected from a library of pre-computed synthetic
spectra based on the latest model atmospheres by R. Kurucz
\footnote{Available at: {\tt http://cfaku5.harvard.edu}.}
(Nordstr\"{o}m et al. 1994), covering a large range of stellar
parameters and calculated for the precise wavelength window and
resolution of the spectra. As described in more detail below, the
effective temperature of $\rho$~Cas varies considerably over the
pulsation cycles. For the purpose of deriving radial velocities from
these spectra, however, we adopt a fixed temperature of 6250~K for
our template. Limited changes around this value have only a small effect 
on the results. Similarly, to match the line broadening of $\rho$~Cas we
adopt a $v \sin i$-value of 40~km~s$^{-1}$, as well as a surface
gravity of $\log{g}$ = 2.5 (the nearest value in our template library)
and [m/H] = 0.0.  The effect of the gravity and metallicity
parameters on the radial velocities is negligible. The one $\sigma$-error
computed from the cross-correlation method is also shown in Figure 4, and 
listed in column [5] of Table 3. 

There is good correspondence between the radial velocity values 
obtained from the bisector of the Fe~{\sc i} $\lambda$5572 line in the 
high-resolution spectra, and from cross-correlating the ORO spectra.
Both methods independently yield the large Doppler shifts observed 
during the pre-outburst cycle and the outburst. The velocity 
curves agree to within 5~$\rm km\,s^{-1}$. Only for the observations 
around JD 2449600 does the cross-correlation yield $V_{\rm rad}$-values
of 9$-$10~$\rm km\,s^{-1}$ smaller than obtained from the Fe~{\sc i } line 
bisector measurements. A detailed comparison of the ORO spectra of this epoch,
with the high-resolution spectra in the same wavelength range, shows that 
the difference results from the many deep and broad absorption lines
in this part of the spectrum. During this epoch, the deep lines develop 
enhanced absorption line wings, indicating a variability phase with stronger wind 
expansion. The deep lines are more asymmetric toward shorter wavelengths     
than the weaker Fe~{\sc i} $\lambda$5572 line. Consequently, the cross-correlation method  
samples more of the outflow by the larger wind opacity in the strong and 
broad lines. We therefore think that the Fe~{\sc i} radial velocity curve is more sensitive to
the oscillations of the deeper photosphere, because absorption 
contributions from extra opacity in the variable wind are less for this weaker line. 
This is further supported by the behavior of the Fe~{\sc i} radial 
velocity curve which closely matches the $V$-brightness changes (Fig. 1), 
although the line was monitored by three different spectrographs during this epoch.       
Stronger blue-shifts for the deep absorption lines were for example measured from the bisector
of the Y~{\sc ii} $\lambda$4900 line during this epoch (Fig. 8 of Lobel et al. 1998).
Note further that during the pre-outburst and outburst cycles the entire Fe~{\sc i} line profile, 
and the absorption line spectrum, strongly Doppler shift (Fig. 2), which explains why the 
measurements from both methods agree within a few $\rm km\,s^{-1}$ during these events.    

We also computed the radial velocity changes based on the method of line moments. 
The moments of a line are defined by $M_{i}= \int_{\rm line}{ (v-v_{0})^{i}\, (1-F(v))\, dv}$, 
with $v_{0}$ the stellar rest velocity, and $F(v)$ the normalized line flux in the velocity scale. 
The center of gravity velocity of the Fe~{\sc i} $\lambda$5572 line is computed
from the first moment $M_{1}$ with $v_{1}=M_{1}/M_{0}+v_{0}$, where the zeroth moment 
$M_{0}$ is the equivalent width of the line. The radial velocity values determined from the 
line half width at half intensity minimum in Figure 4 are almost identical 
to the $v_{1}$-values. The open circles plot the 
velocity variations of the line determined from the second moment $M_{2}$ with $v_{2}=-\sqrt{M_{2}/M_{0}}+v_{0}$. 
The periodicity and phase of the $v_{2}$-curve are very similar to the radial velocity curve 
({\it solid dots}), which indicates that the higher moments of the Fe~{\sc i} line do not 
reveal photospheric oscillation modes of a higher frequency.
The velocities determined from the higher line moments do not show correlations with the 
$V$-brightness curve, which is unlike the strong correlation observed with the zeroth (\textsection\, 6) and 
the first moment. We think that the method cannot detect possible higher oscillation frequencies, 
based on detailed profile changes from the higher line moments, because the photospheric lines of 
the hypergiant are broadened by supersonic macrobroadening velocities (of $\sim$21~$\rm km\,s^{-1}$), 
which strongly increases the line breadth. Small changes of the line shape, caused 
by intrinsic variability, when integrating over the stellar disk (e.g. due to possible
non-radial atmospheric movements), therefore become smoothed out by the overall large line broadening.
It is however also possible that the moment method cannot detect higher oscillation frequencies
because the time resolution of our dataset appears too coarse for shorter time intervals. 
The application of the line moment method is further complicated by the fact that our dataset 
is not fully homogeneous (four spectrographs were used) in both S/N and spectral resolving power, 
which influences the velocity curves obtained from the higher moments.

A remarkable aspect of the radial velocity curve is that the photosphere 
reveals an average variability period of $\sim$300 d between consecutive velocity maxima from  
JD 2449200 to JD 2450450, while the H$\alpha$ absorption core only strongly red-shifts
during this period of time. It signals a strongly velocity stratified dynamic atmosphere in which 
the deeper photospheric layers oscillate at a higher frequency, indicating smaller 
mean radius displacements than the upper H$\alpha$-atmosphere. After this period, 
both atmospheric regions strongly red-shift, indicating a global collapse 
(by $\sim$13~$\rm km\,s^{-1}$ from the stellar rest velocity) during less      
than 200 d. This phase is followed by a blue-shift of the photosphere during 
which the moderate brightness maximum of JD 2450750 occurs. However, H$\alpha$ 
remains strongly redshifted, indicating a continuous collapse of the upper 
atmosphere. Over the next variability phase the photosphere strongly 
blue-shifts during the pre-outburst cycle. This phase is followed by the 
development of a very extended red line wing in the absorption core of H$\alpha$,
together with an unusual strong redshift of the photosphere (by $\sim$18~$\rm km\,s^{-1}$ 
from the stellar rest velocity), which precedes the very large brightness
maximum of JD 2451600, and the subsequent outburst minimum. These observations 
indicate substantial changes in the thermal conditions of the stellar ($V$) continuum formation region
with the growing pulsation amplitudes. The amplified photospheric oscillations lead up to the 
outburst event, which we discuss in the next section.

\section{Changes of Atmospheric Conditions}

Figure 5 compares the visual light curve with the equivalent width ($W_{\rm eq}$) measurements 
of Fe~{\sc i} $\lambda$5572 ({\em open circles}), listed in column [4] of Table 2.  
The $W_{\rm eq}$-values range between 282 m\AA\, and 642 m\AA, with typical errorbars of
$\pm$15 m\AA. The errors are computed by placing the continuum level for the 
line intensity normalization 1\% higher and 1\% lower. Similar error values are obtained 
for H$\alpha$ ({\em crosses}; net equivalent width),
and for Fe~{\sc ii} $\lambda$5325 ({\em open triangles}), also listed in column [8] of Table 2. 
We find a very strong correlation between the $V$-magnitude and the $W_{\rm eq}$-value of the Fe~{\sc i} line.
It shows that this neutral absorption line is very sensitive to variations of the local
excitation conditions, caused by changes in the thermal circumstances of the deeper 
photosphere with pulsation, as we discussed in \textsection\, 3 from the radial velocity curve 
of the line. We find a tight relationship between the $V$-magnitude, and $W_{\rm eq}$(Fe~{\sc i} 5572 \AA) 
expressed in m\AA:
\begin{equation}
V = 1.88\,\times\,10^{-3} \times W_{\rm eq} + 3.75 \,\, {\rm mag.}
\end{equation}    
  
On the other hand, we also find that the $W_{\rm eq}$-changes of the Fe~{\sc ii} line are 
anti-correlated with $V$. This reveals that the population of the energy levels of 
neutral and singly-ionized iron lines in the atmosphere of $\rho$ Cas is dependent 
of the variable local excitation conditions, and strongly sensitive of the abundance of the iron ions. 
In the low-density atmospheres of cool supergiants with $T_{\rm eff}$ around 7000~K,
the iron ionization balance is very sensitive to changes of the local thermal conditions, 
which causes the strength of Fe~{\sc i} and Fe~{\sc ii} lines to vary oppositely with 
small variations of $T_{\rm eff}$.

In the lower panel of Figure 5 we estimate the variation of $T_{\rm eff}$, based on the 
$W_{\rm eq}$-changes of the Fe~{\sc i} line with pulsation. The atmospheric conditions 
for the optical spectrum of 1993 December 20, and of 1995 April 18 have been determined 
in Lobel et al. (1998) from $W_{\rm eq}$ measurements of 23 Fe~{\sc i} and 11 Fe~{\sc ii} lines.
The spectrum of 1993 corresponds to $T_{\rm eff}$=7250$\pm$200 K, and the spectrum of 
1995 to $T_{\rm eff}$=6500$\pm$200 K, whereby $\log{g}$ decreases from 1.0 to 0.5,
from (near) maximum to minimum $V$-brightness. The decrease of $T_{\rm eff}$ corresponds to
an increase from 393 m\AA\, to 551 m\AA\, in Fe~{\sc i} $\lambda$5572, which we utilize
to estimate the $T_{\rm eff}$ for other variability phases by a linear interpolation:
\begin{equation}
T_{\rm eff} = 9115 - W_{\rm eq} \times 750\, /\, 158 \,\, {\rm K}\,.
\end{equation} 
The $T_{\rm eff}$-estimates from the $W_{\rm eq}$-values of the Fe~{\sc i} line are marked with 
open circles. We note however that the direct correlation of the equivalent width values with 
the $V$-curve of equation (1) breaks down for the deep outburst minimum observations of 2000 September 17 and 20. 
During the outburst, the strong decrease of $T_{\rm eff}$ causes the population of this neutral 
line to shift from the linear to the flat part of the curve of growth. In these conditions the 
line depth saturates and becomes no longer directly correlated with the steep brightness decrease. 
An estimation of $T_{\rm eff}$ during the outburst can therefore not be obtained from an 
extrapolation of the $\rm W_{\rm eq}$-values observed during the outburst. 

On the other hand, the strong correlation of the Fe~{\sc i} line with $V$ during the quiescent 
pulsation phases enables us instead to directly estimate $T_{\rm eff}$ from $V$, by inserting 
equation (1) in equation (2):
\begin{equation}
T_{\rm eff} = 9115 - 2525 \times (V-3.75)\,\, {\rm K}\,.     
\end{equation} 
The correlation reveals how $T_{\rm eff}$ varies by $\sim$3000 K during the outburst, 
from a very high value $\simeq$8000 K around JD~2451600 (2000 April), to a low 
$T_{\rm eff}$$\simeq$5000 K during the deep brightness minimum of 2000 September and December.        
The extrapolation with equation (3), based on $V$, applies because we further check from spectral
synthesis calculations that the spectra observed during the deep outburst minimum 
correspond to $T_{\rm eff}$-values below 5000 K. 

It is of note that the Fe~{\sc i} measurements after JD 2450250  
show a {\em decrease} of $W_{\rm eq}$ to a minimum value around JD 2450409. However,
the $V$-brightness estimates suggest an ongoing dimming of the star, which contradicts 
the strong $W_{\rm eq}$-$V$ correlation observed during other quiescent variability phases.
We think however that the visual brightness estimates during this period (spring term) could be
biased by the low altitude of the star for observers in the northern hemisphere.
This is likely the case because our $W_{\rm eq}$-values are measured from spectra of
two different spectrographs, yielding a continuous sequence with a local minimum around JD 2450409,
while the more accurate photo-electric $V$ measurements were only obtained after this date
(Fig. 5, {\em solid dots in the upper panel}).
This is further supported since the $W_{\rm eq}$-curve, with a local minimum around this epoch, also 
confirms the variability period of $\sim$300 d, that is observed between other
consecutive brightness maxima, and also noticeable in the $V_{\rm rad}$-curve (\textsection\, 5).
Alternatively, when taking all visual brightness estimates at face value it would indicate 
that the correlation equations are not always perfect for this semi-regular variable star.      

We further tested the large $T_{\rm eff}$ decrease during the outburst with the ORO spectra. 
The spectra are run through a grid of Kurucz models with 4500 K $\leq$ $T_{\rm eff}$ $\leq$ 8000 K.
For each of the spectra, the cross-correlation method selects a template spectrum that 
best fits the observed spectrum. The $T_{\rm eff}$ and line broadening are varied during 
the iterations, while the model gravity is set to a constant value. A rotational convolution
is performed for the line broadening, but without changing the line equivalent width 
(i.e. the microturbulence broadening is assumed to remain constant). This rather simplified 
method of determining atmospheric changes with phase yields however a clear 
correlation between $T_{\rm eff}$ and the line broadening, in the sense that the 
spectral lines are broader when $T_{\rm eff}$ is higher. The correlation is
very strong during the pre-outburst cycle and the actual outburst when $T_{\rm eff}$
strongly varies. The iterations reveal that $T_{\rm eff}$ increases
from $\simeq$6000~K on JD 2451371 to $\simeq$7600 K on JD 2451598 
(around the large $V$-brightness maximum before the outburst). After this period, 
the deeper photosphere rapidly expands, while the star quickly dims by more than a magnitude in $V$.
Around the brightness minimum of JD 2451800 we compute that $T_{\rm eff}$ 
assumes the smallest value in the model grid of 4500 K. The temperature 
minimum lasts for only $\sim$100 d, while the photosphere quickly contracts
from maximum expansion velocity to the stellar rest velocity (a $V_{\rm rad}$ decrease by 
$\simeq$18~$\rm km\, s^{-1}$). The star subsequently brightens up by $0^{\rm m}.8$ within
the next 100 d, returning to $T_{\rm eff}$ above 6000~K before JD 2452000. 

The rapid $T_{\rm eff}$ excursion by more than 3000 K for the outburst of $\rho$ Cas, 
obtained from both methods described above, indicates that it results from 
the exceptionally large and fast mean atmospheric displacements we observe before and during the event. 
In the next section we investigate the changes of the optical spectrum 
in more detail. We show that the above derived $T_{\rm eff}$-values are in 
close agreement with the detailed spectral synthesis models in which 
the gravity acceleration and the turbulence broadening are also varied. 
During the outburst minimum $T_{\rm eff}$ decreases so rapidly that 
TiO absorption bands can form, which are typically only observed in
M-type (or very late K-type) stars as $\mu$ Cep and Betelgeuse.

\section{Spectral Variability During Outburst}

\subsection{Optical Emission Lines}

We observe that during phases of fast photospheric expansion 
a prominent emission line spectrum develops with line intensity maxima above
the local continuum level ({\em vertical dashed lines labeled a and b in Fig. 1}).
The spectrum of 1993 December ({\em labeled a}) corresponds to a phase around maximum brightness
with $T_{\rm eff}$=7250 K, $\log{g}$=0.5, and a projected microturbulence 
velocity of $\zeta_{\mu}$=12~$\rm km\,s^{-1}$ (Lobel et al. 1998).  
Strong emission lines of low-excitation energy metals
as Fe~{\sc i}, Ni~{\sc i}, and Ca~{\sc i} have been identified
(see Lobel 1997, p. 45). The presently observed strong 
correlation of emission line appearance in the spectrum of $\rho$ Cas
with the radial velocity curve indicates that their 
formation is related to enhanced excitation in the accelerating outflow          
during very fast photospheric expansions. The latter is supported by 
the far violet extended absorption line wings we observe during 
these phases, signaling the development of an optically 
thick expanding wind from the supergiant when $T_{\rm eff}$ is high.
These very extended violet line wings were for example not observed 
in 1995 April, when $T_{\rm eff}$ decreased to 6500 K, and 
the star reached the $V$-brightness minimum around JD 2449820. 
Since the emission line spectrum remains static (and slightly blueshifted around the sound speed) 
it can emerge from a steady shock interface between the fast stellar wind that collides 
with denser circumstellar, or perhaps interstellar material, expelled during the 
violent (eruptive) mass-loss history of the hypergiant.

Figure 6 compares the spectral changes during the outburst
for a small wavelength region around the Fe~{\sc i} $\lambda$6358 line. 
This line is sometimes observed in emission, for example during the pulsation 
phase of 1993 December when the atomic absorption lines 
developed far violet extended wings, indicating a phase 
of enhanced mass-loss. During this epoch, several other 
permitted low-energy Fe~{\sc i} and Ca~{\sc i} lines were 
observed in emission (Lobel 1997, Chap. 4).

The high-resolution spectral monitoring between 
1994 and 1997 does not reveal prominent optical emission lines. 
The radial velocity curve (Fig. 1, {\em lower panel}) shows 
an overall decrease of atmospheric outflow during this period from
JD 2449400 to JD 2450700. However, in 1998 October (Fig. 6, {\em spectrum labeled b}) 
we detect the optical emission lines again above the stellar continuum level. 
Similar to 1993 December, the absorption lines develop extended violet 
wings due to enhanced opacity in the stellar wind. During these phases of 
increased $T_{\rm eff}$ with fast wind expansion, $\rho$ Cas' optical 
absorption line spectrum, mainly composed of neutral atomic species, becomes 
rather weak, and the peculiar emission line spectrum can therefore appear 
above the level of the continuum. 

The synthetic spectrum is shown by the upper dotted line in Figure 6.
The blend of Fe~{\sc i} and V~{\sc i} lines at 6358~\AA\,
becomes very weak, computed for a Kurucz model of $T_{\rm eff}$=7250~K 
and $\log{g}$=1.0. After 1998 October, $V$ dims by at least half
a magnitude during $\sim$5 months, where after the star brightens up
to the very bright maximum of 2000 April. 
The spectrum of 1999 May in Figure 6 ({\em solid line labeled c}), 
observed before this maximum, indicates $T_{\rm eff}$=5750 K and $\log{g}$=0.5 
from a best fit ({\em dotted line}). Hence, $T_{\rm eff}$ decreases by $\sim$1500 K 
after 1998 October, while the wind expansion strongly decelerates after early 1999
during the pre-outburst cycle. 

The high-resolution spectrum of 2000 July 19 in Figure 6 ({\em solid line 
labeled d}) is observed within a month after maximum photospheric expansion velocity. 
Note that the continuum level of the spectrum is shifted down for display purposes.
We observe that the entire absorption line spectrum displaces 
toward the shorter wavelengths. It reveals an average expansion 
of a large fraction of the deeper atmosphere, where the bulk of the photospheric 
spectrum forms. The synthetic spectrum fit ({\em middle dotted line}) shows that 
$T_{\rm eff}$ decreases to 4250 K with $\log{g}$=0. Hence, our detailed spectral modeling 
also reveals a decrease of at least 3000 K during the outburst of $\rho$~Cas.

A comparison with the spectrum of the M-type supergiants Betelgeuse 
({\em dashed line}) (with $T_{\rm eff}$=3500 K),
and $\mu$ Cep ({\em lower solid line}) in Figure 6 shows that 
the cores of neutral V~{\sc i}, Fe~{\sc i}, Ti~{\sc i}, 
and Cr~{\sc i} lines in $\rho$ Cas become considerably deeper during 
the strong $V$-brightness decrease. 
We model the outburst spectrum by neglecting molecular lines in 
the synthesis. The good fit reveals that molecular absorption remains weak, 
and the local continuum level is mainly determined by blends of atomic
lines. On the other hand, the synthetic spectrum 
of Betelgeuse is computed ({\em lower dotted line}) by including many molecular 
lines as TiO, $\rm C_{2}$, CN, etc. (Kurucz 1996). However, molecular opacity 
contributes substantially to the decrease of the local continuum level in 
Betelgeuse and $\mu$ Cep (to $\sim$0.5 in normalized intensity). 
Contributions from TiO opacity are very important for an accurate synthesis 
of these M-type spectra (for a discussion see Lobel \& Dupree 2000b). 
Our spectral observations and modeling indicate that the strong brightness 
decrease observed for $\rho$ Cas between 2000 April and October 
corresponds to a decrease of $T_{\rm eff}$ by at least 3000~K, and probably 
by as much as 3500~K to the deep brightness minimum of late 2000.
In less than half a year after the deep minimum the optical spectrum 
returns to $T_{\rm eff}$$\simeq$5750~K around 2001 February ({\em solid line labeled e}).      

\subsection{H$\alpha$ Absorption and Emission}

The upper panel of Figure 5 shows that 
the changes of the H$\alpha$ profile are uncorrelated with the variability of $V$.
However, we observe a strong decrease of the (total) equivalent line width, while 
the absorption portion of the line strongly red-shifts between JD 2449200 and 
JD 2450450 (see Fig. 3). During this period, the violet extension of the H$\alpha$ absorption core
diminishes (Fig. 2), indicating that the wind in the upper H$\alpha$-atmosphere 
stalls, and begins to collapse onto the deeper photosphere. The $W_{\rm eq}$-changes 
of H$\alpha$ also occur much slower than observed for the iron lines, which confirms 
the rather different mean formation region of H$\alpha$. 

Figure 7 shows a number of high-resolution spectra around H$\alpha$ for the epochs labeled 
{\em a} to {\em d} in Figure 1. We observe that the Ti~{\sc ii} absorption line, shortward of H$\alpha$, becomes
weaker in the outburst spectrum ({\em d}), due to a decrease of the excitation 
temperature in the line formation region. On the other hand, the neutral Fe~{\sc i} and V~{\sc i} lines,
also labeled in the graph, become deeper and assume intensities comparable to these lines observed
in Betelgeuse ({\em long dashed line}) and $\mu$~Cep ({\em long dash-dotted line}).
During the outburst the H$\alpha$ absorption core does not blue-shift, but
instead remains centered around the stellar rest velocity, while the absorption 
core becomes weaker. The line also develops weak asymmetric emission wings 
at both sides of the central absorption core, reminiscent of the strongly self-absorbed 
emission lines observed in the near-UV spectra of Betelgeuse, and of other cool luminous 
stars (i.e. Lobel \& Dupree 2001).  

It is interesting to note that laser pulse laboratory experiments in hydrogen 
plasma produce strong Balmer emission lines during ionization recombination.  
Fill et al. (2000) argue that the observed emission production time-scales and 
population balance calculations reveal that three-body recombination is the main 
mechanism populating the levels that emit the Balmer lines during these experiments. 
Their measurements involve however electron densities $n_{e}$$\simeq$$10^{17}$~$\rm cm^{-3}$
at gas pressures of $P$=$10^{3}$~$\rm dyn\,cm^{-2}$, while hydrostatic photospheric models
of $\rho$ Cas assume $n_{e}$$\leq$$10^{13}$~$\rm cm^{-3}$ for $P$$\leq$10~$\rm dyn\,cm^{-2}$.
Nevertheless, conditions of larger $n_{e}$ and $P$ can result by strong shock compression 
during the outburst, producing enhanced line emission in a cooling and expanding wake.
The emission wings observed for H$\alpha$ in the outburst spectra of $\rho$ Cas 
could therefore result from the recombination of partially ionized hydrogen gas.

The filling in of H$\alpha$ by line emission due to recombination during the outburst is also 
supported by the lack of evidence for a steady chromosphere in $\rho$ Cas. 
In Betelgeuse and $\mu$ Cep the stellar chromosphere strongly populates the H$\alpha$ transition, which
produces the deep H$\alpha$ absorption cores in these M-type stars. However, if these supergiants 
had no chromosphere, the H$\alpha$ absorption core would become invisible against the photospheric spectrum. 
This can be illustrated with a detailed synthesis of their optical spectra (which are mainly formed by
TiO opacity) ({\em dotted spectrum}), excluding the model for the chromosphere (see Lobel \& Dupree 2000b).
The outburst spectrum of $\rho$ Cas, with a comparable low $T_{\rm eff}$, shows however 
much weaker H$\alpha$ absorption with prominent emission line wings. It indicates 
that the line does not become excited in a stellar chromosphere. If a chromosphere were present during 
the outburst, much stronger H$\alpha$ absorption would have to develop. 
This would prevent the recombination of partially ionized hydrogen in the outer atmosphere, 
and hence would suppress the formation of emission line wings.           

The strongly self-reversed shape of the H$\alpha$ emission line during the outburst can be produced 
if the H$\alpha$ transition is sufficiently optically thick. Since we observe that the central 
absorption core remains almost static during the outburst, 
this would indicate that the central line scattering region is formed far above the H$\alpha$ emission formation region,
and does not share the dynamics of the outburst. This implies that $\rho$~Cas' H$\alpha$ atmosphere is considerably
more extended than the dimensions derived from radial displacements based on the photospheric radial velocity changes during 
the outburst (\textsection\, 8.2). Detailed semi-empirical modeling of the Balmer lines of 
$\rho$ Cas will be presented elsewhere. 

It is of note that our NOT-Sofin observations of the G2$-$G5 supergiant 
HD 179821 also show variable red- and blueshifted emission in the H$\alpha$ absorption core.
This supergiant has recently been proposed as a low-mass post-AGB candidate, 
based on a LTE abundance analysis by Th\'{e}venin et al. (2000).
Its far-IR excess signals a cool detached dust shell. There is however no 
consensus on the evolutionary status.
Its high expansion velocity of 34~$\rm km\,s^{-1}$ 
is much larger than the value of 10$-$15~$\rm km\,s^{-1}$ for low-mass
post-AGB stars, which rather suggests a massive supergiant. 
The emission humps in H$\alpha$ result from NLTE-effects in a fast and 
spherically expanding wind (Dupree et al. 1984), inside the dust cavity.
The emission sporadically appears above the stellar continuum level, as is 
observed in $\rho$~Cas.        

\subsection{Ca~{\sc ii} H \& K Absorption}

The absence of a permanent chromosphere in $\rho$ Cas is further supported by 
the spectral monitoring of Ca~{\sc ii} H \& K. Emission reversals in the cores of 
these resonance lines are classic indicators of chromospheric 
activity in cool stars. However, near-UV observations with WHT-UES (1996 December 26) and
NOT-Sofin (1998 October 6) do not show Ca~{\sc ii} emission in the broad absorption 
cores. In Figure 8 the Ca~{\sc ii} H \& K lines only reveal a deep central core with 
zero intensity, within the limited spectral noise levels. This is unlike the prominent 
core emission observed in the Ca~{\sc ii} K line of Betelgeuse ({\em dashed line lower panel}),
modeled in Lobel \& Dupree (2000b) with a kinetic temperature of the chromosphere not in excess of 5500~K. 
Note however that the broad wings of Ca~{\sc ii} H \& K show permanent emission lines 
of Fe~{\sc i} ({\em labeled RMT 45 and 43}).
Recent observations with NOT-Sofin (2002 January 29) of the Ca~{\sc ii} H line after 
the outburst of $\rho$ Cas neither reveal a central emission core ({\em dashed line in upper panel}).   
The absence of a steady chromosphere in $\rho$ Cas is also apparent 
by the absence of central emission in the Mg~{\sc ii} $h$ \& $k$ lines, discussed and 
modeled in Lobel et al. (1998). It is of note that we neither observe central emission 
in Ca~{\sc ii} K of HD 179821, indicating a massive Ia-supergiant like $\rho$ Cas. 

\subsection{Na $D$ Emission}

We observe a remarkable evolution in the split Na~$D$ lines of $\rho$ Cas 
during the 2000-01 outburst. The high-resolution spectra in Figure 9 show that the central 
portion of the lines, between the adjacent black absorption troughs,
turns in emission above the local continuum level ({\em solid lines labeled d}) 
on 2000 July 19 and August 2. The prominent central emission is also observed 
in the lower resolution spectra of September 17, 20, \& 21 (Table 1, profiles not shown in Fig. 9).
The lines are plotted in the heliocentric velocity 
scale, centered at the long-wavelength line of the doublet. 
Note that the central emission maxima do not coincide with the stellar
rest velocity of $-$47~$\rm km\,s^{-1}$ ({\em vertical dotted line}), 
but peak longward by $\sim$10~$\rm km\,s^{-1}$.
Another weaker feature is also observed shortward of stellar rest velocity in both lines 
of the doublet. 

It is of importance to point out that the contamination of interstellar absorption
in the Na~$D$ lines of $\rho$ Cas (at a distance of $\sim$3.1 kpc), and of 
other distant hypergiants, has always cast doubt on an interpretation which attributes 
the curious splitting to a central emission reversal. However, the high-resolution 
outburst spectra clearly show that central emission does (at least partly) involve 
the formation of these very complex resonance lines.

In order to investigate interstellar line contributions 
we compare with the Na~$D$ lines observed in HR~8752.
The thin solid lines of the small panel show both doublet lines of $\rho$ Cas 
in the same velocity scale, compared to these lines in HR~8752 ({\em thin dashed lines}),
observed with NOT-Sofin in 1998 October, with very high resolution ($R$$\sim$160,000).    
Both hypergiants are only 7$\arcdeg$\,22$\arcmin$ apart in the sky, and reveal narrow absorption 
components around $-$50~$\rm km\,s^{-1}$ and $-$30~$\rm km\,s^{-1}$  
(with a similar FWHM of $\simeq$10~$\rm km\,s^{-1}$). These distinct absorption features 
can therefore be attributed to high-velocity clouds of the interstellar medium (ISM)
in the direction of the Cas OB5 association. The interstellar absorption partly 
distorts the shape of the stellar central emission line, so that its actual intensity cannot 
accurately be determined. Moreover, we also observe two even narrower absorption features 
in both Na~$D$ lines of $\rho$ Cas. The features, with velocities around heliocentric 
rest velocity, can possibly be attributed to weaker contributions from the local interstellar medium 
(LISM).          

The increase of the central emission reversals above the local continuum level 
results from the decrease of the optical continuum flux with 
the decrease of $T_{\rm eff}$ during the outburst. The flux of 
the central emission is therefore independent of the photospheric conditions,
indicating that Na~$D$ emission emerges from an extended gas envelope around the 
hypergiant. We observe that the deep absorption portions of the 
Na~$D$ lines do not reveal significant Doppler shifts during the outburst,
similar as for H$\alpha$. These static lines are unlike the weaker 
photospheric absorption spectrum which strongly blue-shifts. However,
it should be remembered that the Na~$D$ lines are strongly intensity 
saturated. The absorption saturation distorts the line shape,
which renders the lines useless for radial velocity measurements.     
On the other hand, the violet wings of both lines are excellent 
indicators for enhanced opacity in an expanding wind. The stellar wind
is observed by the violet wing extensions during the outburst ({\em solid lines}), which
are stronger than in the spectrum before the outburst ({\em dashed line labeled c}).
Note for example that these violet line wings were very strong in the spectrum 
of 1993 December ({\em dash-dotted line labeled a}). In both cases extensions
out to $\sim$130~$\rm km\,s^{-1}$ from the stellar photospheric values can be 
identified.    

\subsection{TiO Absorption Bands}

We observe that TiO absorption bands develop in the outburst 
spectra of $\rho$ Cas in 2000 July, August, and September. 
Figure 10 compares the high-resolution spectra around 5450~\AA\, 
of 2000 July 19 and August 2, with the spectra of 2000 September 17 
and 20, and of 2001 February 5 and October 2 ({\em thin solid lines}). 
The spectra are shifted upward by 15\% of the normalized stellar continuum flux.
This wavelength region contains two split Fe~{\sc i} lines 
($\lambda$5446.9 and $\lambda$5455.6) of RMT 15. 
The cores of these Fe~{\sc i} lines are also observed in the spectra 
of Betelgeuse ({\em dashed line}) and $\mu$~Cep ({\em dash-dotted line})
where they do not split. The local normalized continuum level in the 
spectra of these M-type supergiants is strongly reduced (to $\sim$40\%) 
due to photospheric TiO opacity. 

In the summer of 2000 the local continuum level in the $\rho$ Cas spectra 
longward of 5447~\AA\, decreases by 10$-$15\% with respect to the mean continuum 
level shortward of Fe~{\sc i} $\lambda$5446.9. The intensity jump
results from the development of TiO ($C-X$) $\alpha$-bands of R(11) 0$-$1 transitions,
with bandheads at 5447.912~\AA, 5448.980~\AA, and 5450.729~\AA.
To demonstrate the influence of these bands on the atomic spectrum 
we calculate the TiO spectrum for a Kurucz model atmosphere with 
$T_{\rm eff}$=4000 K and $\log{g}$=0. The spectrum is computed 
with only TiO lines ({\em bold solid line}), using a macrobroadening velocity 
of 21~$\rm km\,s^{-1}$. The three TiO bandheads are visible, and 
overplot in the heliocentric frame. The overlapping 
bands extend and degrade longward over several tens of angstroms, where they merge 
with other TiO bandheads. The extended TiO line list was computed by Kurucz (1999), 
based on semi-empirical calculations by Schwenke (1998). The list includes lines 
from the five $\rm ^Z$Ti\,$\rm ^{16}O$ isotopomers ($Z$=46$-$50), with Earth 
abundance fractions.  
             
Although the TiO bands at 5447~\AA\, strongly develop in the outburst 
spectra of $\rho$ Cas, we found it difficult to compute good fits to 
determine the atmospheric parameters. This is because these bandheads blend
with strong atomic lines complicating our fit procedure.
The absorption feature observed between 5447~\AA\, and 5448~\AA\, (heliocentric wavelengths) 
in the summer of 2000 is due to TiO, but its shape appears rather distorted due to the adjacent 
atomic lines. We therefore identified other `cleaner' TiO bands that are closer 
to the stellar continuum level in the echelle spectra.                   
In Figure 11 we identify TiO bands from ($A-X$) $\gamma$-band R(16) 0$-$0 
transitions, with a prominent bandhead at 7052 \AA. The bands appear with 
stronger contrast in Betelgeuse, where they are permanent. 
The solid vertical lines  mark a number of sharp characteristic TiO features 
that are weakly blended with other atomic and molecular lines. 
The synthetic spectrum ({\em dotted lines}) strongly fits the complex 
TiO spectrum of Betelgeuse with an atmospheric model of 
$T_{\rm eff}$=3500 K and $\log{g}$=$-$0.5. The line list typically 
contains $\sim$1500 TiO lines per \AA, besides many other atomic and molecular species. 

The high-resolution spectra show that the distinct TiO absorption 
features around 7070 \AA\, are broader than sharp telluric lines (of water vapor and $\rm O_{2}$), 
also marked in Figure 11. We compute a best fit to the distinct TiO bands for a model 
atmosphere with $T_{\rm eff}$=3750 K and $\log{g}$=0. For comparison 
we plot the synthetic spectrum computed with only TiO lines 
({\em upper dotted line}). Models with higher or lower atmospheric 
temperatures yield intensities for these bands that are too weak 
or too strong, and do not match the relative intensity of the TiO features.
The spectrum is computed with solar abundance for TiO, and scaled to the observed 
spectrum. The best model obtained from the TiO synthesis is 500~K cooler than the 
photospheric model computed from the atomic lines in the outburst spectrum 
(Fig. 6). 

The upper panel of Figure 12 compares the high-resolution WHT-UES spectra of $\rho$ Cas
during outburst (2000 July 19 and August 2), and a spectrum observed 
with NOT-Sofin on 2002 January 29, after the outburst. The spectra show 1.2 \AA\,
around the TiO absorption band at 7069.2~\AA\, in the heliocentric wavelength scale.
The spectra are continuum normalized, and plotted without vertical shifts.    
The spectrum of Betelgeuse is also shown ({\em lower solid line}), together
with the synthetic spectrum of Figure 11 ({\em lower dotted line}). A microturbulence
velocity of 2~$\rm km\,s^{-1}$, and a geometric mean value of 12~$\rm km\,s^{-1}$ for 
the macrobroadening has been utilized in Betelgeuse (see Lobel \& Dupree 2000b). 
The vertical lines mark only the strongest TiO lines in the list with $\log{gf}$-values above $-$1.

The outburst spectrum of $\rho$ Cas of 2000 July is best fit after broadening 
the synthetic TiO spectrum ({\em upper dotted lines}) with a microturbulence velocity 
of 11~$\rm km\,s^{-1}$, and a mean value for the macrobroadening of 21~$\rm km\,s^{-1}$.
The broadening values are obtained after degrading the computed spectrum with the instrumental 
profile. The cross-correlation with the synthetic spectrum reveals that this 
TiO band assumes a heliocentric radial velocity of $-$82$\pm$2 $\rm km\,s^{-1}$. 
It signals an expansion velocity of the TiO line formation region by $\sim$15 to 
20 $\rm km\,s^{-1}$ faster than we observe for the atomic line spectrum
during the outburst. This indicates that the atomic lines form on average deeper in the 
photosphere, below the TiO formation region. 
The spectrum of 2002 January 29 (JD 2452303) is observed around the recent 
$V$-brightness maximum with $T_{\rm eff}$$>$ 6500 K. The weak TiO bands are 
therefore not detected. The spectrum also shows that this wavelength region 
is very close to the stellar continuum level, and does not 
reveal weak absorptions that could blend with the TiO band.       
Hence, we can use the band at 7069~\AA\, to reliably estimate 
the mean gas density in the TiO line formation region during the outburst.

\section{TiO Formation Region}

\subsection{Model Density}

The lower panel of Figure 12 shows that the TiO lines are formed in 
the selected model atmosphere over a range of optical depths with 
$-$3 $\leq$ $\log{\tau_{\rm Ross}}$ $\leq$0 ({\em vertical dashed lines}). 
Deeper in the atmosphere the kinetic temperature $T$ ({\em dashed line}) increases to 
above 4000 K, where the TiO molecules dissociate in the chemical equilibrium calculations.
On the other hand, TiO molecules do form at $\log{\tau_{\rm Ross}}$ $\leq$ $-$3,
but these outer layers are very optically thin and contribute little to 
the TiO absorption line formation. The TiO line contribution functions 
assume largest values in this model for 3000~K $\leq$ $T$ $\leq$ 4000~K, 
with negligible contributions outside the corresponding gas density regions. 

Hence, we compute that the density $\rho$ ({\em solid line}) 
in the TiO line formation region during outburst ranges between 
$-$9.8 $\leq$ $\log{\rho}$ $\leq$ $-$8.5. It is interesting to note that 
the limited range of gas densities, derived from modeling newly formed TiO 
bands, are useful to also estimate the gas mass-loss rate during the eruption. 
Further tests show that a mean formation density during the outburst is 
much harder to constrain from modeling atomic absorption lines. The atomic
lines develop strong asymmetries, with very extended blue line wings that 
form in a fast expanding low-density wind. The atomic lines thereby 
form over a much larger density range than the TiO bands. To infer the mass-loss rate
from modeling the atomic line profiles would require detailed 
(semi-empiric) time-dependent models of the velocity- and density-structure of 
the atmosphere during the outburst, which is outside the scope of the present investigation.

\subsection{TiO Mass-loss Rate}

Based on the outburst spectra we can assume a lower density limit for the TiO line 
formation region of $\rho_{\rm min}$ = $10^{-10}$ $\rm gr\,\,cm^{-3}$. 
The distance over which this region expands (assuming spherical geometry) 
is computed from the radial velocity we observe for the TiO bands. 
The star velocity of $-$47 $\rm km\,s^{-1}$ yields a mean velocity for the TiO envelope expansion 
of $-$82 + 47 = $-$35 $\rm km\,s^{-1}$. It corresponds to a travel distance 
of $d_{\rm s}$ = 869 $\rm R_{\odot}$ ($\sim$2 $R_{*}$) over a period of $\sim$200 d. 
The latter is the time which elapsed between the maximum and minimum 
of the radial velocity curve, observed before the deep brightness minimum (Fig. 1). 

During the fast expansion the temperature of 
the outer atmosphere rapidly plummets to below 4000 K, thereby continuously 
forming TiO molecules. The photosphere accelerates to very fast expansion
velocities ($\sim$35 $\rm km\,s^{-1}$), probably causing a circumstellar shock wave. 
During the brightness decrease the outer atmosphere expands over 
about twice the stellar radius. 
This corresponds to a total gas mass lifted above $R_{*}$
of $M_{\rm s}$=3$\times$$10^{-2}$ $\rm M_{\odot}$  (= 4 $\pi$ $R_{*}^2$ $d_{\rm s}$ $\rho_{\rm min}$) 
during 200 d. We thus compute a mass-loss rate of $\dot{M}$=5.4$\times$$10^{-2}$ $\rm M_{\odot}\, yr^{-1}$ 
during the eruption. This value is 587 times larger than 
the mass-loss rate of 9.2 $\times$ $10^{-5}$ $\rm M_{\odot}\,yr^{-1}$, derived by
Lobel et al. (1998) for the pulsation phase of 1993 December.
For comparison, during the 200 days the hypergiant sheds
the same amount of mass it normally would over a period 
of 3000 years by quiescent pulsation with $\dot{M}$ below $10^{-5}$ $\rm M_{\odot}\,yr^{-1}$. 

The value of $\dot{M}$=5.4$\times$$10^{-2}$ $\rm M_{\odot}\,yr^{-1}$  
we calculate for the 2000-01 outburst is a {\it lower limit} since the lower limit of the
density-range and $R_{*}$=400~$\rm R_{\odot}$ are adopted to estimate a spherical mass-loss rate. 
We emphasize that the combination of the density- and kinetic temperature-range obtained 
in \textsection\, 8.1 is required to correctly compute the intensity ratios observed 
across the newly formed TiO bands. Although we utilize a hydrostatic model atmosphere 
to perform the radiative transport calculation, this combination of density- 
and kinetic temperature-range, providing the correct relative band intensities, is also required in more 
realistic dynamic models. During the outburst the atmosphere likely assumes a very 
different density structure than that of a hydrostatic model, but the TiO fit procedure
will require similar mean gas densities and kinetic temperatures in an expanding atmosphere.
On the other hand, we should also mention that the geometric 
position of the TiO formation layers, having the required thermal conditions
in a hydrostatic model, is very likely different from the location that these 
layers will assume in a hydrodynamic model. Dynamic models that 
are more extended than static models (i.e. with TiO opacity at larger atmospheric radii), 
yield larger mass-loss rates in the TiO formation region. This is also the case when using
plane-parallel models instead of spherically symmetric models. 
The TiO bands computed in spherical symmetry yield smaller line depths 
(for the same atmospherical model parameters), because the light rays that graze 
the atmosphere escape earlier from a curved model than from a plane-parallel model. 
The best fit calculated with spherical geometry hence requires larger gas densities 
in the TiO formation region than for a plane-parallel model, assuming that the mean 
optical depth is not appreciably different for both types of models. The latter 
assumption is justified here because the best fit indicates that the optical depth of 
the weak TiO bands is large ($-$3$<$$\log{\tau_{Ross}}$$<$0), or that these
bands form at (sub-)photospheric levels where the bulk of the atomic line spectrum 
emerges. Larger differences are expected toward smaller optical depths with
$\log{\tau_{Ross}}$$<$$-$3, where the extended violet wings of the atomic lines are formed
in a much faster expanding wind. Best fits to the new TiO bands using more extended dynamic
and spherical models are therefore expected to provide larger $\dot{M}$-values than 
the lower limit we estimate from static plane-parallel models.  

The present outburst observations, together with the outbursts 
of 1893, 1945-47, and 1985, indicate that the eruptions of $\rho$ Cas 
occur on time-scales of about half a century. For an average 
mass of 3$\times$$10^{-2}$ $\rm M_{\odot}$ expelled during 
these events, the star would lose 20 $\rm M_{\odot}$ during a period 
of only 33,750 years (over 675 outbursts). Since the current 
mass of $\rho$ Cas is expected not to exceed 40 $\rm M_{\odot}$ from 
evolutionary calculations, and since this  
time span (or an order of magnitude longer) is comparable to 
the evolutionary life time during the post-red supergiant phase, 
we infer that the recurrent eruptions, causing punctuated mass-loss episodes,
are likely the major mass-loss mechanism for this massive hypergiant.

\section{Is $\rho$ Cas Bouncing Against the Yellow Hypergiant Void?}

De Jager \& Nieuwenhuijzen (1997) suggested that the recurrent eruptions 
in yellow hypergiants occur when these stars approach the cool boundary of 
the `yellow void', and bounce off redward. The evolutionary void is an 
area in the upper HR-diagram around $T_{\rm eff}$$\sim$10,000~K with $\log{L/L_{\odot}}$$\geq$5.7, 
which is nearly devoid of (white) supergiants. Their detailed calculations 
revealed that hydrostatically stable solutions cannot be computed 
for the atmospheres of {\em blueward} evolving yellow hypergiants at a 
low boundary of $T_{\rm eff}$$\simeq$8300~K. The effective acceleration 
becomes negative, or directed outward, causing an unstable atmosphere 
above 8300~K. In a review on hypergiants de Jager (1998) suggested 
that the mechanism for outbursts may be related to increased 
pulsation velocities during a period of instability when the star,
during rapid blueward evolution, increases its pulsation amplitudes. 
This `bouncing against the void' may explain why most 
of the cool hypergiants cluster near this low-temperature boundary.

Another evolutionary aspect of yellow hypergiants is their possible 
connection with the S Dor variables. In a recent review on these 
hotter luminous stars, van Genderen (2001) addressed the question 
if fast evolving yellow hypergiants are in fact proto-S Dor variables,
as proposed by de Jager (1998). This can be supported by 
noting that both types of variables are evolving blueward, and that 
the void may provide a temporal obstacle at which the unstable atmosphere 
of the yellow hypergiant is fastly transformed into that of a more 
stable (less luminous) S Dor variable. The void could thus be considered 
as a stellar atmospherical phenomenon, during rapid blueward evolution, which 
causes a natural separation between two distinct classes of luminous stars 
-the Luminous Blue Variables (LBV) and the cool hypergiants- with essentially 
the same, or comparable, evolutionary life-times. The segregation 
possibly only occurs during the post-red supergiant phase after a yellow 
hypergiant, at the cool border of the void, has shed sufficient amounts 
of mass during recurring phases of punctuated mass-loss (outbursts)    
that quickly strip off its extended atmosphere. Such a scenario
for the evolution of massive stars would imply that the instantaneous
stellar mass is a fundamental parameter which determines at what 
point during blueward evolution a yellow hypergiant makes the transit 
across the void. In other words, the bouncing against the void occurs 
only if the blueward evolutionary track is sufficiently fast to quickly 
approach the cool border of the void. Less luminous (i.e. less massive) 
yellow hypergiants require more time with quiescent mass-loss before 
the recurrent outbursts can develop. It can explain why 
recurrent outburst have not been observed in many other galactic 
yellow hypergiants so far, although they bear the spectroscopic 
signatures of a cool hypergiant. Presumably, $\rho$ Cas is therefore 
one of the most massive known yellow hypergiants, quickly heading to
the point of transiting the void.        

The above considerations in this section are rather tentative, but it
is of note that advanced evolutionary calculations of massive supergiants   
do adopt average mass-loss rates during the yellow hypergiant 
phase that are compatible with the observed values of about
20~$\rm M_{\odot}$ during 33,750 yr. The blueward tracks 
in these simulations (Meynet et al. 1994) last for $\sim$49,000 yr    
with $\log{L/L_{\odot}}$=5.33 during this period, and 
only 6500 yr when  $\log{L/L_{\odot}}$=5.77. In other words, 
the high mass-loss rates in these calculations yield very 
short-lived blue loops over which the yellow hypergiants cross the 
upper portion of the HR-diagram, thereby considerably diminishing
the stellar mass, by about halving it. The recurrent outbursts of $\rho$ Cas could therefore
be a major cause which quickly transforms the yellow hypergiant
into a hotter less massive hypergiant. We note that 
mass-loss estimates for the Giant and Lesser eruptions of 
the LBV $\eta$ Car, by $\sim$50 years apart, of 7 $\times$ $10^{-2}$ $\rm M_{\odot}\,yr^{-1}$
(Davidson \& Humphreys 1997) are of the same order of magnitude as we observe for the outburst of 
$\rho$ Cas. 

To conclude this section, we should point out 
that cool hypergiants as $\rho$ Cas are the prime candidates 
for progenitors of type II supernovae (SN). Their actual evolutionary phase
remains poorly understood. The He-core burning phase of cool hypergiants 
is likely accompanied by tremendous episodic mass-loss events as presented here. 
If these stars lose half of the current mass in only a few tens of 
thousand years, it will lead to an active reconstruction of the stellar interior. 
Lessons from SN 1987A teach that we should monitor, by all possible means, a 
number of galactic type II SN progenitor candidates (such as the cool hypergiants), 
if we aim to understand the relationship between the mass of the progenitor and 
the mass of the compact SN remnant. 
       
\section{Comparison With Previous Outbursts of $\rho$ Cas}

\subsection{$V$-band Brightness Curves}

Figure 13 compares the $V$-brightness curve observed during the 
three outbursts of $\rho$~Cas in the last century. 
The upper panel shows the 2000-01 outburst, the middle panel 
the moderate outburst of 1985-87, and the lower panel the 
famous eruption of 1945-47. The three panels are plotted 
for the same time intervals and $V$-magnitude range, 
so that the outburst properties can be compared. 
Photo-electric $V$ observations in 1985-87 ({\em solid dots in middle panel})  
are from Leiker \& Hoff (1987), Leiker et al. (1988), and Leiker, Hoff, \& Milton (1989).
Visual magnitudes during this period ({\em solid boxes in middle panel}) are offered by AFOEV$^{5}$.
The light curve during 1945-47 is from photographic and visual observations ({\em solid dots in lower panel}) 
listed in Gaposchkin (1949). The light curve before the 
brightness decrease of 1945 August is obtained from Payne-Gaposchkin \& Mayall (1946) 
({\em open symbols in lower panel}). The open circles are 20-day 
means with five or more photographic estimates, while the open boxes show
two to four estimates.   
Visual brightness estimates from AAVSO during the 1946 outburst are 
also shown in Figure 1 of Beardsley (1961). 

The brightness decrease lasted for $\sim$660 d during the entire 
outburst of 1945-47, with a very steep increase and rather symmetric 
decrease of $1^{\rm m}.5$ in $V$ (Gaposchkin 1949). 
A lesser brightness decrease was observed during the outburst of 1986, with 
a (visual) dimming of only $\sim$$0^{\rm m}.5$, and a (photo-electric) brightness increase 
of $\simeq$$0^{\rm m}.7$. This event lasted for 550 d, but with a shallower brightness decrease 
and increase. The recent outburst of 2000-01 is more pronounced as the previous one,
but a remarkable strong brightness maximum precedes the deep minimum in the fall of 2000. 
However, the entire outburst occurred over a much shorter time span of less 
than 500 d. The phases with the deep brightness minima therefore appear to have considerably 
shortened over the three outbursts; from $\sim$400 d in 1946, 
to $\sim$200 d in 1986, to $\sim$100 d in late 2000 (see also Hassforther 2001).

After the deep minimum of 1986 the $V$-brightness declined, along with a steady 
decline of the variability amplitude over time (see also Zsoldos \& Percy 1991). 
The brightness maximum after the deep minimum is followed by three declining 
brightness cycles, while returning to the mean magnitude level before the outburst. 
It signals pulsation phases over which the photosphere `relaxes' to the initial 
conditions. The outbursts are followed by cycles of decreasing pulsation compression
while the atmosphere gradually settles to the star's gravitational potential well
by dissipating the excess energy released by the explosion into the circumstellar environment 
(possibly somewhat similar to a bouncing ball losing height by friction).

\subsection{Recurrent TiO bands}

The vertical lines in the upper panel of Figure 13 mark our high-resolution spectroscopic
observations, while spectroscopic observation dates quoted in the literature are
marked for the other outbursts in the lower panels. The dates when TiO
was observed in the spectrum are connected with a horizontal line and labeled with TiO.
Popper (1947) was the first to report the observation of TiO bands in the spectrum of 1945 October
of $\rho$~Cas (see also Keenan 1947), shortly {\em before} the decline in light started.  
Interestingly, he also mentions that: ``A spectrogram obtained by Struve in 
August, 1944, before the light had started to fade, shows that the 
spectroscopic changes were already well underway, although the TiO bands had not yet 
appeared.'' It is important to point out that the outburst of 1945-47 was 
preceded by a pre-outburst cycle (JD 2431000 $-$ JD 2431600), similar as we observe 
before the outburst of 2000-01 between JD 2451000 and JD 2451600. 
During the latter period, we do not find indications for TiO in the spectra
({\em labeled with numbers 39 $-$ 63 in the upper panel}), although the spectra 
observed around the brightness minimum appear to be of a later spectral type (G$-$K).
Our observations confirm Popper's statement that TiO is not observed during 
the pre-outburst cycle of 1944 ({\em labeled `no TiO' in the lower panel}). 

We detect TiO bands in the WHT-UES spectra of 2000 July and August ({\em Nos. 64 \& 65}), 
{\em before} the deep brightness minimum is reached in late September 2000. 
The presence of TiO before the deep minimum of 1946 is also stated by Beardsley (1961). 
A spectrum obtained by Bidelman in 1945 October, a short time after the decline to the 
deep minimum had commenced shows TiO bands, although the star was still fairly bright.  
On the other hand, the WHT-UES spectrum of 2001 February 5 ({\em No. 69}) does not 
contain discernible TiO bands. During the four months before this observation 
the star quickly brightens up by $\sim$$0^{\rm m}.3$, while the expansion velocity of the 
deeper photosphere strongly decelerates, and begins to collapse during the deep 
brightness minimum ({\em solid dots connected by the dotted line 
in upper panel are $V_{\rm rad}$-values from Fe~{\sc i}; the other solid dots are values 
from ORO}). The spectra obtained between 2001 February and 2002 February ({\em Nos. 69 $-$77}) 
do not show TiO bands. These observations confirm the observations by
Beardsley (1961) that TiO was no longer present in his spectra observed after 1947 May. 
He also mentions that Thackeray (1948) found that in 1947 April (around JD 2432280), 
when the star was on the rise to full brightness, the TiO bands were still present, 
although ``almost gone''. Thackeray (1948) makes the remarkable statement that by 
1947 April the absorption lines resumed their sharp(er) supergiant character of 1939 
(observed before the outburst by Tai \& Thackeray 1948). By comparing different wavelength 
portions of the spectrum with $\mu$ Cep (M2) and $\beta$ Cam (G1) he concludes that: 
``The temperature corresponding to the lines spectrum seems to vary 
from 5500 deg. in the green to 3500 deg. in the red.''
In other words, the outburst spectra of 1945-47 and 2000-01 show that TiO bands
occur throughout the period of brightness decline {\em and} the period of 
rise as well. 

These observations show that the deep brightness minimum during the outbursts is not 
the primary cause for the formation of molecular bands in the spectrum.
They rather indicate that the bands can form when only the upper atmosphere 
expands beyond a certain threshold velocity. For example, the TiO bands 
do not form during the pre-outburst cycle of 1999 because the photospheric 
expansion velocity does not exceed 10~$\rm km\,s^{-1}$ (with respect to the stellar rest frame),
while in the summer of 2000 the expansion velocity increases to 18~$\rm km\,s^{-1}$,
and TiO is observed. The molecular bands are produced by rapid cooling of their 
outer formation region, which results from the unusual fast expansion of the entire atmosphere. 
It follows from our observations that the time-scale for the formation of the 
TiO bands is less than 100 d (between JD 2451650 and JD 2451744). 
Conversely, the collapse of the deeper photosphere raises the kinetic temperature in the 
molecular line formation region, which ceases to produce metal-oxides. 
The disappearance of these peculiar molecular bands after the deep minimum 
is not the result of the strong brightness increase only. 
Greenstein (1948) and Beardsley (1961) observe
AlO in spectra of 1947 September and November (after JD 2432430), 
when TiO was no longer present. The AlO bands also disappear after 
that time, but they were observed during a period when the stellar brightness
had returned to normal levels (Fig. 13, {\em lower panel}). It confirms
that the appearance of newly formed molecules with the outbursts of $\rho$ Cas 
chiefly results from the exceptional dynamics of the atmosphere during these events.     
Note further that we do not observe AlO in the outburst spectra of 2000-01. 
On the other hand, many broad absorption features develop in these complex spectra 
suggesting newly formed molecular bands, but which however still defy identification.   
 
\subsection{Blue H$\alpha$ Emission Before Outburst and Na~$D$}

Boyarchuk, Boyarchuk, \& Petrov (1988) also observe the appearance of 
the TiO band at 5448~\AA\, (\textsection\, 7.5) in 1986 February (JD 2446450 in Fig. 13). 
The band becomes gradually deeper and assumes maximum strength in 1986 May. 
It is still present in 1986 July, but disappears completely in 1986 August.  
They also monitored H$\alpha$ during the shallow brightness minimum 
(see their Fig. 5). The high-resolution spectra show that a very 
strong emission wing is present in the blue H$\alpha$ absorption core in 
1984 November (JD 2446000), before the brightness decline. 
Strong emission in the blue wing of H$\alpha$ is also observed during 
the pre-outburst cycle of 1999. They further show that the H$\alpha$ absorption 
core becomes conspicuously weak during the deep brightness minimum
of mid 1986, as we observe during the deep outburst minimum of late 2000 
(\textsection\, 7.2). Note however that these authors do not observe central 
emission above the continuum in Na~$D$, which also confirms a smaller 
continuum (hence $T_{\rm eff}$) decrease during the event of 1986.  
During the minimum of 1946 the broad absorption cores of Na~$D$ completely 
vanish, and weak emission lines appear above the continuum level 
(Fig. 1 of Thackeray 1948). Weak H$\beta$ and H$\gamma$ lines are 
observed during the deep minimum, with the presence of core emission in 
H$\alpha$ (Beardsley 1961).

The spectroscopic observations during the three outbursts of $\rho$ Cas in 
the last century reveal remarkably similar evolutions of the detailed line profiles.
These similarities indicate that the outbursts result from 
a driving mechanism which produces phases of fast photospheric expansion
and rapid decreases of $T_{\rm eff}$, which causes a substantial cooling of the outer atmosphere. 
In the next section we investigate the physical origin of this driving mechanism. 

\section{Outburst Mechanism}

\subsection{Dynamic Atmospherical Stability}

In a study of atmospheric dynamic instability of red supergiants 
Paczynski \& Ziolkowski (1968) were the first to simulate the ejection 
of a stellar envelope as a possible formation mechanism of planetary nebulae. 
They found that at sufficiently high luminosity, convective envelopes develop
that are dynamically unstable, due to the low value of the first generalized 
adiabatic index $\Gamma_{1}$$\equiv$($d$\,ln\,$P$/$d$\,ln\,$\rho$$)_{\rm ad}$ in the thick zones of partial ionization of 
hydrogen and helium. The total energy of the envelope becomes positive 
or unbound when recombination of the ionized hydrogen and helium 
is taken into account. This was first conjectured by Lucy (1967),
who suggested that planetaries may form when the mass contained 
in the hydrogen ionization zone is sufficient to reduce the specific 
heat ratio of the envelope to below 4/3. In more sophisticated 
numerical simulations Tuchman, Sack, \& Barkat (1978) found 
that the envelope becomes dynamically unstable, and mass is being 
expelled as a series of short-time events. The mass-loss process 
results from repetitive shock ejections, where the red giant loses 
about 3\% of its prevailing envelope per ejection. During the event 
the mean envelope density decreases by an order of magnitude, 
and the region with $\Gamma_{1}$$<$4/3 displaces toward the deeper 
mass layers with the release of recombination energy ({\em see their Figs. 3 \& 4}).
Interestingly, in a subsequent paper Tuchman, Sack, \& Barkat (1979)
simulated dynamically stable, but pulsational unstable envelopes, 
and demonstrated how a shell ejection event can occur 
after about 10 to 15 stable pulsation cycles. The radial pulsation amplitudes
increase over time, up to a point where 20\% of the outer envelope attains the
escape velocity and is ejected (see their Fig. 12). The time interval 
between successive ejections is $\sim$30 years, which is close to the 
envelope's thermal time-scale (Tuchman 1983).          
Wagenhuber \& Weiss (1994) computed with a hydrostatic code, fast envelope 
ejection in Asymptotic Giant Branch models, caused by the recombination of hydrogen. 
Whether an ionization-recombination mechanism can eject planetary nebulae 
remains presently unanswered because a time-independent treatment of convection 
is assumed in these simulations. More recently, Harpaz (1998) also showed  
that mass ejections from giants could be complicated by the decrease of electron 
opacity during hydrogen recombination. Other hydrostatic calculations by 
Jiang \& Huang (1997) indicate possible effects of turbulent
pressure for mass ejections from surface layers. 
All these studies appear to suggest that an ionization-recombination driving
mechanism is an important candidate for the origin of outbursts 
in cool luminous stars as $\rho$ Cas. Note that more recently, 
Stothers \& Chin (1997) also computed that an ionization-induced 
dynamic instability mechanism (with $\Gamma_{1}$$<$4/3) can 
cause strong eruptive mass-loss for massive hot stars (i.e. LBVs
as S Dor that also exhibit sudden spectral changes; see Massey 2000),
and argued against a supercritical radiative acceleration for such events. 
A number of other outburst mechanisms for hot supergiants have also been suggested in the 
literature. `Geysers models' and `violent mode-coupling instabilities' 
have been reviewed and compared by Humphreys \& Davidson (1994). 

In the next paragraph we investigate the ionization-recombination 
mechanism for the outburst of $\rho$ Cas. Based on hydrostatic 
calculations we provide arguments in favor of driving by a thermal 
runaway mechanism that causes the eruption. The main goal is to derive an
analytic expression which correctly predicts the time-scale for the changes 
of $T_{\rm eff}$ during the outburst, based on the observed stellar parameters, 
and the hydrogen recombination energy. We show that our equation is also valid 
for other variable eruptive cool stars.      

\subsection{Ionization-recombination Driving}

We demonstrate first that pure thermal expansion 
of an ideal hydrogen gas cannot provide sufficient energy 
to accelerate the atmosphere to an outburst. 
This results from the conservation of total energy. Consider a gas element 
with flow velocity $v$ at a distance $r$ in the atmosphere. The total energy of this
element per unit mass is  
\begin{equation}
\frac{1}{2}v^{2} + e + \frac{P}{\rho} - \frac{G\,M_{*}}{r} = {\rm const.}\,,
\end{equation}
where $e$ is specific  
internal energy (expressed per unit mass), $P$ the local gas pressure, and $\rho$ the mass density.
For the specific internal energy of an ideal gas $e=3/2\,N\,k\,T$, due to the translational 
motion of the gas particles in the parcel, with an equation of state $P=N\,k\,T\,\rho$, 
the differential of equation (4) yields
\begin{equation}
- v\, \Delta v= \frac{5}{2}N\,k\,\Delta T + g_{*} \,\Delta \,r  \,,
\end{equation}
where $k$ is the Boltzmann constant, $N$ the number of 
particles per unit mass, and $g_{*}$ the atmospheric gravity acceleration 
(at $r$=$R_{*}$). Equation (5) provides an estimate of the 
acceleration of the gas element due to the change of the local temperature,
assuming that its total energy per unit mass remains constant. 
Since we observe that the kinetic temperature in the atmosphere 
decreases by $\Delta$$T$$\simeq$$-$3000~K during the outburst,
with an outflow velocity of $v$=35~$\rm km\,s^{-1}$ for the TiO bands in \textsection\, 8.2, 
we compute that its corresponding acceleration
$\Delta$$v$ would not exceed 2~$\rm km\,s^{-1}$ (when neglecting 
the gravity acceleration). We observe, however, that 
the radial velocity increases to an expansion velocity above 30~$\rm km\,s^{-1}$ during the outburst. 
When including the gravity term in equation (5), the outward acceleration due to pure thermal expansion of 
the gas element, decreases to $\Delta$$v$$<$2~$\rm km\,s^{-1}$.  

However, if we consider instead the internal energy of a partially
ionizing hydrogen gas $e=3/2\,N\,k\,T(1+x)$ + $N\,I_{\rm H}\,x$, with $x$ the 
local ionization fraction (0$\leq$$x$$\leq$1), and an equation of state 
that includes the electron pressure of the gas element 
$P=N\,k\,T\,\rho(1+x)$ (hence the number of electrons 
per unit mass is $N_{e}$=$x$\,$N$), the differential of equation (4) yields
\begin{equation}
 - v\, \Delta v = \frac{5}{2}N\,k\,\Delta T \, (1+x) + \frac{5}{2}N\,k\,T\, \Delta x + N\,I_{\rm H}\, 
\Delta x  + g_{*} \,\Delta \,r  \,,
\end{equation} 
where $I_{\rm H}$= 13.6 eV is the ionization energy of hydrogen.
We assume that all translational energy of the free electrons is 
converted into accelerating the gas element due to ionization-recombination during cooling, 
and the hydrogen gas thereby almost completely recombines, hence $x$$\simeq$0 and $\Delta$$x$$\simeq$$-$1.    
For an average local temperature of $T$=2500~K, with $v$=35~$\rm km\,s^{-1}$, 
we compute with equation (6) for $\log{g_{*}}$=$-$0.5 and $\Delta$$r$$\simeq$$R_{*}$ 
that $\Delta$$v$$\simeq$19~$\rm km\,s^{-1}$. If the mean atmospheric gravity acceleration decreases 
further to $g_{*}$=0.1 $\rm cm\,s^{-2}$ during the outburst, the acceleration of the gas element 
increases to $\Delta$$v$$\simeq$37~$\rm km\,s^{-1}$.
This value corresponds to the observed change of expansion velocity during the outburst of 
$\rho$ Cas. It results from the release of ionization energy stored 
in the gas. The third term on the r.h.s. of equation (6) is the ionization 
energy content, which provides the major contribution to the change of total internal 
energy at relatively low atmospheric temperatures. These approximate calculations show that 
the conversion of thermal (translation) energy only is insufficient to increase 
the outflow velocity to the observed values. The release of hydrogen
ionization energy is required to adequately drive the outburst. 

\subsection{Temperature Dependence}     

In this section we investigate the ionization-recombination 
driving mechanism of $\rho$ Cas' outburst with a more detailed numerical calculation.
We show that the observed time-scale, and the expansion velocity of the outburst  
can be well predicted from an adiabatic hydrostatic calculation in spherical geometry.     
We emphasize however that this discussion does not involve 
a detailed solution of the momentum equation to compute the time-evolution of the 
dynamic and thermal wind structure. The presented calculations 
rather consider the driving properties of different atmospheric layers, that result from the 
conservation of total energy and mechanical flow momentum at different temperatures. 
The full hydrodynamic treatment of the outburst would require a time- {\em and} 
temperature-dependent solution of the conservation equations. This would demand, however,  
the detailed knowledge of the physical mechanism by which ionization-recombination energy is 
converted into wind momentum (possibly due to electron scattering), besides the adiabatic expansion
of the atmosphere. The driving mechanism is opacity dependent, which therefore also depends on the 
local gas temperature and density. Consequently, the atmospheric opacity also determines the detailed 
cooling properties during the outburst. With hydrogen recombination the (electron) opacity falls rapidly, 
thereby diminishing the driving force by the free flow of radiation.
We also emphasize that the calculations do not consider radiative damping,
or that we neglect possible radiation loss terms in the energy balance equation.
This approximation assumes that the energy released by the recombining gas elements 
is entirely converted into flow momentum which drives an expansion. In this respect, 
we adopt a `fully optically thick' limit to evaluate the energy and momentum balance 
of the outflow, and demonstrate that this assumption provides good agreement with the observations.
The amount of radiative damping can possibly be determined by
modeling the H$\alpha$ line profile, to determine its optical thickness 
and the energy in the emission of the transition. The radiative losses    
are dependent on the atmospheric opacity, which includes the density dependent
electron opacity. Presumably, laboratory experiments performed at smaller electron densities 
could be useful to determine the importance of increased radiative recombination of hydrogen,
compared to the three-body recombination observed for high densities (Fill et al. 2000).

We assume that the total energy of every gas (mass) element at temperature $T$ in the expanding atmosphere 
is constant and given by equation (4), with the material gas pressure $P=N\,k\,T\,\rho(1+x)$.
The constant equals $e_{0}$, or the total energy the element assumed before the acceleration.
Since we can suppose that the outburst is preceded by a compressed atmosphere which is almost static 
(or $v_{0}=0$),  the initial internal energy $e_{0} = 5/2\, N\,k\,T_{0}\,(1+x_{0}) + N\, I_{\rm H} \,x_{0}$. $T_{0}$
is the initial temperature of the gas element, and $x_{0}$ its corresponding ionization fraction.   
For further calculations we can neglect the small contribution of the potential energy of gravity.
Hence, the energy conservation equation for atmospheric gas elements is given by
\begin{equation}
\frac{1}{2}v^{2} +  \frac{5}{2}\,N\,k\,T\,(1+x) + N\,I_{\rm H}\,x = \frac{5}{2}\, N\,k\,T_{0}\,(1+x_{0}) + N\,I_{\rm H}\,x_{0} \,,
\end{equation}
and the r.h.s. remains constant with the changes of $T$, $x$, and $v$, during the expansion.
We further assume that the total mechanical momentum 1/2$\rho\,v^{2}+P$ of every gas parcel remains 
constant along a flow line in the expanding atmosphere, and equals the layer pressure of the 
static atmosphere $P_{0}$ before the expansion. Hence the Bernoulli equation yields  
\begin{equation}
\frac{1}{2}\,\rho\,v^{2}+P=P_{0}\,.
\end{equation}
The LTE Saha equation is used to compute the changes of the thermal ionization fraction $x$ 
during the expansion. Since $x$ is a function of $T$ and $P$, the ionization equation, together with 
the conservation of energy and momentum (Eqns. 7 \& 8), define a set of coupled equations 
that can iteratively be solved with an equation of state of partially ionized hydrogen gas.    

Figure 14 shows the expansion velocity $v$ of the gas element computed for different temperatures $T$
({\em bold lines}). The gas acceleration curves are calculated for three different gas parcel momenta 
of 0.1 $P_{0}$ ({\em dashed line}), $P_{0}$ ({\em solid}), 10 $P_{0}$ ({\em dash-dotted}), 
assuming $T_{0}$=8000 K. For 6000 $\leq$ $T$ $\leq$ 8000 K the gas remains nearly fully 
ionized ($x$=1) ({\em thin lines}), and $v$ does not exceed $\sim$17~$\rm km\,s^{-1}$.
Note that the addition of gravitation would decrease the computed flow velocities, 
although the computed $v-T$ dependence remains similar. 
For 4000 $\leq$ $T$ $\leq$ 6000 K, we find that $v$ rapidly increases 
to $\sim$70~$\rm km\,s^{-1}$, and assumes a constant value toward lower kinetic temperatures.
The fast acceleration in this temperature range results from the release of ionization 
energy by hydrogen recombination. An important property of these curves is that the fast acceleration 
occurs over a kinetic temperature range of $\sim$3000~K, independent of the 
total parcel momentum. This range corresponds to the mean atmospheric temperature change 
we observe for the outburst of $\rho$ Cas. An increase (decrease) of the momentum 
displaces the region of fast acceleration toward higher (lower) temperatures, but the 
width of the temperature range remains unaltered. This limited range is determined by the 
kinetic temperature dependency of thermal hydrogen ionization, which supports that the outburst is 
driven by hydrogen recombination. The bold dotted line is the velocity computed for a gas element 
without ionization energy ($x$=0) and $e_{0}$=$5/2\,N\,k\,T_{0}$. It reveals that pure thermal expansion
cannot provide the observed outflow velocities above 30~$\rm km\,s^{-1}$ in atmospheric layers
of lower temperature.      

For $T$$\leq$4000 K the fast velocity acceleration levels off due to the 
conservation of total energy. In the partial hydrogen ionization zone (around $x$$\simeq$0.5)
gas parcels become very expandable because the local adiabatic indices $\Gamma_{1}$$-$1 and $\Gamma_{3}$$-$1
assume minimum values. The inverse values are shown in Figure 14 for hydrogen recombination. 
However, when the atmospheric layers nearly fully recombine ($x$$\simeq$0), 1/($\Gamma_{1}$$-$1) 
and 1/($\Gamma_{3}$$-$1) decrease, or the adiabatic expansivity diminishes. 
At low temperatures the outflow stalls with the decrease of recombination driving, and the smaller
adiabatic expansivity. The outward driving forces therefore vanish, and the recombined hydrogen gas 
assumes conditions of maximum gravitational potential energy at low kinetic temperatures. 

Figure 15 shows the solution of the above considered system of equations for every mass layer of the photospheric
model with $T_{\rm eff}$=7250~K and $\log{g}$=1.0, determined for the pre-outburst 
condition of $\rho$~Cas (\textsection\, 7). The upper panel shows the model kinetic temperature structure ({\em bold solid line}),
which together with the gas pressure structure, provides the input values $T_{0}$ and $P_{0}$ to calculate the 
ionization-recombination acceleration of the model layers. For Rosseland optical depths 
$-$0.5 $\leq$ $\log{\tau_{\rm Ross}}$ $\leq$ 0.5 the hydrogen ionization fraction $x$
({\em thin solid line}) steeply increases from nearly zero to unity. 
For this model $\Gamma_{1}$ ({\em thin short-dash dotted line}) 
assumes values below the value for atmospheric dynamic stability of 4/3 ({\em horizontal dotted line})
over the entire upper fraction at $\log{\tau_{\rm Ross}}$ $\leq$ 0.5. It signals
an enhanced adiabatic expansivity for the upper atmospheric layers. 
Since the third adiabatic index is defined by $\Gamma_{3}$$-$1=$-$$(d {\rm ln}T / d {\rm ln} V)_{\rm ad}$,
and $\Delta\,V$/$V$=3\,$\Delta\, r$/$r$ in spherical geometry, the relative radius increase for 
every mass layer with spherical adiabatic expansion is    
\begin{equation}
\frac{\Delta\,r}{r} = \frac{1}{3} \frac{(- \Delta\,T)}{T}\, \frac{1}{\Gamma_{3}-1} \,.    
\end{equation}    
Detailed LTE expressions of $\Gamma_{3}-1$ for singly-ionizing monatomic gas mixtures with equilibrium 
radiation are provided in Lobel et al. (1992). We presently limit the calculations to 
thermal hydrogen ionization without equilibrium radiation, for which (e.g. Mihalas \& Weibel Mihalas 1984)
\begin{equation}
\Gamma_{3}-1 = \frac{c_{p}-c_{v}}{c_{v}}\frac{\chi_{\rho}}{\chi_{T}}= \frac{1+ (x/2)(1-x)(5/2+ I_{\rm H}/(k\,T))}{3/2+(x/2)(1-x)(3/2 + (3/2 + I_{\rm H}/(k\,T))^{2} )  } \,,
\end{equation}   
where $c_{p}$ and $c_{v}$ are the heat capacities, and $\chi_{\rho}$ and $\chi_{T}$ the isothermal and 
isochoric factor, respectively.  

The upper panel of Figure 15 shows the temperature structure of the photospheric model
with $T_{\rm eff}$=3750~K and $\log{g}$=0, obtained from the TiO outburst spectrum 
({\em bold long-short dashed line}). This temperature structure is assumed after outburst, 
and can be well approximated by lowering the kinetic temperature of the gas layers in the pre-outburst 
model (with $T_{\rm eff}$=7250~K) by $\Delta T_{\rm eff}$/ $T_{\rm eff}$, with $\Delta\,T_{\rm eff}$=$-$3500~K
({\em bold dash-dotted line}). Hence, we can assume that the outburst diminishes the 
temperature $T_{0}$ of each layer by $\Delta\, T = T_{0}\, {\Delta T_{\rm eff}}/{T_{\rm eff}}$,
which yields with equation (9) an adiabatic radius change of
\begin{equation}
\Delta\, r  = \frac{r}{3} \frac{(-\Delta T_{\rm eff})}{T_{\rm eff}} \frac{1}{\Gamma_{3}-1}\,.
\end{equation} 
With the cooling of the layer temperatures to $T$ = $T_{0}+ \Delta\,T$, 
we compute the corresponding layer accelerations from the conservation of total energy and momentum. 
The velocity increase from static initial conditions is shown by the bold solid 
line in the lower panel of Figure 15. We calculate a maximum
layer expansion velocity of 36$-$37~$\rm km\,s^{-1}$ at $\tau_{\rm R}$$\simeq$1. Around these 
optical depths, the hydrogen ionization fraction $x$ strongly decreases during the outburst because
the gas is partially ionized over this region of the atmosphere 
({\em thin solid line in upper panel}). Hydrogen gas can therefore 
recombine more by the decrease of the local temperature than in the neutral or 
fully ionized atmospheric regions. 
The decrease of $x$ is shown by the thin long-dash dotted line in the upper panel. 
The adiabatic expansion velocity in the lower panel peaks in 
layers where $\Delta$$x$ is largest. It is of note that toward the deeper atmosphere at $\tau_{\rm R}$$>$10,
the adiabatic expansion velocity increases, although hydrogen remains nearly fully ionized. 
It results from pure thermal expansion with the decrease of the kinetic temperature in conditions 
of larger density. However, in these layers the adiabatic expansivity 
1/($\Gamma_{3}$$-$1) is much smaller than for layers at $\tau_{\rm R}$$<$10 
({\em thin short-dash dotted line in lower panel}). With equation (11) we calculate that these deeper 
atmospheric layers expand over much shorter distances during the outburst. For example, 
the relative radius increase $\Delta\,r$/$r$ ({\em thin dashed line}) at $\tau_{\rm R}$=1 is $\simeq$2.5 
near the maximum of expansion velocity, whereas $\Delta\,r$/$r$ decreases to below 30\%
at $\tau_{\rm R}$$>$10. 
We compute that the adiabatic expansion time $t_{\rm ad}$ = $\Delta\,r$/$v$
({\em thin solid line}) of atmospheric layers at a distance $r$=$R_{*}$$\simeq$400~$\rm R_{\odot}$ ranges between 
170 and 267 d for $-$3$<$ $\log{\tau_{\rm R}}$$<$0.5. We therefore  conclude that the close 
correspondence of $t_{\rm ad}$ with the observed time-scale of $\sim$200 d, over which $T_{\rm eff}$
decreases from 7250~K to $\leq$4000 K, indicates a quasi-adiabatic spherical expansion of the entire upper 
atmosphere during the outburst, driven by the release of hydrogen ionization energy due to complete recombination. 
After the outburst the value of $\Gamma_{1}$ increases to above 4/3 in the region 
where $\Delta\,x$ is maximum, and hence the inflated atmosphere becomes dynamically stable
({\em long dashed line in upper panel}).

\subsection{Outburst Time-scale}

From the considerations in the previous section we derive a more general equation for the time-scale of 
recombination-driven stellar outbursts. Since the maximum expansion velocity 
occurs in the partial ionization layers of the pre-outburst atmosphere 
where $x_{0}$$\simeq$0.5, the adiabatic outburst velocity, due to complete recombination to 
$x$$\simeq$0 in this region, is approximated with equation (7):
\begin{equation}
v^{2} \simeq  5\,N\,k\,(-\Delta\,T) + 5\,N\,k\,(T_{0}\,x_{0}-T\,x) + 2\, N\, I\,(-\Delta\,x)\,,  
\end{equation}
where $I$ is the energy of the recombining ionization stage of the driving element under 
consideration. Hence, for $\Delta\,x$$\simeq$$-$1/2, with a temperature change of 
$\Delta\,T$ in layers of $T_{0}$, equation (12) casts to:
\begin{equation}
v^{2} \simeq 5\,N\,k\,T_{0}\, \left( \frac{1}{2} - \frac{\Delta\, T_{\rm eff}}{T_{\rm eff}} \right) + N\,I\,.
\end{equation}  
The adiabatic index of volume expansion in the partial ionization region with $x_{0}$$\simeq$0.5 
simplifies with equation (10) to          
\begin{equation}
\Gamma_{3} - 1 = \frac{21+2\, I/(k\,T_{0})}{27 + 2 \left(3/2 + I / (k\,\,T_{0})    \right)^{2}}\,.
\end{equation}   
The outburst time (over which $T_{\rm eff}$ decreases by $-$$\Delta\, T_{\rm eff}$), 
driven by the recombining atmospheric region around $T_{0}$, is obtained  
with equations (11), (13), and (14)  
\begin{equation}
t_{\rm burst} = \frac{\Delta R_{*}}{v} \simeq \frac{R_{*}}{3\,\sqrt{N\,k\,T_{0}}}\,\frac{(-\Delta\,T_{\rm eff}/T_{\rm eff})}{\sqrt{5/2-5\,\Delta\, T_{\rm eff}/T_{\rm eff}+I'}}\, 
\left(\frac{27 + 2\,(3/2 + I')^{2}}{21+2\,I'}\right)\,,
\end{equation}   
where we denote $I'$=$I/(k\, T_{0})$.
The first two factors at the r.h.s. of equation (15) result from gas dynamics (total energy conservation and spherical 
geometry), whereas the last factor is of thermodynamic origin (adiabaticity). The last two factors are dimensionless. 
For the outburst of $\rho$ Cas, we compute $t_{\rm burst}$=221 d 
for $R_{*}$=400~$\rm R_{\odot}$, $T_{\rm eff}$=7250 K, $\Delta\, T_{\rm eff}$=$-$3500 K, 
and $T_{0}$=8,000 K in the partial hydrogen ionization zone. For supergiant atmospheres
composed of hydrogen $I'_{\rm H}$=19.7. However, for pulsating supergiants with hydrogen deficient atmospheres, 
for which outbursts can be driven by the first recombination of helium, equation (15) predicts much shorter
time-scales. For example, for R CrB, a Ib-supergiant with $T_{\rm eff}$$\simeq$6900~K and $R_{*}$=80~$\rm R_{\odot}$, 
we compute for $T_{0}$=13,000 K in the first helium recombination zone ($I_{\rm He}$=24.58 eV and $I'_{\rm He}$=21.9)  
an outburst time of 39 d for $\Delta\, T_{\rm eff}$=$-$3500 K. This time-scale corresponds to the brightness declines by 3$-$6 mag. 
observed in R CrB-type variables, which occur in less than 50 d (Clayton 1996). In these smaller stars 
the outbursts occur faster and more violently than in the hypergiants. The burst-time also depends 
on the relative abundances of hydrogen and helium, whereas equation (15) assumes a 
pure hydrogen (or helium) atmosphere. For gas mixtures, the last factor 
of equation (15) requires the multi-component expressions of 1/($\Gamma_{3}$$-$1), computed in 
Lobel et al. (1992) for solar abundance values (Anders \& Grevesse 1989). 
In the first ionization zone of helium with $T_{0}$=13,000 K, we compute that
$1/(\Gamma_{3}-1)$=6.67, which yields for $\rho$ Cas $t_{\rm burst}$=72 d. This time-scale is at least a factor of two
shorter than the observed burst-time, which excludes driving by helium recombination in this luminous hypergiant.
Note that for $\rho$ Cas we compute that the outburst extends the atmosphere with $v$=36 $\rm km\,s^{-1}$ 
over $\Delta R_{*}$=$v\,\times\,t_{\rm burst}$=988 $\rm R_{\odot}$ ($\simeq$2.5 $R_{*}$), while for R CrB the equation 
predicts that $v$=48~$\rm km\,s^{-1}$ and $\Delta R_{*}$=235 $\rm R_{\odot}$ ($\simeq$2.9 $R_{*}$). 

\section{Discussion}

In a study of stellar outburst mechanisms, Sedov (1958) obtained time-dependent solutions for 
the propagation of a spherically expanding shock wave in a gravity field. Analytic solutions are provided
in which the energy of the shock motion remains equal at any time to the initial energy in the 
equilibrium state of the unperturbed atmosphere. He derived a simple 
analytic solution in cases of a constant specific heat ratio $\gamma$=$c_{v}/c_{p}$=7/6, and showed that 
the radial velocity of the shock decreases monotonically with $v_{\rm sh}$=const$\times$$t^{-1/6}$. The equations 
can also be applied to the extended partial hydrogen ionization zone in the upper atmosphere of $\rho$ Cas 
with $\gamma$$<$4/3 before the outburst. This zone provides mechanical shock energy by hydrogen recombination 
to drive the outburst, with the conservation of total energy.                 
     
Since the adiabatic expansion velocities computed for the outburst (and observed also) of $\rho$ Cas
exceed the adiabatic sound speed ($v_{\rm s}$=$\sqrt{\Gamma_{1}\, P/ \rho}$=5$-$13 $\rm km\,s^{-1}$), 
we propose the formation of a circumstellar shock wave that surpasses the gravitational pull of the
supergiant, carrying momentum beyond the sonic point into the circumstellar environment. 
The disrupted and inflated atmosphere reaches a state of reduced total energy,
and adapts to this energy dissipation by contracting globally. 
Hydrogen recombination driving thus provides a limit to the mechanical energy which can be 
released by this thermal outburst mechanism, and may explain why comparable 
atmospheric changes have occurred during the three outbursts of $\rho$ Cas in the last century;
for example, the formation of TiO in outer layers with $T$$<$4000 K, and the decrease of $V$-brightness 
by $\leq$$1^{\rm m}.5$, for total outburst times of 500 to 660 d.     

We think that the outburst results from a synchronization phenomenon whereby 
ionization-recombination energy is released upon decompression, in a global 
cooling of the entire atmosphere. The increase of outflow momentum causes an outward 
acceleration by which the atmosphere further cools, thereby reinforcing
the release of more ionization energy. The cooling and recombination rates  
synchronize, which causes a fast escalation of ionization energy release 
into atmospheric expansion toward the deeper layers that contain more ionization energy.
This unleashes even more energy, and hence produces an explosive runaway event. 

The condition for an outburst is therefore 
strongly dependent of the overall temperature-
and density-structure during the atmospheric compression phase that 
leads up to the eruption. The atmospheric dimensions    
determine whether the recombination rate (and hence the 
wind acceleration) can keep up with the cooling rate, 
and this in turn determines whether a `disruption' front, or a strong expansion shock wave, 
can move down the atmosphere, and consume most of its ionization energy. The full time-dependent 
solution of the hydrodynamic and thermodynamic equations will 
therefore also require the proper treatment of the detailed 
physics of strong shock waves in partially ionizing gas, with variable 
gammas across the shock surface.    
It is known that the ionization of hydrogen due to the
compression by strong shock waves 
can store 5 to 6 times more energy in ionization energy than the amount that is available in 
the translational motion of particles of the shocked gas (Lobel 1997, p. 80).         
Note that partial thermal ionization can raise the compressibility of the shocked gas, 
whereby the density jump ratio across a compression 
front can increase to very high values of $\sim$20. However, the density jump ratio is 
limited to a maximum value of only 4 in cases of ideal
gas, for which partial ionization is neglected. 

The ionization-recombination instability requires a `trigger' that provides the
precise atmospheric structure to this cascade of 
energy release. The outburst of $\rho$ Cas
is preceded by a conspicuously long period between 
1993 and 1997 over which the stellar wind practically 
vanishes. The wind acceleration 
appears to stall (observed from the disappearance of violet extended wings 
in the photospheric lines), while the (photospheric) radial velocity  
increases by 30 $\rm km\,s^{-1}$ (Fig. 1). The $V_{\rm rad}$-curve,
shows four quasi-periodic oscillations, with  
$T_{\rm eff}$-changes of 500 to 800~K during
the pulsation cycles. 
After that time the atmosphere expands rapidly, with the strong 
$V$-brightness decrease of the pre-outburst cycle. It signals the sudden 
release of flow momentum that has accumulated in the 
atmosphere over the four preceding years, for example in a denser 
mass shell that has been swept up in the atmosphere by a subsonic wind. 

During the large oscillation preceding the actual outburst, 
the atmosphere decompresses and cools, thereby changing 
its thermodynamic structure, so that the next phase of compression 
can effectively convert the kinetic energy from a
ballistic supersonic downfall (by $\sim$13 $\rm km\,s^{-1}$) into ionization energy. 
The atmosphere therefore strongly collapses, becoming denser and warmer, 
while partly storing this kinetic energy into ionization energy. 
What follows is a phase of unusual high $T_{\rm eff}$$\sim$ 8000 K. 
Cool supergiant atmospheres become very compressible (or expandable) around $T_{\rm eff}$=7000$-$8000 K 
by the decrease of $\Gamma_{1}$, which is related to the atmospheric 
elasticity (or bulk modulus), to below the stability value of 4/3 over a large geometric fraction. Lobel (2001) 
shows (see Fig. 6) that these atmospheres 
can even assume $\Gamma_{1}$-values below unity for $\log{\tau_{\rm R}}$$\leq$0, 
due to partial LTE- and non-LTE-ionization of hydrogen. For these atmospheric conditions
the volume-averaged $<$$\Gamma_{1}$$>$ assumes values below 4/3 down 
to the base of the atmosphere, beyond the hydrogen and helium ionization zones.
Local mechanic perturbations, i.e. gas rarefactions, 
then become not sufficiently balanced by the counterworking thermodynamic forces. The latter 
rather enhance the rarefactions and release more recombination energy, which causes 
their displacements $\delta$$r$/$r$ from hydrostatic equilibrium to grow exponentially with time.            

For the conditions of supergiant atmospheres with $T_{\rm eff}$=7000$-$8000 K, 
the atmospheric structure is very dynamically unstable, and tuned so that a  
self-sustained hydrodynamic-thermal avalanche can occur. The entire atmosphere rapidly 
de-ionizes and expands from the top to the very bottom, while driving an accelerating wind. 
This mechanism is time-critical and comes to a halt only when most 
of the atmospheric ionization energy has been released by the outburst, and 
$<$$\Gamma_{1}$$>$ returns to above 4/3 in the outer atmosphere. 
The atmosphere cools down to below $T_{\rm eff}$=4500 K with
the fast expansion, whereby the upper layers with $T_{\rm kin}$ $<$ 4000~K 
form TiO bands that characterize the spectra of M-type stars. 
      
The outburst trigger can be compared to lifting a heavy hammer
over a long period of time, that suddenly drops and hits a lighter 
rubber ball on the floor. The ball jumps up very high because
its elasticity conserves linear momentum. For the outburst of $\rho$ Cas, 
thermal ionization-recombination ensures that the high compressibility 
of the atmosphere `bounces back' in a resonant amplification of a superwind 
that further cools and releases more energy from the decompressing ionization reservoir. 
Think of it as if the ball bounced hard enough to also release internal chemical energy, 
giving it an additional kick (e.g. jump higher on a trampoline by timely bending and 
stretching your legs). In a study of gravitational collapse of stellar cores, 
which can eject the stellar envelope in a supernova-like explosion, 
Bruenn, Arnett, \& Schramm (1977) compute that the kinetic energy for 
a reflected shock wave is maximized when a large inward acceleration is 
rapidly changed to large outward acceleration during the bounce. They suggest 
that this is accomplished with an adiabatic index below 4/3, 
that goes suddenly well above 4/3. Arnett \& Livne (1994) calculate that 
when the first adiabatic index is only slightly above 4/3, the released (here thermo-nuclear) 
energy will drive a pulsation with large amplitude.      

The conversion of ionization-recombination energy into expansion energy may 
also explain why the outbursts of $\rho$ Cas are rather sporadic.  
The progressive built-up of momentum in the atmosphere by a 
stalled wind over a long period of time, which can trigger the coherent gravitational collapse, 
is frequently disrupted by the usual quasi-periodic oscillations
of the atmosphere. Ordered atmospheric oscillations effectively transport momentum upward 
which inhibits activation of the trigger, and does not allow sufficient momentum to built up 
(e.g. $T_{\rm eff}$ above 7250 K has 
previously not been reported for $\rho$ Cas). 
This occurs most of the time, but occasionally     
the quiescent pulsations become very disordered
(or incoherent) due to the weak force of gravity and the 
enormous atmospheric extension (Lobel et al. 1994; 
e.g. Soker \& Harpaz 1992 examine the location of the `transition region'
relative to the ionization zones which determines if non-radial pulsation modes are excited 
in AGB stars). Because these random modes cease to effectively convey pulsation 
momentum upward, the stellar wind driving vanishes, and the piston
momentum becomes temporarily stored in the atmosphere. 
If momentum can accumulate over a period of more than four pulsation 
cycles (four years say), a subsequent average radial 
pulsation is proceded by very high atmospheric compression, 
which produces the required thermal conditions of $<$$\Gamma_{1}$$>$ $<$ 4/3,   
that yield the ionization-recombination driven runaway. The outburst 
occurs when the atmospheric cooling rate synchronizes with the 
displacement of the driving recombination front toward the deeper 
atmosphere. In these low-gravity atmospheres the recombination acceleration is 
almost independent of $g_{*}$ (e.g. Figure 1 is computed 
for $g_{*}$=0). The latter of course determines the magnitude of 
the expansion velocity during the outburst, and whether the effective atmospheric 
acceleration $g_{\rm eff}$ becomes negative (Nieuwenhuijzen \& de Jager 1995). 
An inward directed $g_{\rm eff}$ over a 
long period of time plays a crucial role for the initial built-up of momentum,
and a subsequent activation of the trigger. 
Outbursts that are initiated by a similar trigger mechanism are therefore expected 
in other pulsating stars that assume comparable thermal conditions at the base of the 
atmosphere as $\rho$ Cas. 

It is interesting to note that the time-scale over which the 
total ionization-recombination energy, released into the circumstellar environment by 
the outburst (assuming complete hydrogen recombination), is 
dissipated by the stellar radiation field over a period following the fast envelope expansion of 
\begin{equation}
t_{\rm brightness-min} \simeq \frac{I_{\rm H}\,N\,{\rho_{\rm s}}\, d_{\rm s}}{\sigma\, T_{\rm eff}^{4}} \,,
\end{equation}
where $\sigma$ is the Stefan-Boltzmann constant. For an atmospheric expansion of 
$d_{\rm s}$ = 869 $\rm R_{\odot}$ during outburst, and an envelope density of
$\rho_{\rm s}$ = $10^{-10}$ $\rm gr\,cm^{-3}$ (determined from TiO bands in \textsection\, 8.2), 
we compute with equation (16) that $t_{\rm brightness-min}$$\simeq$86~d for $T_{\rm eff}$=3500 K. 
This time corresponds very well to the period over which the deep brightness minimum of 
$\sim$100 d was observed in late 2000.  

\section{Summary and Conclusions}

1. Spectroscopic monitoring of the yellow hypergiant $\rho$ Cas with 
high-resolution over the past 8.5 years reveals a strong correlation between 
the radial velocity curve and the $V$-magnitude variations. 
The latter lag the photospheric oscillations by $\sim$100 d. 
Between 1994 and 1997 the entire upper and lower 
atmosphere collapses, resulting in the pre-outburst cycle of 1999, when a very 
large brightness maximum of $V$$\sim$$4^{\rm m}.0$ is assumed. During 
the $V$-maximum, the photosphere expands unusually fast, and $V$ subsequently 
dims by $1^{\rm m}.2$$-$$1^{\rm m}.4$ to the very deep outburst minimum during
$\sim$100 d in late 2000. The profile shape and velocity variations observed in 
selected optical Fe~{\sc i} lines, formed in the stellar photosphere, 
occur on markedly shorter time-scales of $\sim$300 d than the changes of H$\alpha$. 
It signals a strongly velocity-stratified pulsating atmosphere. 

2. The outburst is preceded by a variability phase whereby enhanced 
absorption develops in the long-wavelength wing of H$\alpha$, 
together with a conspicuously strong short-wavelength emission.
During the outburst the H$\alpha$ absorption core becomes very weak and develops satellite 
emission lines. It indicates that the line is filled in by emission due to 
recombination of hydrogen with the global cooling of the atmosphere by $\sim$3000 K.
The equivalent width variations of the Fe~{\sc i} $\lambda$5572 line reveal that the $T_{\rm eff}$-changes 
are limited to $\sim$500$-$800~K during the quiescent pulsation phases. Spectral 
synthesis calculations reveal however that $T_{\rm eff}$ decreases 
from $\geq$7250~K to $\leq$4250 K during the outburst. 

3. During the outburst the optical absorption spectrum strongly blue-shifts,
and resembles the spectra of the early M-type supergiants Betelgeuse 
and $\mu$ Cep.  Central emission appears above the continuum level in the split 
Na~$D$ lines with the decrease of the optical continuum flux. The hypergiant 
is surrounded by a tenuous gas envelope producing permanent line emission in 
these broad resonance lines, and in forbidden optical [Ca~{\sc ii}] lines.  
Prominent permitted emission lines of neutral metals appear above the local 
continuum level during exceptional pulsation phases when $T_{\rm eff}$ is 
large, and the photosphere rapidly begins to expand.
It indicates that the peculiar emission lines are excited by the supersonic stellar wind.   
On the other hand, high-resolution observations do not reveal central emission 
in the broad Ca~{\sc ii} H \& K lines, indicating the absence of a (classic) permanent 
chromosphere in the hypergiant.

4. Most remarkably, weak electronic TiO bands develop in the summer of 2000 
before the $V$-brightness minimum. Detailed spectral synthesis calculations show that 
the bands form over atmospheric regions with 3000~K $\leq$ $T$ $\leq$ 4000~K 
and $-$9.8 $\leq$ $\log{\rho}$ $\leq$ $-$8.5, and assume an expansion velocity 
of 35~$\rm km\,s^{-1}$. These molecular bands develop within $\sim$100~d after 
fast outward acceleration, due to the rapid decline of the kinetic 
temperature in the line formation region. During the outburst, the TiO
envelope expands faster than the photosphere, with the photospheric expansion 
determined from the radial velocity of the atomic absorption line spectrum. 

5. We compute an exceptionally large mass-loss rate of $\dot{M}$=5.4$\times$$10^{-2}$~$\rm M_{\odot}\,yr^{-1}$
for spherical expansion of the TiO envelope with a minimum density of 
$\rho_{\rm min}$=$10^{-10}$ gr~$\rm cm^{-3}$ at $R_{*}$=400~$\rm R_{\odot}$.
The cool gas shell expands to $\sim$2.5~$R_{*}$ during an outburst period of $\sim$200 d.
The large mass-loss rate indicates that the recurrent outbursts of $\rho$ Cas, about every 
half a century, are the main mass-loss mechanism of this cool star in the post-red supergiant
evolutionary phase.   

6. Within half a year after the deep brightness minimum $T_{\rm eff}$ increases 
to $\simeq$5750~K, and the peculiar TiO bands vanish. Optical and near-IR TiO
bands were also observed during the previous outbursts of $\rho$ Cas in 1986 and 1946,
when the star dimmed and brightened up over comparable time-scales.
TiO is also observed before the deep brightness minimum of 1946,
indicating that the TiO formation is dependent on the atmospheric 
expansion velocity, which is largest before the deep minimum.
Strong emission in the short-wavelength wing of H$\alpha$ is also observed before
the outburst of 1986, indicating that the wind dynamics in the outer 
atmospheric envelope plays a crucial role for an activation of the 
trigger that causes the outbursts of $\rho$ Cas.     
 
7. We compute that the outburst mechanism can be driven 
by the release of ionization energy due to the recombination 
of hydrogen as the atmosphere cools. When $T_{\rm eff}$
increases to $\geq$7250 K, the supergiant atmosphere becomes dynamically unstable 
because the first generalized adiabatic index $\Gamma_{1}$ decreases to below 4/3 in the partial ionization 
zone of hydrogen. Atmospheric expansion in this region becomes further amplified 
with the adiabatic cooling and the release of more driving energy by hydrogen 
recombination, until $T_{\rm eff}$$\leq$3750~K and $\Gamma_{1}$ increases above 4/3.
The activation of the trigger therefore requires a phase of 
exceptionally high $T_{\rm eff}$ before the outburst, which 
can result from enhanced global atmospheric compression during 
a cycle of strong coherent pulsation in the pre-outburst cycle of 1999.
Based on these simple considerations we derive an equation that 
correctly predicts the time-scale for the rapid brightness
decrease in terms of observable stellar parameters $R_{*}$, $T_{\rm eff}$,
the change of $T_{\rm eff}$, and the ionization energy of the recombining
element under consideration.     
      
\acknowledgments
This research was supported in part through grant number GO-5409.02-93A to 
the Smithsonian Astrophysical Observatory from
the Space Telescope Science Institute, which is operated by the AURA, under NASA
contract NAS 5-26555. A. L. would like to thank S. Bagnulo at European Southern 
Observatory (Chile), and A. E. Rosenbush at the Crimean Astrophysical Observatory 
(Ukraine) for providing optical spectra of $\rho$ Cas. We thank R. Kurucz
at the Smithsonian Astrophysical Observatory for discussions and help with 
the spectral synthesis calculations. A. L. gratefully acknowledges 
support by the SRON-Utrecht (The Netherlands) over the years for maintaining the 
spectral data base of $\rho$ Cas. We thank numerous observers
of the WHT-UES service programs who have contributed to this long-term project.

\newpage
\clearpage

\begin{figure}
\figcaption[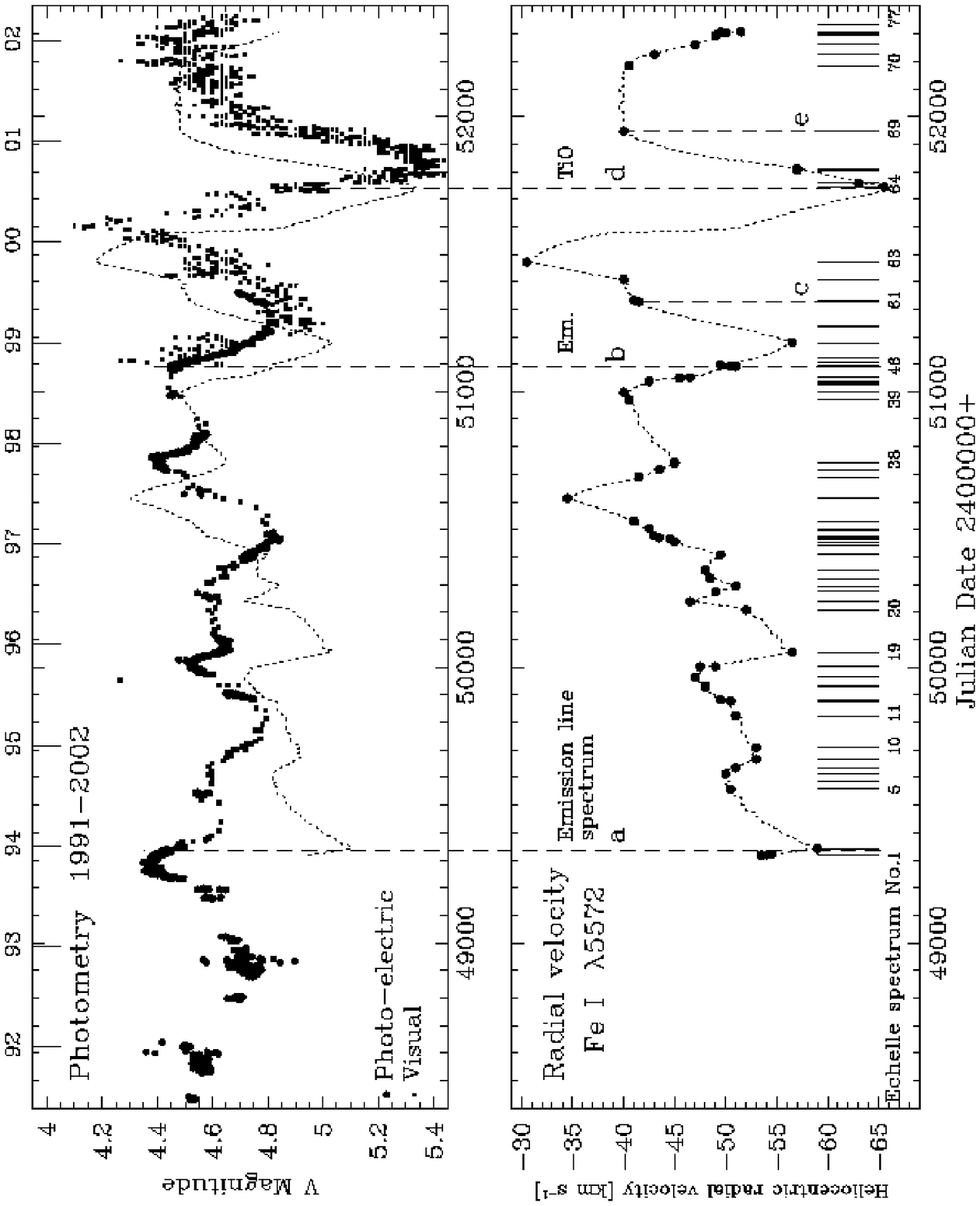]{The $V$-brightness curve of $\rho$ Cas
({\em solid symbols}) is compared  in the upper panel to the radial velocity curve ({\em dotted line}),
observed over the past 8.5 years. Observation dates of echelle spectra are marked with vertical lines in 
the lower panel. The radial velocity curve ({\em connected solid dots}) of Fe~{\sc i} $\lambda$5572 
shows a strong increase of the photospheric pulsation amplitude before the outburst of fall 2000 (JD 2451800$-$JD 2451900),
when TiO bands develop ({\em marked TiO}). The peculiar emission line spectrum of $\rho$ Cas is observed during phases with fast
atmospheric expansion ({\em and is marked by Em.}). The vertical dashed lines, 
labeled a to e, are compared and modeled in this paper. Note that the dotted line connecting the radial velocity 
values is determined from the half width of the Fe~{\sc i} line at half absorption minimum
by linearly interpolating between the profile shapes. \label{fig1}  }
\end{figure}
\begin{figure}
\plotone{f1.eps}
\end{figure}

\newpage
\clearpage

\begin{figure}
\figcaption[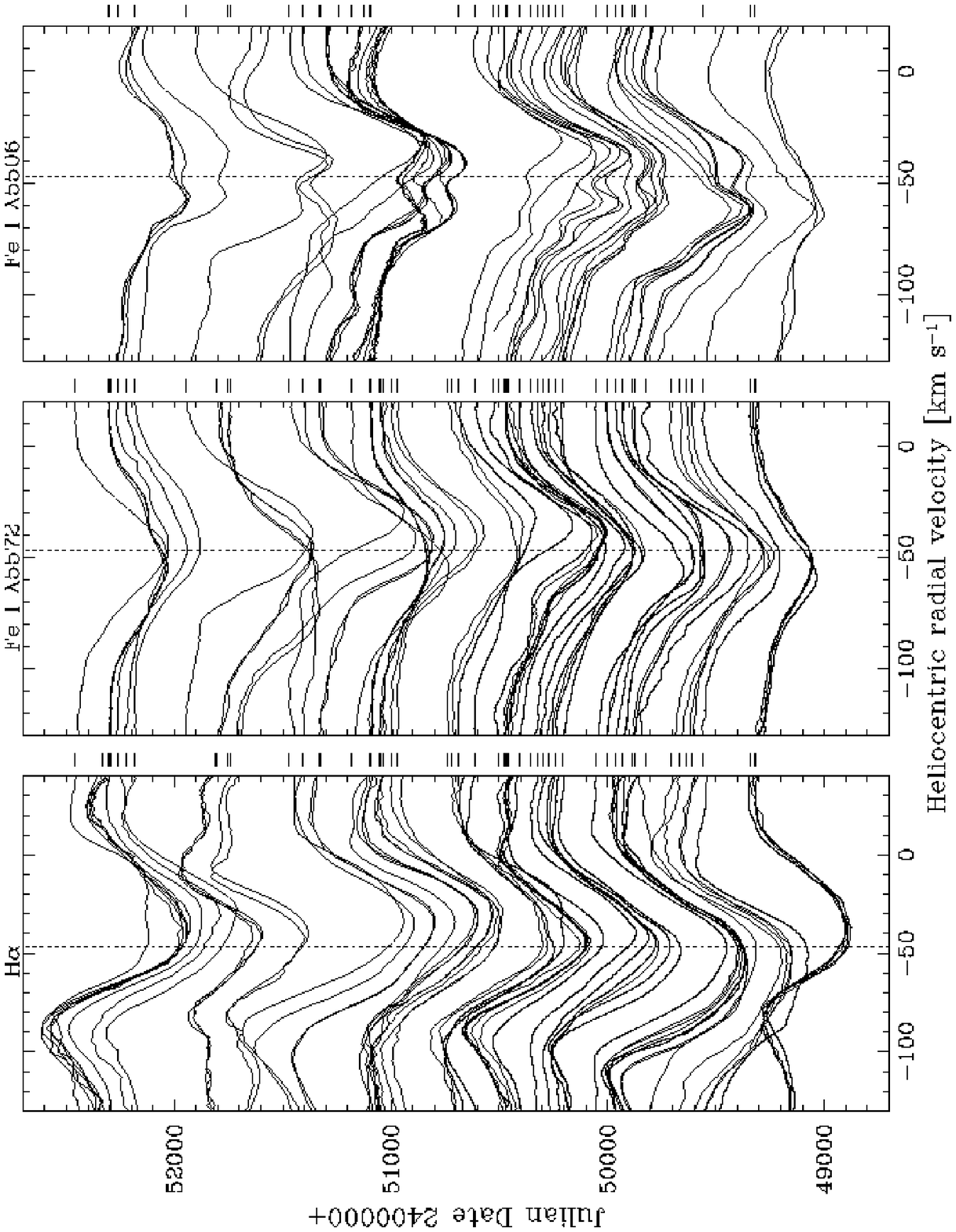]{Detailed evolution of the line profiles of H$\alpha$, 
Fe~{\sc i} $\lambda$5572, and Fe~{\sc i} $\lambda$5506 ({\em panels left to right}),
observed between 1993 November and 2002 July. The right-hand tickmarks ({\em 
bottom to top}) indicate the continuum level. The vertical dotted lines mark the stellar rest 
velocity of $\rho$ Cas. The H$\alpha$ line shows variable emission wings, 
with a central absorption core that becomes very weak during the outburst.
The Fe~{\sc i} lines show Doppler shifts, and variable extended violet line wings during the photospheric
pulsations and the outburst. The Fe~{\sc i} $\lambda$5506 is split due to a static central emission core (see text).  \label{fig2}}
\end{figure}
\begin{figure}
\plotone{f2.eps}
\end{figure}

\newpage
\clearpage

\begin{figure}
\figcaption[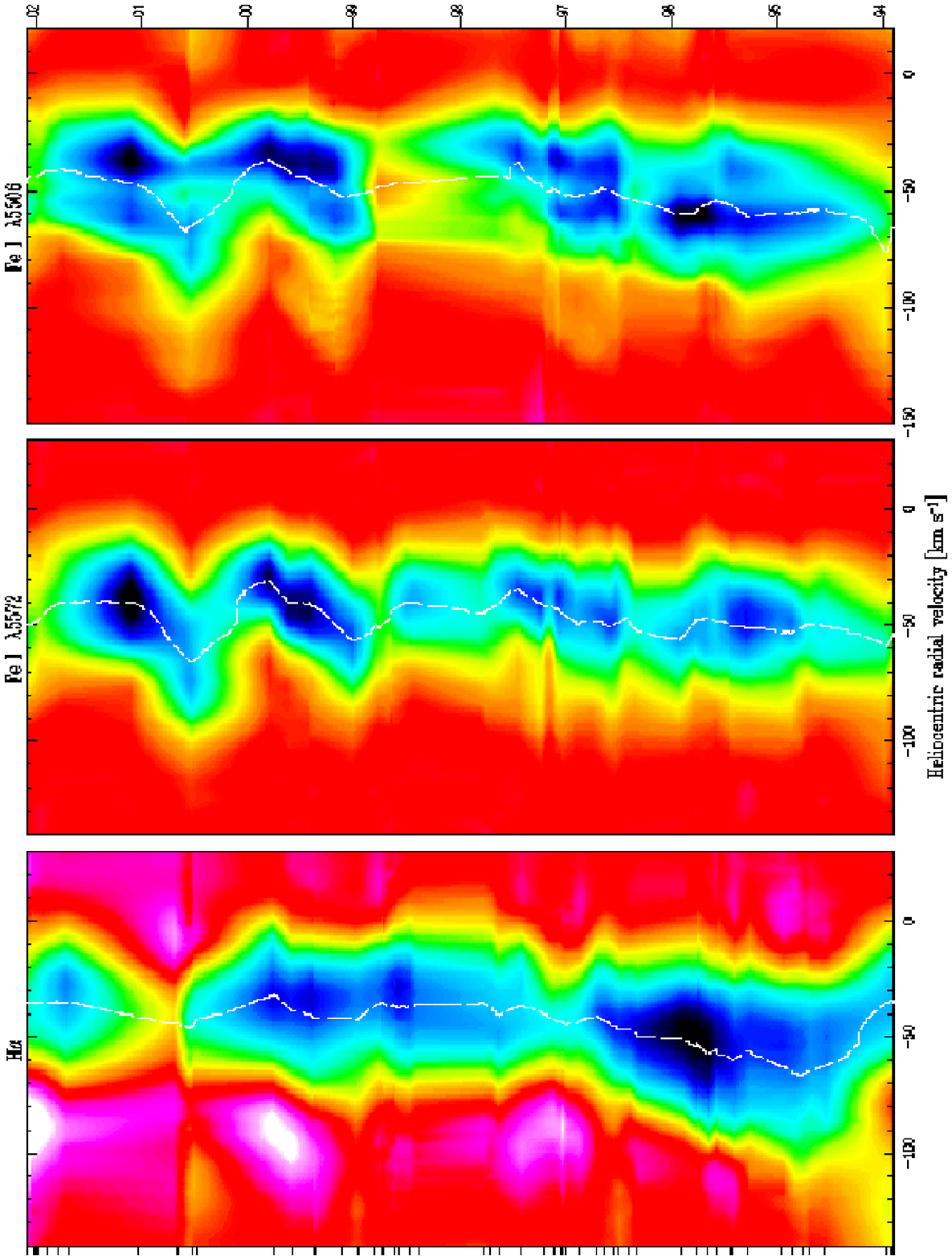]{Dynamic spectra of the three lines of Figure 2. The grayscale is linear
(online figure: red color indicates the local continuum level, and blue the depression in the line cores). 
The white spots in H$\alpha$ are emission above the continuum level. 
The line profiles are linearly interpolated between subsequent observation nights, marked
with the left-hand tickmarks. Time runs upward, indicated for each new calendar year with 
the right-hand numbers. The dashed white lines trace the radial 
velocity curve, determined from the bisector at half intensity minimum of the absorption portion of the line profile.
The curves reveal a strongly velocity-stratified dynamic atmosphere. 
Notice the strong blue-shift of the Fe~{\sc i} lines during the outburst (mid 2000). 
The outburst is preceded by very strong emission in the short-wavelength wing of H$\alpha$, 
while the absorption core extends longward, and the photospheric Fe~{\sc i} lines strongly red-shift. 
A strong collapse of the entire atmosphere precedes the outburst (see text).\label{fig3}}
\end{figure}
\begin{figure}
\plotone{f3.eps}
\end{figure}

\newpage
\clearpage

\begin{figure}
\figcaption[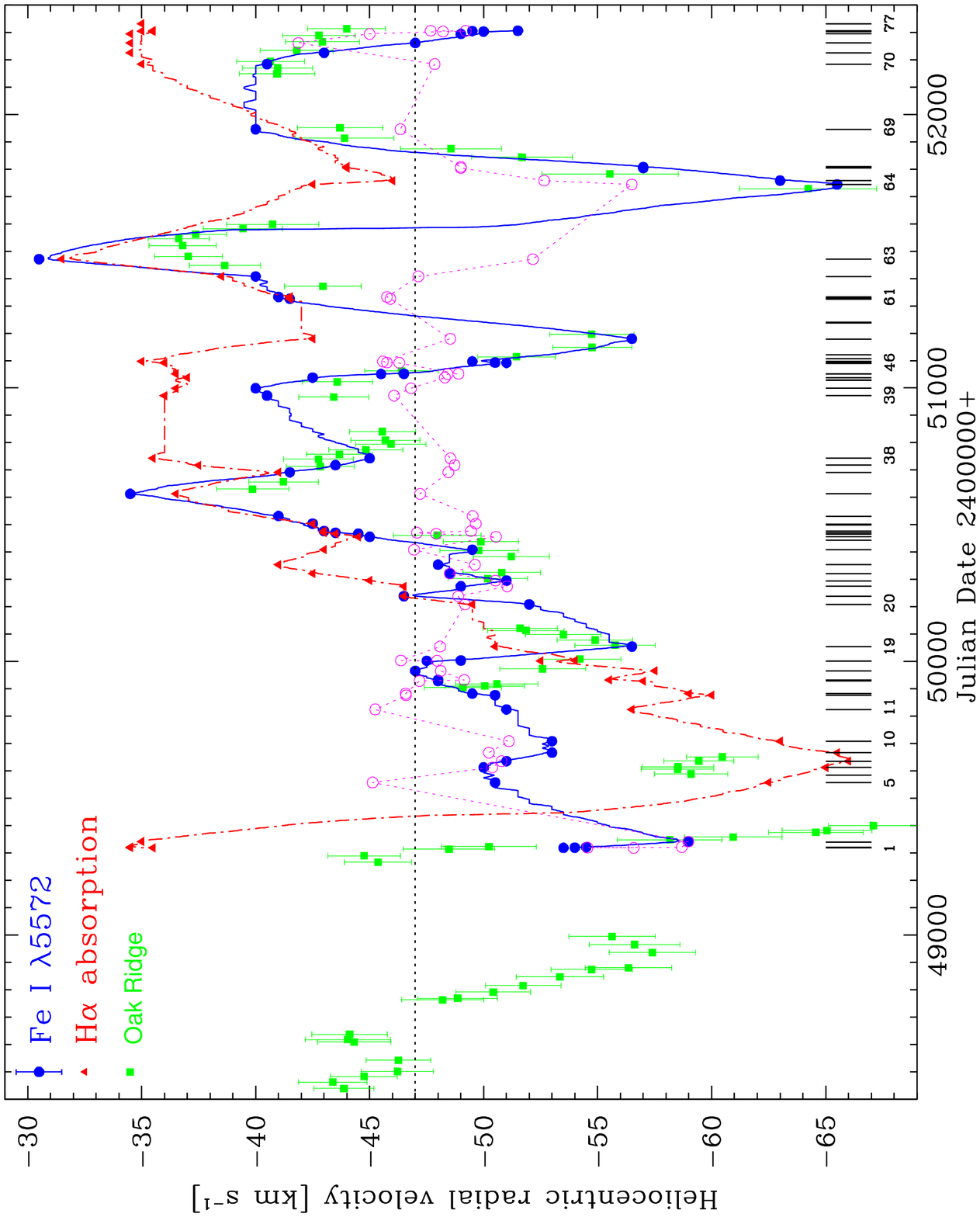]{Comparison of the radial velocity curve measured from
Fe~{\sc i} $\lambda$5572 ({\em dots}), and from cross-correlating the  
spectrum observed between 5166 \AA\, and 5211 \AA\, ({\em boxes}). The photospheric
expansion decelerates over a long period of time between JD 2449300 and JD 2450400.
After this time, the photosphere and the H$\alpha$-envelope collapse
until JD 2450600. The global contraction phase is followed by the pre-outburst cycle 
in 1999 during which the photosphere rapidly expands and contracts, resulting in the 
very large $V$-brightness maximum of JD 2451600. The maximum is proceded 
with the outburst and a subsequent $V$-magnitude decrease by $1^{\rm m}.2$$-$$1^{\rm m}.4$ 
during $\sim$200 d. Both radial velocity measurement techniques are in good agreement 
during the strong velocity excursions. The open circles are velocity variations
determined from the second moment of the Fe~{\sc i} $\lambda$5572 line (see text).\label{fig4}}
\end{figure}
\begin{figure}
\plotone{f4.eps}
\end{figure}

\newpage
\clearpage

\begin{figure}
\figcaption[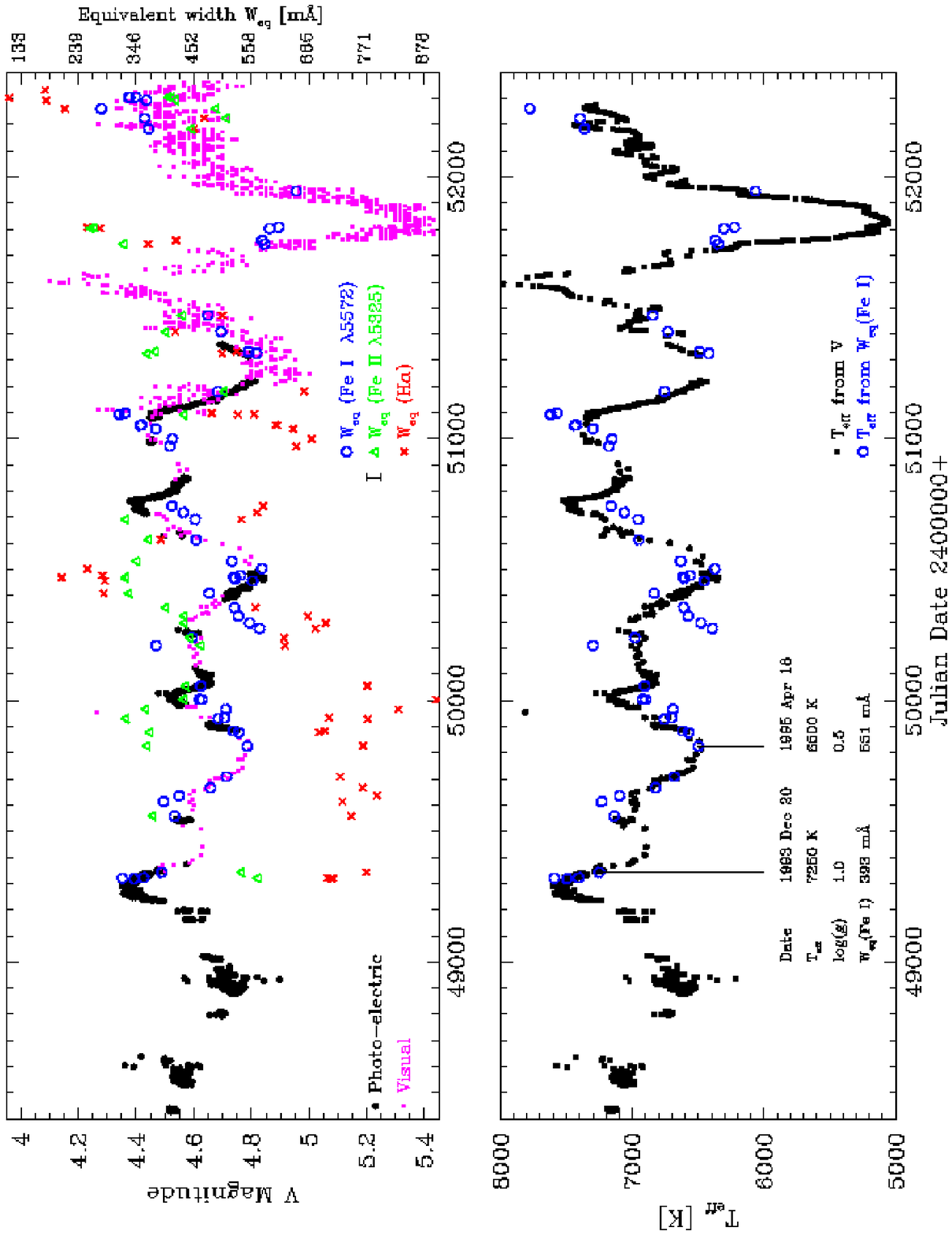]{A comparison of the $V$-brightness curve with
equivalent width values measured from Fe~{\sc i} $\lambda$5572 ({\em circles}),
Fe~{\sc ii} $\lambda$5325 ({\em triangles}), and H$\alpha$ absorption and emission 
({\em crosses}) reveals a strong correlation of the Fe~{\sc i} equivalent width 
with $V$. It signals temperature changes in the line formation 
region, determined by variations of $T_{\rm eff}$ with the photospheric pulsations.
The Fe~{\sc ii} line is anti-correlated due to changes of the iron ionization 
balance. H$\alpha$ is uncorrelated with a longer variability period, indicating 
a mean line formation region different from the photospheric iron lines. The lower panel shows 
the changes of $T_{\rm eff}$ with pulsation, computed from the equivalent width 
values of Fe~{\sc i} ({\em circles}), and from the $V$-magnitude values ({\em boxes}). 
A decrease of $T_{\rm eff}$ by $\sim$3000 K occurs during the outburst.\label{fig5}}
\end{figure}
\begin{figure}
\plotone{f5.eps}
\end{figure}

\newpage
\clearpage

\begin{figure}
\figcaption[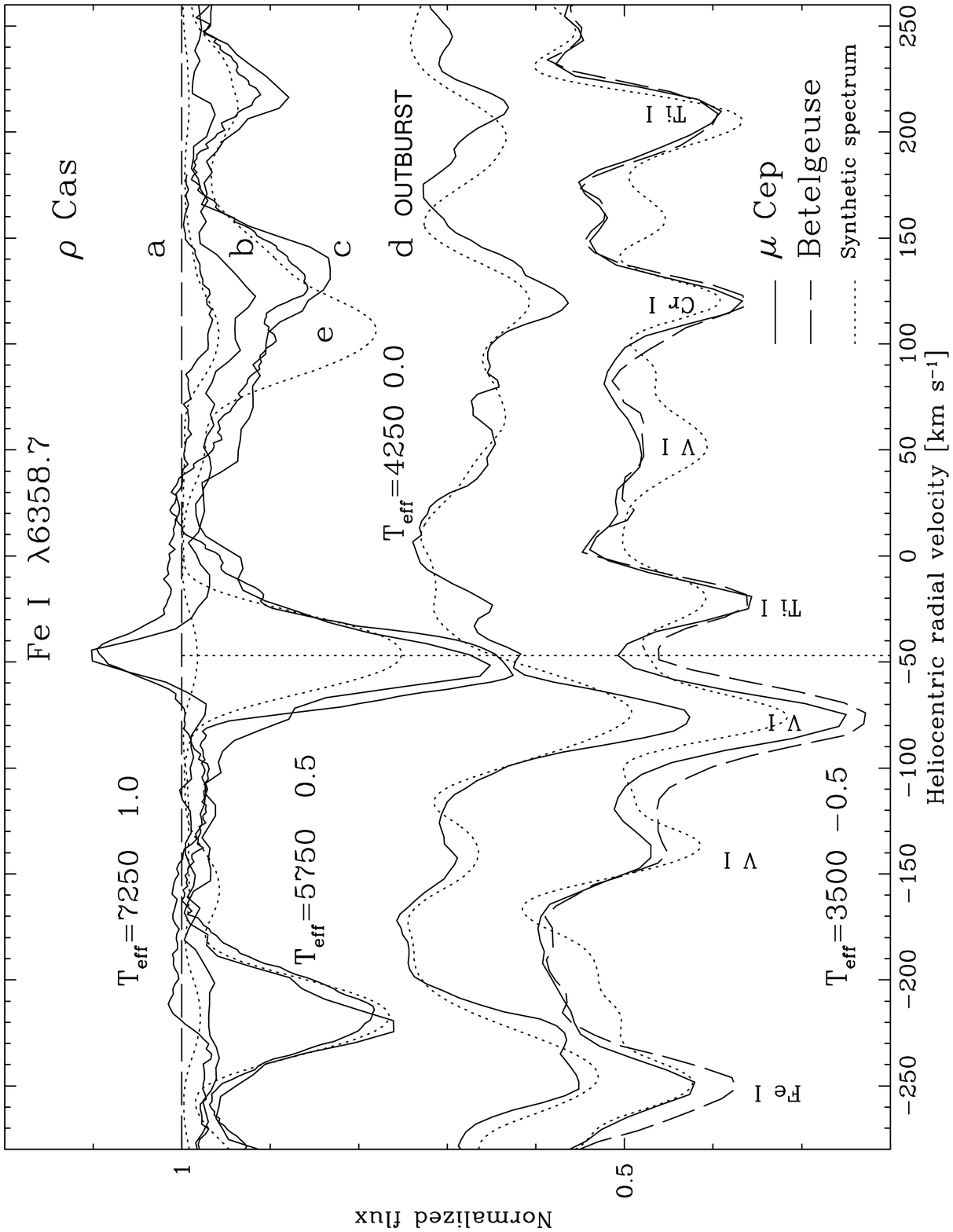]{High-resolution spectra of 
$\rho$ Cas around Fe~{\sc i} $\lambda$6358 ({\em solid lines labeled a to e for the dates of Fig. 1}) 
show that the photospheric spectrum strongly 
blue-shifts during the outburst. Synthetic spectrum calculations ({\em dotted lines})
indicate that $T_{\rm eff}$ decreases to $\simeq$4250~K, and the spectrum becomes comparable to the 
early M-type supergiants $\mu$~Cep ({\em lower solid line}) and Betelgeuse ({\em dashed line}),
with $T_{\rm eff}$=3500~K. A prominent emission line spectrum is observed in $\rho$ Cas 
during phases of high $T_{\rm eff}$=7250~K, when the atmosphere rapidly accelerates outward. Note that the 
Fe~{\sc i} emission line blends with an absorption line of V~{\sc i}. Other neutral absorption 
lines are also indicated. The vertical dotted line is drawn at the stellar rest velocity. \label{fig6}}
\end{figure}
\begin{figure}
\plotone{f6.eps}
\end{figure}

\newpage
\clearpage

\begin{figure}
\figcaption[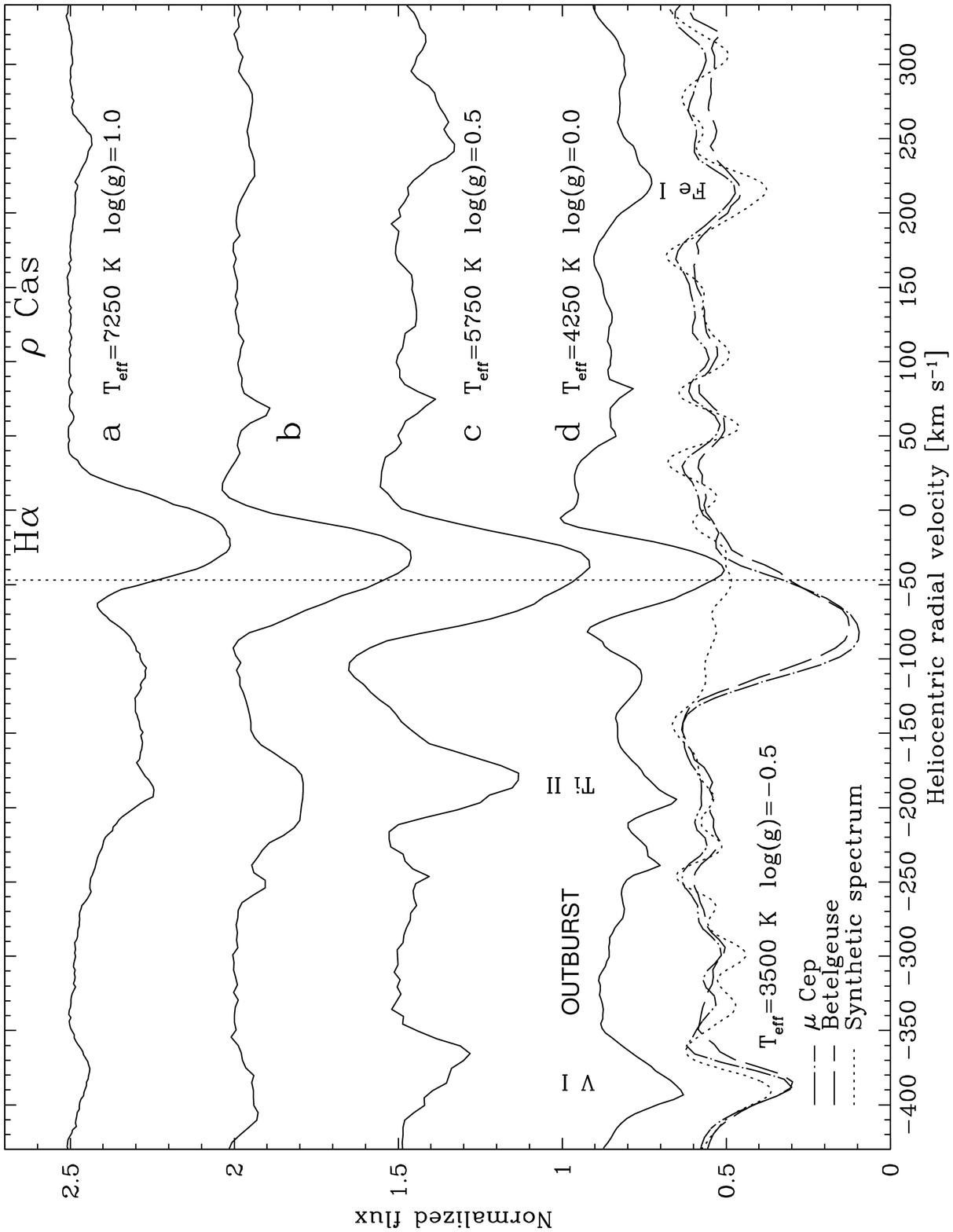]{High-resolution spectra of $\rho$ Cas
reveal that the H$\alpha$ absorption core becomes very weak during the outburst ({\em solid 
line labeled d}). The line core does not blue-shift, and develops strong emission line 
wings. The line becomes filled in by recombination emission with the cooling 
of the H$\alpha$-envelope during outburst. 
Strong emission is also observed in the blue wing of H$\alpha$ ({\em long-short dashed line})
for the pre-outburst cycle ({\em c}), during the atmospheric collapse that precedes the large
$V$-brightness maximum before the outburst. The H$\alpha$ line core of $\rho$ Cas is much weaker than in 
Betelgeuse and $\mu$ Cep ({\em solid dashed and dash-dotted lines}), where it is excited by 
a permanent chromosphere. Note that the spectra of Betelgeuse and $\mu$ Cep are blueshifted to align the
photospheric absorption lines with the outburst spectrum of $\rho$ Cas.
 The synthetic spectrum of Betelgeuse ({\em dotted line}) is computed without 
a model of the chromosphere (see text).\label{fig7}}
\end{figure}
\begin{figure}
\plotone{f7.eps}
\end{figure}

\newpage
\clearpage

\begin{figure}
\figcaption[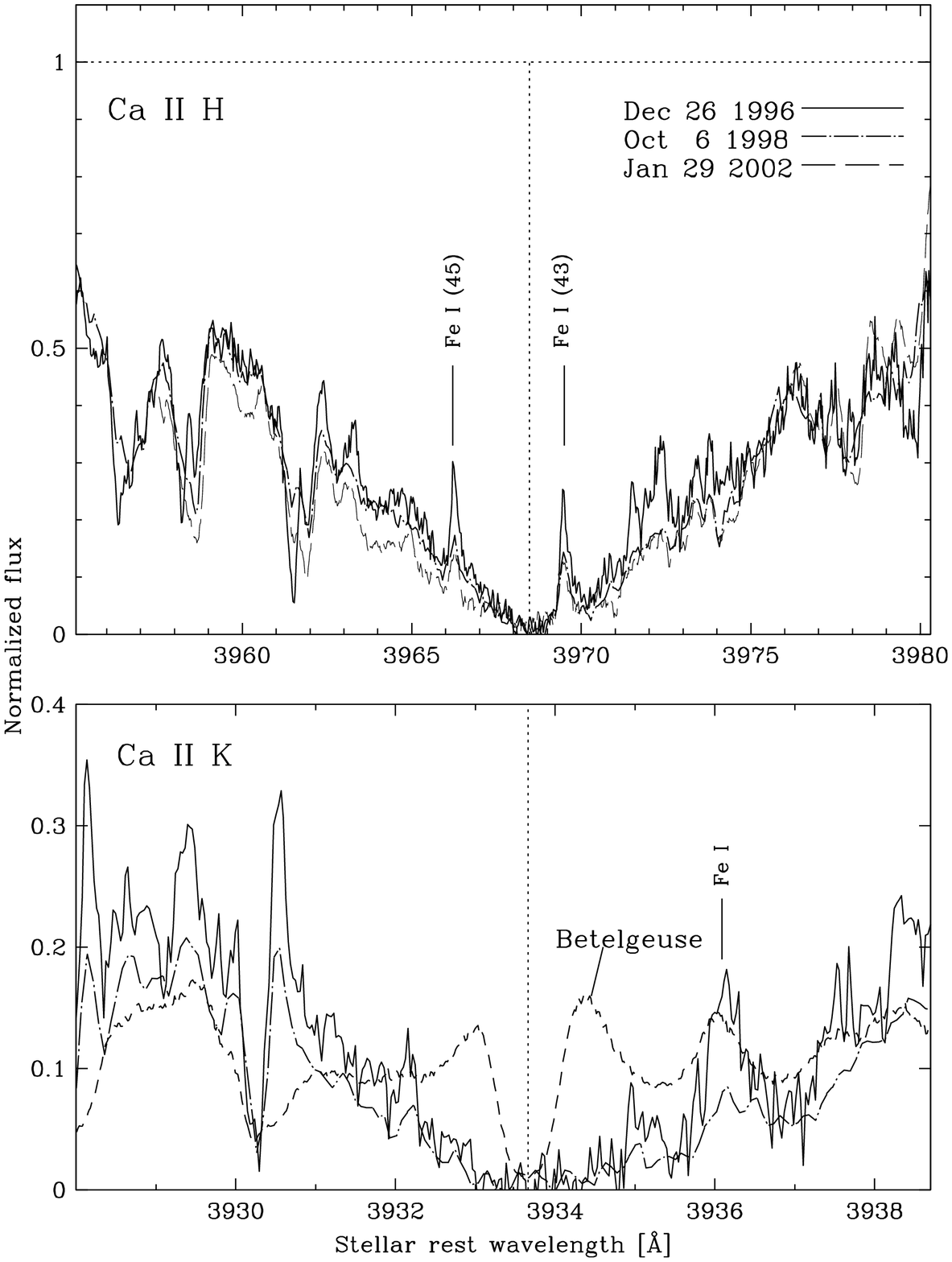]{High-resolution spectra of $\rho$ Cas
in the upper panel do not reveal central Ca~{\sc ii} H emission in 1996 December ({\em solid line}),
1998 October ({\em dash-dotted line}), and 2002 January ({\em dashed line}) after the outburst.
Permanent Fe~{\sc i} emission lines of RMT 45 and 43 are observed in the broad Ca~{\sc ii} H line wings. 
Central emission is neither observed around the stellar rest wavelength ({\em vertical dotted lines}) 
of Ca~{\sc ii} K in the lower panel. Central emission observed in Betelgeuse ({\em short dashed line})
emerges from the extended and permanent chromosphere. The Ca~{\sc ii} H \& K lines indicate the absence of 
a classic chromosphere in $\rho$ Cas.\label{fig8}}
\end{figure}
\begin{figure}
\plotone{f8.eps}
\end{figure}

\newpage
\clearpage

\begin{figure}
\figcaption[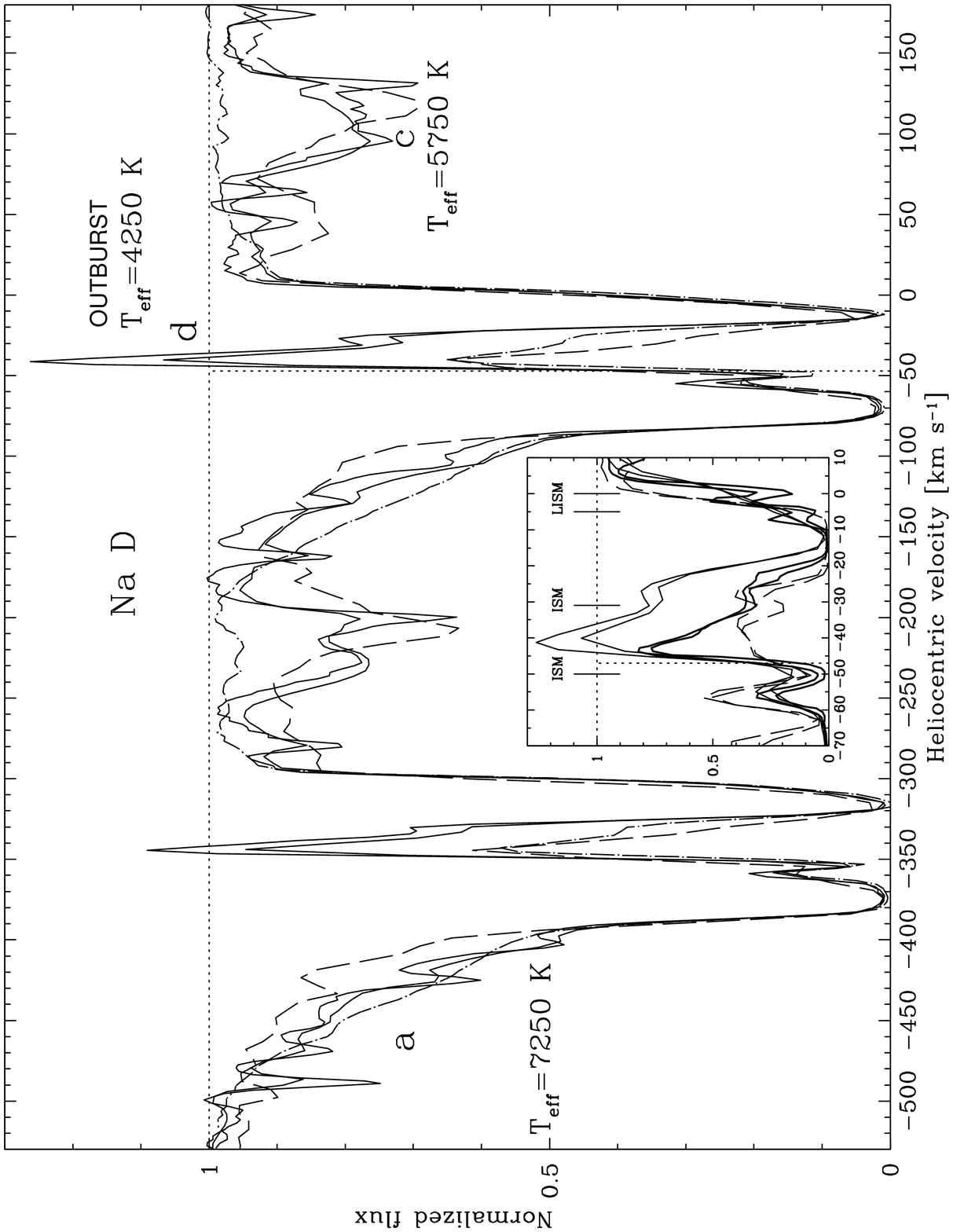]{During the outburst central emission appears 
above the local continuum level ({\em horizontal dotted line}) in the Na~$D$ lines ({\em solid lines labeled d} 
of 2000 July 19 and August 2). The broad absorption portions of the lines are strongly intensity 
saturated, and do not show Doppler shifts with time ({\em dash-dotted and dashed lines labeled a and c}). 
The maximum of the central emission line is red-shifted with respect to the stellar 
rest velocity ({\em vertical dotted line}), and a blue-shifted narrow feature is observed in both lines of the doublet.
The smaller figure shows a comparison of the doublet lines in $\rho$ Cas ({\em bold solid lines})
with the lines observed in HR~8752 ({\em thin dashed lines}). 
Two interstellar absorption components are observed around $-$50 $\rm km\,s^{-1}$ and $-$30~$\rm km\,s^{-1}$ (see text).
The outburst spectra clearly show that central emission contributes to 
the complex Na~$D$ line formation, besides the interstellar absorption (see text).\label{fig9}}
\end{figure}
\begin{figure}
\plotone{f9.eps}
\end{figure}

\newpage
\clearpage

\begin{figure}
\figcaption[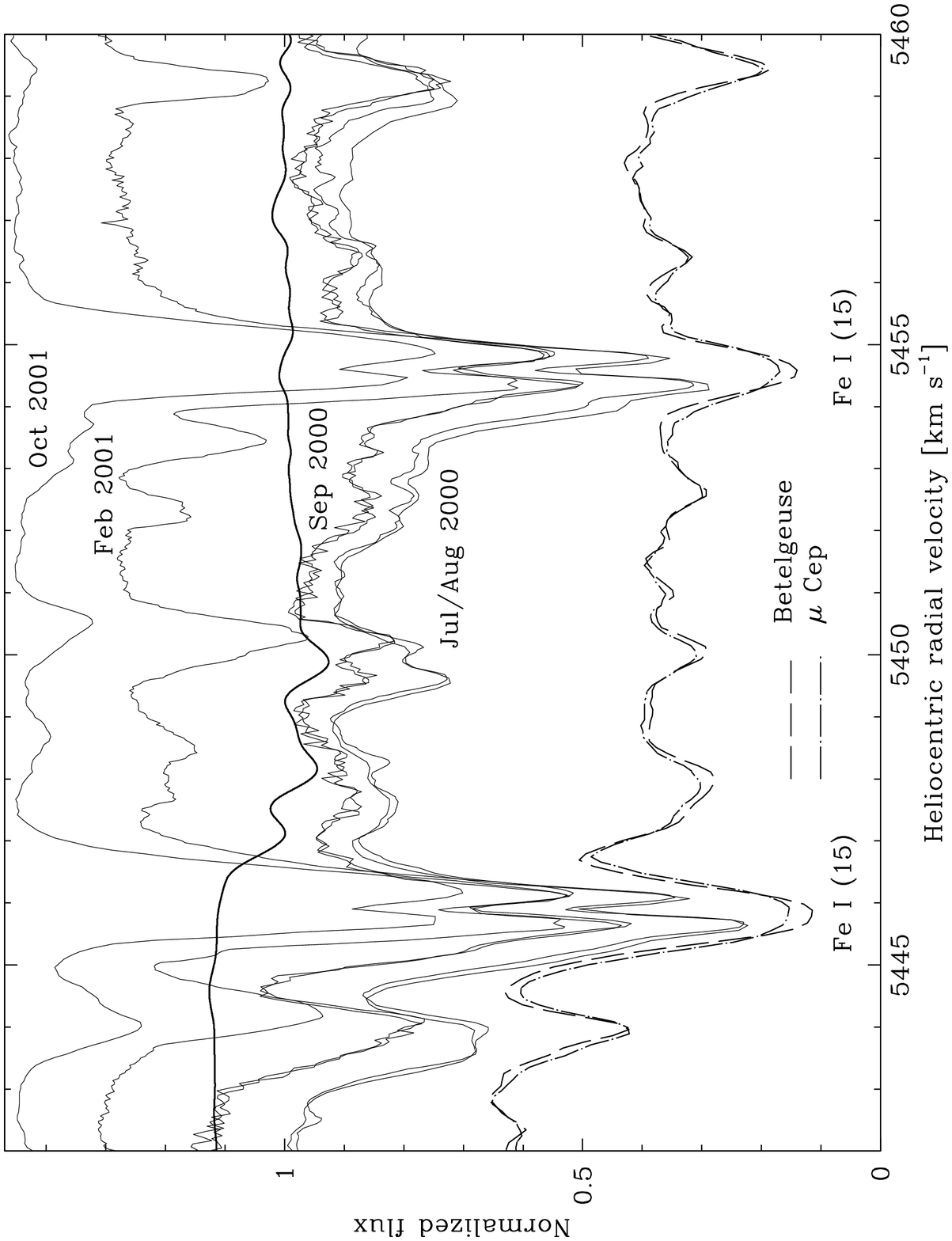]{
The outburst spectra of $\rho$ Cas of 2000 Juny, August, and September show the development 
of optical titanium oxide (TiO) bands with three bandheads around 5446~\AA. 
For display purposes, the spectra are shifted upwards by 15\% of the normalized stellar continuum level. 
The TiO bands disappear in the spectra of 2001 February and October when $T_{\rm eff}$
increases above 4500 K ({\em upper thin solid lines}). The TiO band is permanent in the early M-type 
supergiants Betelgeuse and $\mu$~Cep ({\em lower broken lines}). A spectrum synthesis with only TiO 
lines shows how these bands significantly decrease the continuum flux level longward of the bandheads
({\em bold solid line}). Note the typical split cores of Fe~{\sc i} lines in $\rho$ Cas, caused by central 
emission, which is absent in Betelgeuse and $\mu$ Cep. Note also how the component ratio of the 
split lines varies over time with the atmospheric dynamics.\label{fig10}}
\end{figure}
\begin{figure}
\plotone{f10.eps}
\end{figure}

\newpage
\clearpage

\begin{figure}
\figcaption[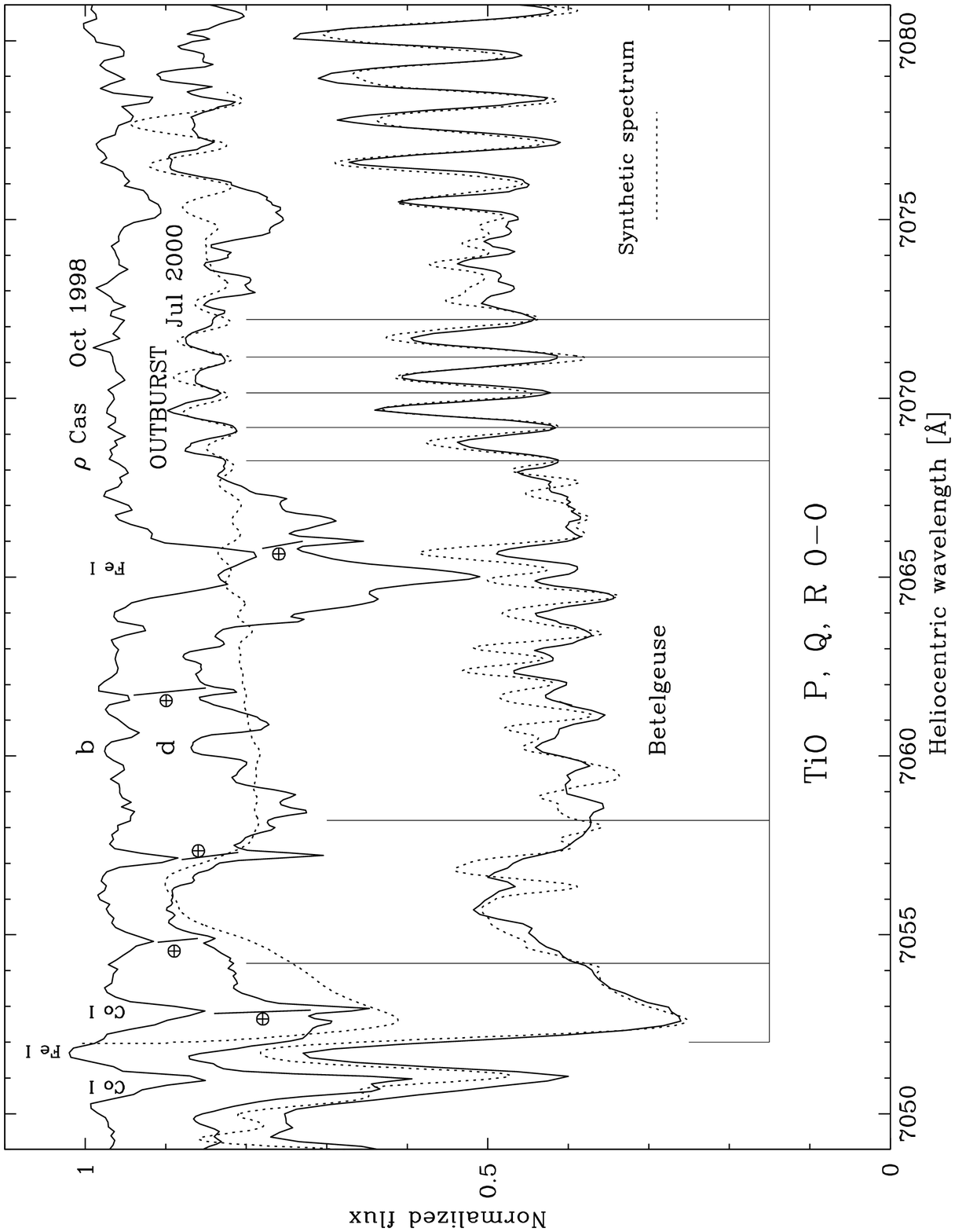]{Near-IR TiO absorption
bands are observed during the outburst of $\rho$ Cas in 2000 July ({\em middle solid line}).
The weak TiO bands around 7070~\AA\, have distinct shapes ({\em at vertical solid lines}),
also observed in Betelgeuse ({\em lower solid line}). The synthetic 
spectrum of $\rho$ Cas is computed with only TiO lines (of the P, Q, \& R branches) ({\em upper dotted line}), 
while that of Betelgeuse ({\em lower dotted line}) also includes molecular and atomic lines. 
Note that the TiO bands are considerably broader than the telluric lines ({\em marked with $\oplus$}).\label{fig11}}
\end{figure}
\begin{figure}
\plotone{f11.eps}
\end{figure}

\newpage
\clearpage

\begin{figure}
\figcaption[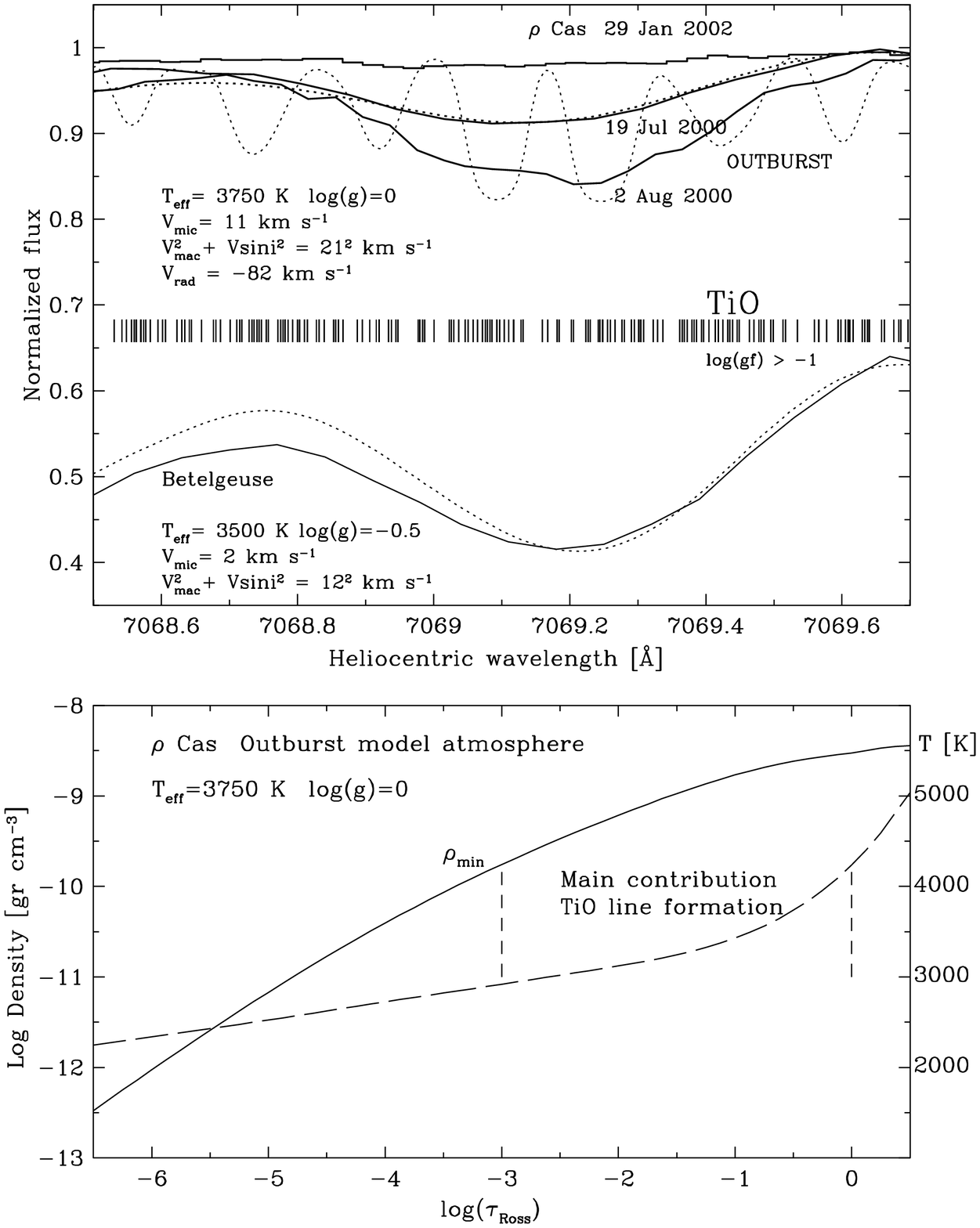]{The TiO band at 7069.2~\AA,
observed during the outburst of $\rho$ Cas on 2000 July 19 in the upper panel, 
is best fit ({\em dotted line}) for a model atmosphere with $T_{\rm eff}$=3750~K 
and $\log{g}$=0 in the lower panel. The spectrum of 2002 January with higher $T_{\rm eff}$
does not show the TiO bands. A microturbulence velocity of 11~$\rm km\,s^{-1}$, 
and macrobroadening of 21~$\rm km\,s^{-1}$ are required to broaden the 
synthetic spectrum ({\em dotted line}) of $\rho$ Cas to the observed shape of the TiO band. The best fit
yields a radial velocity of $-$82~$\rm km\,s^{-1}$, or an expansion velocity of 35~$\rm km\,s^{-1}$.
The strongest TiO lines for the synthesis, with $\log{gf}$-values $>$$-$1, are marked ({\em vertical lines}).
The synthetic spectrum for Betelgeuse of Fig. 11 ({\em lower dotted line}), and the fit parameters 
are also shown. The model reveals that the main contribution to the TiO band formation occurs at 
$-$3 $\leq$ $\log{\tau_{\rm Ross}}$ $\leq$ 0, with a lower gas density limit of
$\rho_{\rm min}$=$10^{-10}$~gr~$\rm cm^{-3}$ (see text).\label{fig12}}
\end{figure}
\begin{figure}
\plotone{f12.eps}
\end{figure}

\newpage
\clearpage

\begin{figure}
\figcaption[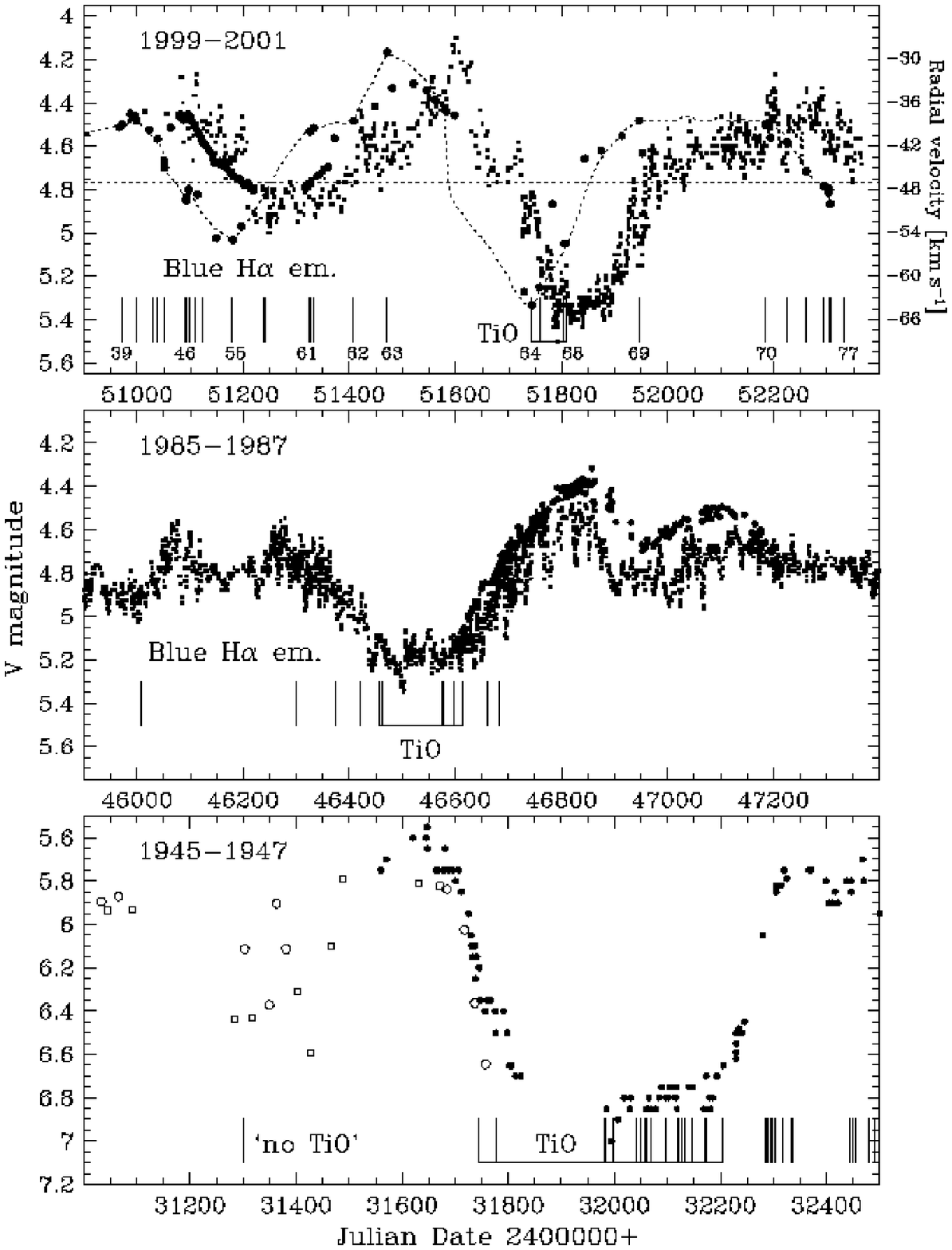]{The panels compare the $V$-brightness changes 
during the three outbursts of $\rho$ Cas in the last century. The strong 
$V$-brightness decrease by $1^{\rm m}.2$$-$$1^{\rm m}.4$ in 1946 and 2000 
is preceded by a pre-outburst cycle and a bright maximum.
The deep brightness minima during the three outbursts have considerably shortened 
from $\sim$400 d in 1946, $\sim$200 d in 1986, to $\sim$100 d in 2000. 
TiO bands are not observed in 1945 August ({\em vertical line labeled `no TiO'}),
before the brightness maximum. Neither are they observed in the high-resolution 
pre-outburst spectra of 1999 ({\em vertical lines No. 39 to 63}).
The TiO bands develop in the summer of 2000 ({\em Nos. 64 \& 65}), before the deep minimum, 
and disappear during the brightness increase ({\em No. 69}). 
TiO bands are also observed before the deep minimum of 1946. 
It indicates that TiO forms only during phases of very fast atmospheric expansion,
when the atmosphere sufficiently cools to $T_{\rm eff}$$\leq$4250 K.  
The outburst of 1986 is rather moderate. Optical TiO bands are only observed 
during the deep minimum. Strong emission in the blue wing of H$\alpha$ 
develops before the outbursts of 1986 and 2000, when the photosphere collapses
(see text).\label{fig13}}
\end{figure}
\begin{figure}
\plotone{f13.eps}
\end{figure}

\newpage
\clearpage

\begin{figure}
\figcaption[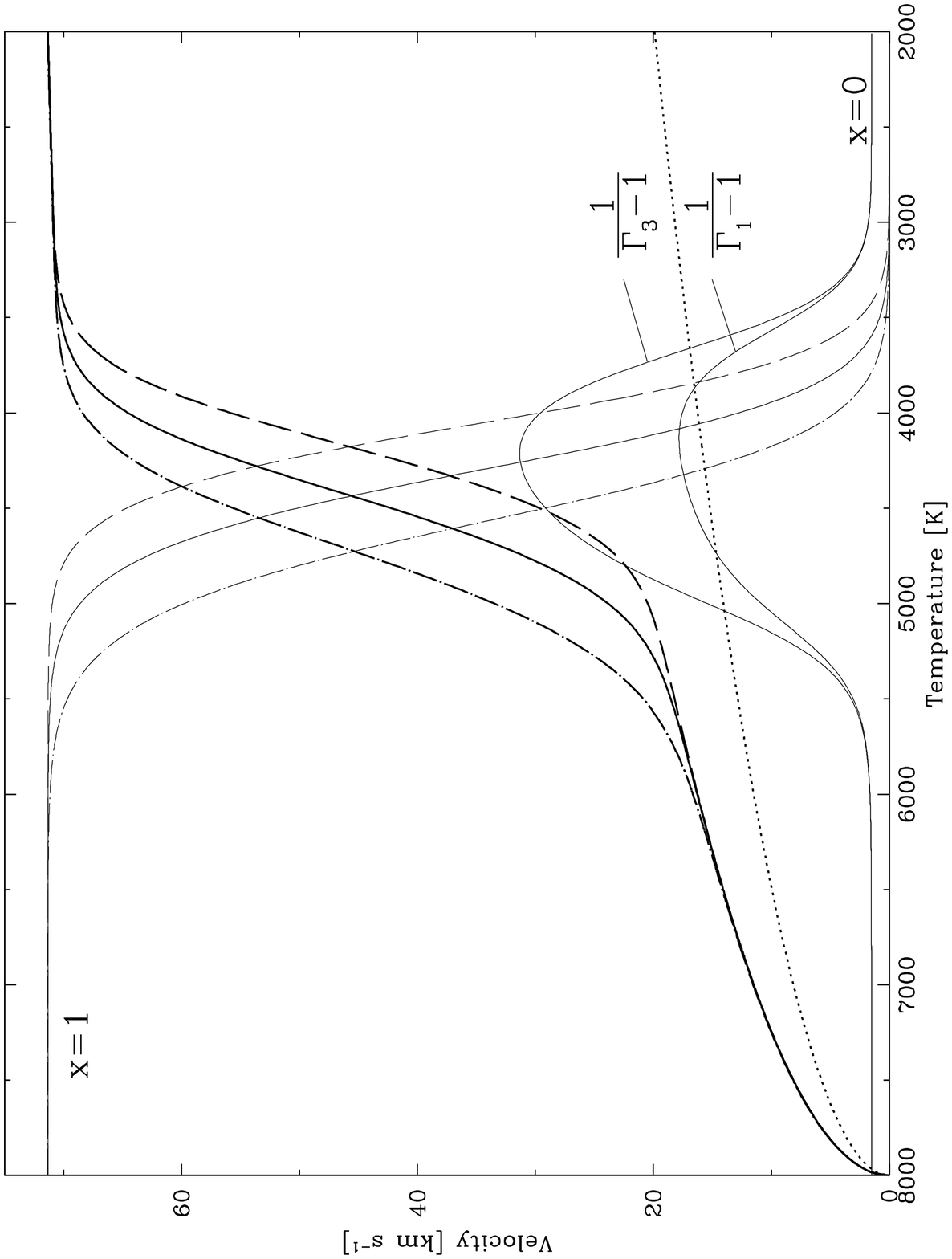]{
The expansion velocity of recombining hydrogen gas with an initial kinetic 
temperature of 8000~K, is computed as a function of the final gas temperature $T$ ({\em bold lines}).
With the recombination during cooling, the velocity of gas parcels that cool to  
4000 K $\leq$$T$$\leq$6000 K, with an ionization fraction of $x$$\sim$0.5 ({\em thin lines}), 
strongly increases, because the recombination energy drives the expansion with the conservation 
of total energy and flow momentum. The gas parcels are initially at rest, and their acceleration
is computed for flow momenta of 0.1 ({\em dashed lines}), 1 ({\em solid lines}), 
and 10 ({\em dash-dotted lines}) times the initial gas pressure. In the partial
hydrogen ionization-recombination zone the gas parcels become very expandable or compressible, 
because 1/($\Gamma_{1}$$-$1) and 1/($\Gamma_{3}$$-$1) assume maximum values. The dotted 
line is computed without recombination (i.e. pure thermal expansion), which
does not yield sufficiently large photospheric expansion velocities observed for 
the outburst of $\rho$ Cas.\label{fig14}}
\end{figure}
\begin{figure}
\plotone{f14.eps}
\end{figure}

\newpage
\clearpage

\begin{figure}
\figcaption[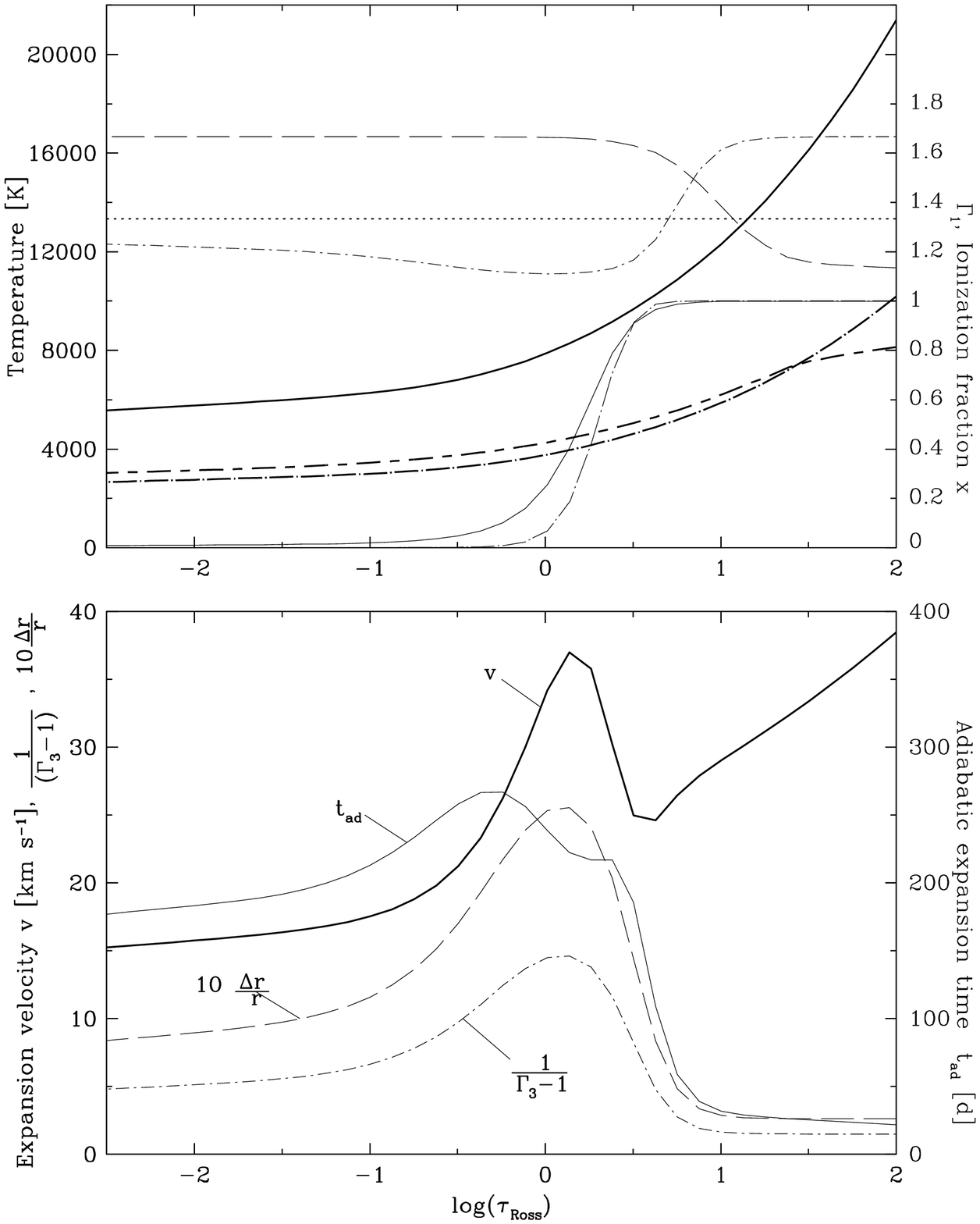]{Top: Kinetic temperature structure 
$T$ with optical depth ({\em thick solid line}), the ionization fraction $x$ ({\em thin solid line}), 
and $\Gamma_{1}$ ({\em thin dot short-dashed line}) of the model atmosphere
before outburst ($T_{\rm eff}$=7250 K). 
The changes of $T$ ({\em thick long-dashed short-dashed line}), $x$ ({\em thin dot long-dashed line}), 
and $\Gamma_{1}$ ({\em thin long-dashed dotted}) after the outburst (to $T_{\rm eff}$=3750 K) 
are also plotted. $\Gamma_{1}$ increases from a value below 4/3 for dynamic 
atmospheric stability ({\em horizontal dotted line}), to well above 4/3 beyond 
the partial recombination zone of hydrogen ($-$0.5 $\leq$ $\log{\tau_{\rm Ross}}$ $\leq$0.5). 
Bottom: The cooling of 3500 K increases the photospheric outflow 
velocity $v$ ({\em bold solid line}) in the partial ionization zone  
to above 35~$\rm km\,s^{-1}$, observed for the TiO bands. These atmospheric layers are 
very expandable, where 1/($\Gamma_{3}$$-$1) assumes maximum values ({\em thin dash-dotted line}),
and yield a relative radius increase $\Delta$$r$/$r$$\simeq$2.5 ({\em thin dashed line}).
The time-scale for spherical adiabatic expansion $t_{\rm ad}$ ({\em thin solid line})
in this region at $r$=$R_{*}$$\simeq$400~$\rm R_{\odot}$ ranges between 
170 and 267 d, comparable to the observed time of $\sim$200 d for 
the rapid $V$-brightness decrease during the outburst of $\rho$ Cas (see text).\label{fig15}}
\end{figure}
\begin{figure}
\plotone{f15.eps}
\end{figure}

\clearpage

\begin{deluxetable}{rcccclccrl}
\tabcolsep0.07cm 
\tablecaption{Near-UV, optical, and near-IR high-resolution echelle spectra of $\rho$~Cas in 1993-2002\label{tbl-1}}
\tablehead{
\colhead{No.}  & \colhead{Telescope-} & \colhead{Obs Date} & \colhead{Jul. Date} & \colhead{Time} & \colhead{Air-} & \colhead{Wav. range} & \colhead{Detector} & \colhead{$R$} & \colhead{Exp. times} \\
\colhead{}     & \colhead{Instr.}     & \colhead{}          & \colhead{2400000$+$} & \colhead{(UT)}       & \colhead{mass} & \colhead{($\rm \AA$)} & \colhead{/Grating}   & \colhead{}    & \colhead{(s)} 
}
\startdata
 1   & Ritter      & 1993 Nov 28  & 49319  & 04:28   & 1.2  & 5473$-$6799 & CCD1    &  26,000 &  1200                \\    
 2   & SAO/AN      & 1993 Nov 29  & 49320  &         &      & 4400$-$6700 &         &  36,000 &                      \\
 3   & Ritter      & 1993 Dec 01  & 49322  & 02:28   & 1.11 & 5473$-$6799 & CCD1    &  26,000 &  1200                \\ 
 4   & WHT-UES     & 1993 Dec 20  & 49342  & 20:28   & 1.18 & 4800$-$7740 & E31     &  50,000 &  100,2$\times$150    \\
 5   & WHT-UES     & 1994 Jul 25  & 49558  & 02:05   & 1.33 & 4800$-$7740 & E31     &  50,000 &  60,3$\times$50      \\
 6   & WHT-UES     & 1994 Aug 20  & 49584  & 03:57   & 1.15 & 4800$-$7740 & E31     &  50,000 &  4$\times$100        \\
 7   & Ritter      & 1994 Sep 18  & 49613  & 05:47   & 1.04 & 5470$-$6796 & CCD1    &  26,000 &  1800                \\
 8   & Ritter      & 1994 Oct 11  & 49636  & 05:48   & 1.08 & 5469$-$6794 & CCD1    &  26,000 &  1800                \\
 9   & Ritter      & 1994 Nov 11  & 49667  & 03:35   & 1.07 & 5469$-$6794 & CCD1    &  26,000 &  1200                \\
 10  & Ritter      & 1994 Dec 23  & 49709  & 01:08   & 2.02 & 5473$-$6799 & CCD1    &  26,000 &   900                \\
 11  & WHT-UES     & 1995 Apr 18  & 49825  & 05:28   & 2.37 & 4490$-$6690 & E31     &  50,000 &  200,400             \\
 12  & SAO/AN      & 1995 Jun 09  & 49877  & 04:11   & 1.1  & 4924$-$6600 &         &  36,000 &                      \\
 13  & SAO/AN      & 1995 Jun 15  & 49883  & 03:51   & 1.1  & 4924$-$6600 &         &  36,000 &                      \\
 14  & SAO/AN      & 1995 Jul 31  & 49929  &         &      & 5496$-$6575 &         &  80,000 &                      \\
 15  & SAO/AN      & 1995 Aug 04  & 49933  & 22:57   & 1.05 & 4925$-$6601 &         &  36,000 &                      \\
 16  & SAO/AN      & 1995 Sep 06  & 49966  &         &      & 5499$-$6578 &         &  80,000 &                      \\
 17  & Ritter      & 1995 Oct 13  & 50003  & 04:32   & 1.04 & 5472$-$6798 & CCD1    &  26,000 &   900                \\
 18  & SAO/AN      & 1995 Oct 13  & 50004  & 21:37   & 1.08 & 4923$-$6598 &         &  36,000 &                      \\
 19  & SAO/AN      & 1995 Dec 03  & 50055  & 18:41   & 1.10 & 4928$-$6605 &         &  36,000 &                      \\
 20  & SAO/A       & 1996 May 06  & 50209  & 02:05   & 1.32 & 4906$-$6634 &         &  36,000 &                      \\
 21  & WHT-UES     & 1996 Jun 05  & 50239  & 05:06   & 1.35 & 4490$-$6690 & E31     &  50,000 & 100,60               \\
 22  & SAO/AN      & 1996 Jul 11  & 50275  & 00:51   & 1.04 & 4902$-$6629 &         &  36,000 &                      \\
 23  & SAO/AN      & 1996 Jul 31  & 50296  & 23:43   & 1.04 & 4902$-$6629 &         &  36,000 &                      \\
 24  & SAO/AN      & 1996 Aug 27  & 50322  & 00:22   & 1.08 & 4901$-$6627 &         &  36,000 &                      \\
 25  & WHT-UES     & 1996 Sep 28  & 50354  & 00:20   & 1.14 & 4490$-$6690 & E31     &  50,000 & 40,60,90             \\
 26  & SAO/AN      & 1996 Nov 21  & 50409  & 18:29   & 1.06 & 4897$-$6622 &         &  36,000 &                      \\
 27  & WHT-UES     & 1996 Dec 26  & 50444  & 23:10   & 1.71 & 3840$-$4990 & E31     &  50,000 &  300                 \\
 28  & Ritter      & 1997 Jan 08  & 50456  & 23:47   & 1.9  & 5341$-$6597 & CCD1    &  26,000 & 3600                 \\
 29  & SAO/AN      & 1997 Jan 18  & 50467  & 16:08   & 1.14 & 4554$-$6760 &         &  80,000 &                      \\
 30  & SAO/AN      & 1997 Jan 22  & 50471  & 15:24   & 1.12 & 4912$-$6641 &         &  36,000 &                      \\
 31  & Ritter      & 1997 Jan 29  & 50477  & 00:38   & 1.25 & 5341$-$6597 & CCD1    &  26,000 &  3600                \\
 32  & WHT-UES     & 1997 Feb 20  & 50500  & 19:26   & 1.68 &  cen 5513   & E79\tablenotemark{a} &  50,000 &  2$\times$30,2$\times$100  \\
 33  & SAO/AN      & 1997 Feb 24  & 50504  & 16:10   & 1.49 & 5499$-$6570 &         & 170,000 &                      \\
 34  & SAO/AN      & 1997 Mar 24  & 50532  &         &      & 5321$-$6150 &         & 170,000 &                      \\
 35  & SAO/AN      & 1997 Jun 14  & 50613  & 00:26   & 1.18 & 4906$-$6634 &         &  40,000 &                      \\
 36  & SAO/AN      & 1997 Aug 31  & 50692  & 02:30   & 1.3  & 4915$-$6624 &         &  40,000 &                      \\
 37  & Ritter      & 1997 Sep 27  & 50718  & 04:03   & 1.06 & 5340$-$6596 & CCD1    &  26,000 &  1200                \\
 38  & Ritter      & 1997 Oct 22  & 50743  & 06:17   & 1.15 & 5339$-$6596 & CCD1    &  26,000 &  1200                \\
 39  & Ritter      & 1998 Jun 08  & 50972  & 08:26   & 1.3  & 5340$-$6596 & CCD1    &  26,000 &  1800                \\
 40  & Ritter      & 1998 Jul 05  & 50999  & 07:57   & 1.15 & 5340$-$6596 & CCD1    &  26,000 &  1800                \\
 41  & WHT-UES     & 1998 Aug 03  & 51029  & 23:00   & 2.01 & 3100$-$3700 & E31     &  50,000 &  600                 \\
 42  & WHT-UES     & 1998 Aug 04  & 51030  & 04:43   & 1.14 & 3100$-$3700 & E31     &  50,000 &  900                 \\
 43  & Ritter      & 1998 Aug 13  & 51038  & 07:30   & 1.05 & 5339$-$6596 & CCD1    &  26,000 &  1800                \\
 44  & Ritter      & 1998 Aug 26  & 51051  & 06:02   & 1.06 & 5339$-$6596 & CCD1    &  26,000 &  2000                \\
 45  & Ritter      & 1998 Aug 27  & 51052  & 08:51   & 1.09 & 5339$-$6596 & CCD1    &  26,000 &  2000                \\
 46  & NOT-Sofin   & 1998 Oct 04  & 51091  & 23:58   & 1.14 & 4507$-$6110 & Cam1    & 160,000 &  2$\times$600        \\
 47  & NOT-Sofin   & 1998 Oct 05  & 51092  & 23:19   & 1.15 & 5060$-$7500 & Cam2    &  80,000 &  200,3$\times$180    \\
 48  & NOT-Sofin   & 1998 Oct 06  & 51092  & 01:17   & 1.18 & 3420$-$11550& Cam3    &  33,000 &  190,110             \\
 49  & NOT-Sofin   & 1998 Oct 08  & 51094  & 00:39   & 1.18 & 4791$-$7240 & Cam1    & 160,000 &  3$\times$600        \\ 
 50  & WHT-UES     & 1998 Oct 08  & 51095  & 22:14   & 1.20 & 4050$-$8800 & E31     &  50,000 &  2$\times$100        \\
 51  & NOT-Sofin   & 1998 Oct 11  & 51097  & 02:38   & 1.43 & 4322$-$5620 & Cam2    &  80,000 &  5$\times$900        \\
 52  & NOT-Sofin   & 1998 Oct 11  & 51097  & 04:12   & 1.81 & 3470$-$11550& Cam3    &  33,000 &  2$\times$300,2$\times$240  \\
 53  & WHT-UES     & 1998 Oct 22  & 51109  & 23:55   & 1.16 & 4050$-$8800 & E31     &  50,000 &  2$\times$100,50            \\ 
 54  & NOT-Sofin   & 1998 Nov 05  & 51123  & 23:51   & 1.24 & 4507$-$6289 & Cam1    & 160,000 &  480,300                    \\ 
 55  & WHT-UES     & 1999 Jan 01  & 51180  & 19:35   & 1.17 & 4050$-$8800 & E31     &  50,000 &  2$\times$100               \\
 56  & NOT-Sofin   & 1999 Feb 28  & 51238  & 19:44   & 1.97 & 4407$-$5660 & Cam2    &  80,000 &  3$\times$900               \\
 57  & NOT-Sofin   & 1999 Mar 02  & 51240  & 19:29   & 2.07 & 4407$-$5660 & Cam2    &  80,000 &  6$\times$900               \\
 58  & NOT-Sofin   & 1999 May 27  & 51325  & 05:52   & 1.33 & 4407$-$5660 & Cam2    &  80,000 &  400                        \\
 59  & NOT-Sofin   & 1999 May 28  & 51326  & 05:07   & 1.45 & 3468$-$11332& Cam3    &  33,000 &  450,230                    \\
 60  & NOT-Sofin   & 1999 May 30  & 51328  & 03:15   & 1.95 & 4407$-$5660 & Cam2    &  80,000 &  725,600                    \\
 61  & NOT-Sofin   & 1999 Jun 04  & 51333  & 04:34   & 1.48 & 3495$-$11332& Cam3    &  33,000 &  2$\times$80,180            \\
 62  & WHT-UES     & 1999 Aug 18  & 51408  & 03:08   & 1.14 & 4050$-$8800 & E31     &  50,000 &  2$\times$100,2$\times30$   \\
 63  & NOT-Sofin   & 1999 Oct 19  & 51471  & 23:19   & 1.14 & 3495$-$11333& Cam3    &  33,000 &  430,2$\times$80            \\
 64  & WHT-UES     & 2000 Jul 19  & 51744  & 02:10   & 1.37 & 4050$-$8800 & E31     &  50,000 &  150                        \\
 65  & WHT-UES     & 2000 Aug 02  & 51758  & 05:12   & 1.15 & 5410$-$8800 & E31     &  50,000 &  60,30                      \\
 66  & Ritter      & 2000 Sep 17  & 51804  & 06:38   & 1.13 & 5339$-$6596 & CCD1    &  26,000 &  3000                       \\
 67  & Ritter      & 2000 Sep 20  & 51808  & 05:28   & 1.17 & 5339$-$6596 & CCD1    &  26,000 &  3000                       \\
 68  & Shajn       & 2000 Sep 21  & 51809  & 22:41   & 1.04 & \tablenotemark{c} & CCD &  30,000 &  1800,1200,3000 \\
 69  & WHT-UES     & 2001 Feb 05  & 51946  & 20:34   & 1.72 &  cen 7090 & E79\tablenotemark{a}   &  50,000 &  300,2$\times$240  \\ 
 70  & WHT-UES     & 2001 Oct 02  & 52184  & 03:14   & 1.33 & 4050$-$8800 & E31     &  50,000 &  100,2$\times90$                \\
 71  & Ritter      & 2001 Nov 12  & 52225  & 04:52   & 1.13 & 5339$-$6596 & CCD1    &  26,000 &  3600                           \\
 72  & TBL         & 2001 Dec 18  & 52262  & 20:28   & 1.15 & 4489$-$6619 & CCD     &  35,000 &  1600                           \\
 73  & Ritter      & 2002 Jan 19  & 52294  & 23:52   & 1.15 & 5339$-$6596 & CCD1    &  26,000 &  3600                           \\
 74  & NOT-Sofin   & 2002 Jan 29  & 52303  & 19:53   & 1.4  & 3758$-$11225& Cam2    &  80,000 &  \tablenotemark{b}              \\
 75  & NOT-Sofin   & 2002 Jan 31  & 52305  & 19:44   & 1.4  & 3620$-$10947& Cam2    &  80,000 &  \tablenotemark{b}              \\ 
 76  & NOT-Sofin   & 2002 Feb 01  & 52306  & 19:36   & 1.4  & 3572$-$11321& Cam3    &  33,000 &  \tablenotemark{b}              \\
 77  & Shajn       & 2002 Feb 26  & 52332  & 16:12   & 1.46 & \tablenotemark{c} & CCD&  30,000 & 900,900,900                    \\
 78  & Ritter      & 2002 Jul 07  & 52462  & 08:07   & 1.55 & 5339$-$6596 & CCD1    &  26,000 &  3000                           \\ 
\enddata
\tablenotetext{a}{ E79 grating with 18 orders; exposures at indicated central wavelength contain no H$\alpha$}
\tablenotetext{b}{ S/N $>$ 300 at echelle orders 6000$-$9000 \AA}
\tablenotetext{c}{ Coud\'{e} spectra of 30~\AA\, and 64~\AA, centered at Ca~{\sc ii} $\lambda$8540, H$\alpha$, and Na $D$}  
\end{deluxetable}

\clearpage

\begin{deluxetable}{cccccccc}
\tabcolsep0.07cm 
\tablecaption{H$\alpha$ and Fe~{\sc i} radial velocities, and equivalent width values 
from echelle spectra of $\rho$ Cas in Table 1\label{tbl-2}}
\tablehead{
\colhead{Spectrum No.}  &  \colhead{Obs Date} & \colhead{Jul. Date}  & Fe~{\sc i} $\lambda$5572   &                     &   H$\alpha$         &                    &   Fe~{\sc ii} $\lambda$5325     \\
\colhead{}              &  \colhead{}         & \colhead{2400000$+$} &        (m\AA)              &  ($\rm km\,s^{-1}$) &  (m\AA) & ($\rm km\,s^{-1}$) &  (m\AA)      
}
\startdata
 1   &  1993 Nov 28  &  49319 &  341  &   $-$53.5  &   707  & $-$35.5 &  -  \\    
 2   &  1993 Nov 29  &  49320 &  321  &   $-$54.0  &   709  & $-$34.5 & 571 \\
 3   &  1993 Dec 01  &  49322 &  360  &   $-$54.5  &   700  & $-$34.5 &     \\ 
 4   &  1993 Dec 20  &  49342 &  393  &   $-$59.0  &   770  & $-$35.0 & 541 \\
 5   &  1994 Jul 25  &  49558 &  416  &   $-$50.5  &   743  & $-$62.5 & 377 \\
 6   &  1994 Aug 20  &  49584 &       &            &        &         &     \\
 7   &  1994 Sep 18  &  49613 &  396  &   $-$50.0  &   726  & $-$65.0 &  -  \\
 8   &  1994 Oct 11  &  49636 &  425  &   $-$51.0  &   790  & $-$66.0 &  -  \\
 9   &  1994 Nov 11  &  49667 &  483  &   $-$53.0  &   763  & $-$65.5 &  -  \\
 10  &  1994 Dec 23  &  49709 &  513  &   $-$53.0  &   722  & $-$63.0 &  -  \\
 11  &  1995 Apr 18  &  49825 &  551  &   $-$51.0  &   764  & $-$56.5 & 366 \\
 12  &  1995 Jun 09  &  49877 &  535  &   $-$50.5  &   684  & $-$60.0 & 371 \\
 13  &  1995 Jun 15  &  49883 &  525  &   $-$49.5  &   694  & $-$59.0 &     \\
 14  &  1995 Jul 31  &  49929 &  497  &   $-$48.0  &   773  & $-$57.0 & 328 \\
 15  &  1995 Aug 04  &  49933 &  508  &   $-$48.0  &   702  & $-$55.5 &     \\
 16  &  1995 Sep 06  &  49966 &  511  &   $-$47.0  &   830  & $-$57.5 & 364 \\
 17  &  1995 Oct 13  &  50003 &  467  &   $-$47.5  &   902  & $-$52.5 & -   \\
 18  &  1995 Oct 13  &  50004 &  462  &   $-$49.0  &   899  & $-$54.0 & 434 \\
 19  &  1995 Dec 03  &  50055 &  465  &   $-$56.5  &   772  & $-$50.5 & 438 \\
 20  &  1996 May 06  &  50209 &  382  &   $-$52.0  &   620  & $-$49.5 & 465 \\
 21  &  1996 Jun 05  &  50239 &  449  &   $-$46.5  &   619  & $-$46.5 & 445 \\
 22  &  1996 Jul 11  &  50275 &  573  &   $-$49.0  &   677  & $-$46.5 &     \\
 23  &  1996 Jul 31  &  50296 &  555  &   $-$51.0  &   695  & $-$45.0 & 435 \\
 24  &  1996 Aug 27  &  50322 &  535  &   $-$48.5  &   662  & $-$42.5 & 435 \\
 25  &  1996 Sep 28  &  50354 &  527  &   $-$48.0  &   566  & $-$41.0 & 401 \\
 26  &  1996 Nov 21  &  50409 &  481  &   $-$49.5  &   286  & $-$43.0 & 333 \\
 27  &  1996 Dec 26  &  50444 &   -   &    -       &    -   &   -     &  -  \\
 28  &  1997 Jan 08  &  50456 &  561  &   $-$45.0  &   288  & $-$44.5 &  -  \\
 29  &  1997 Jan 18  &  50467 &  529  &   $-$44.5  &   208  & $-$43.5 & 326 \\
 30  &  1997 Jan 22  &  50471 &  526  &   $-$43.5  &   210  & $-$43.0 & 327 \\
 31  &  1997 Jan 29  &  50477 &  538  &   $-$43.0  &   283  & $-$43.0 & -   \\
 32  &  1997 Feb 20  &  50500 &  -    &    -       &   -    &   -     &     \\
 33  &  1997 Feb 24  &  50504 &  578  &   $-$42.5  &   256  & $-$42.5 & -   \\
 34  &  1997 Mar 24  &  50532 &  523  &   $-$41.0  &    -   &    -    & 347 \\
 35  &  1997 Jun 14  &  50613 &  456  &   $-$34.5  &   392  & $-$36.5 & 369 \\
 36  &  1997 Aug 31  &  50692 &  455  &   $-$41.5  &   540  & $-$41.0 & 327 \\
 37  &  1997 Sep 27  &  50718 &  433  &   $-$43.5  &   569  & $-$37.5 & -   \\
 38  &  1997 Oct 22  &  50743 &  412  &   $-$45.0  &   580  & $-$35.5 & -   \\
 39  &  1998 Jun 08  &  50972 &  408  &   $-$40.5  &   642  & $-$36.0 & -   \\
 40  &  1998 Jul 05  &  50999 &  413  &   $-$40.0  &   670  & $-$36.5 & -   \\
 41  &  1998 Aug 03  &  51029 &  -    &     -      &    -   &    -    & -   \\
 42  &  1998 Aug 04  &  51030 &  -    &     -      &    -   &    -    & -   \\
 43  &  1998 Aug 13  &  51038 &  382  &   $-$42.5  &   636  & $-$37.0 & -   \\
 44  &  1998 Aug 26  &  51051 &  354  &   $-$45.5  &   605  & $-$36.5 & -   \\
 45  &  1998 Aug 27  &  51052 &  356  &   $-$46.5  &   604  & $-$36.5 & -   \\
 46  &  1998 Oct 04  &  51091 &   -   &    -       &    -   &   -     & -   \\
 47  &  1998 Oct 05  &  51092 &  314  &   $-$51.0  &   534  & $-$36.0 & 434 \\
 48  &  1998 Oct 06  &  51093 &  315  &   $-$50.5  &   563  & $-$36.0 & 439 \\
 49  &  1998 Oct 08  &  51094 &   -   &    -       &    -   &   -     &  -  \\ 
 50  &  1998 Oct 08  &  51095 &  317  &   $-$50.5  &   555  & $-$36.0 & 445 \\
 51  &  1998 Oct 11  &  51097 &  325  &   $-$49.5  &   485  & $-$35.0 &  -  \\
 52  &  1998 Oct 11  &  51097 &  325  &   $-$49.5  &   485  & $-$35.0 &     \\
 53  &  1998 Oct 22  &  51109 &       &            &        &         &     \\ 
 54  &  1998 Nov 05  &  51123 &   -   &     -      &    -   &   -     &  -  \\ 
 55  &  1999 Jan 01  &  51180 &  497  &   $-$56.5  &   655  & $-$42.5 & 508 \\
 56  &  1999 Feb 28  &  51238 &   -   &     -      &    -   &   -     &  -  \\
 57  &  1999 Mar 02  &  51240 &   -   &     -      &    -   &   -     &  -  \\
 58  &  1999 May 27  &  51325 &   -   &     -      &    -   &   -     &  -  \\
 59  &  1999 May 28  &  51326 &  568  &   $-$41.5  &   505  & $-$41.5 & 367 \\
 60  &  1999 May 30  &  51328 &   -   &    -       &    -   &   -     &  -  \\
 61  &  1999 Jun 04  &  51333 &  553  &   $-$41.0  &   532  & $-$41.5 & 381 \\
 62  &  1999 Aug 18  &  51408 &  502  &   $-$40.0  &   418  & $-$38.5 & 401 \\
 63  &  1999 Oct 19  &  51471 &  478  &   $-$30.5  &   505  & $-$31.5 & 431 \\
 64  &  2000 Jul 19  &  51744 &  583  &   $-$65.5  &   368  & $-$42.5 & 323 \\
 65  &  2000 Aug 02  &  51758 &  578  &   $-$63.0  &   419  & $-$46.0 &  -  \\
 66  &  2000 Sep 17  &  51804 &  592  &   $-$57.0  &   279  & $-$44.0 & 263 \\
 67  &  2000 Sep 20  &  51808 &  609  &   $-$57.0  &   256  & $-$44.0 & 279 \\
 68  &  2000 Sep 21  &  51809 &  -    &     -      &  $-$82 & $-$44.0 &  -  \\
 69  &  2001 Feb 05  &  51946 &  642  &   $-$40.0  &    -   &    -    &  -  \\ 
 70  &  2001 Oct 02  &  52184 &  369  &   $-$40.5  &   452  & $-$35.0 & 449 \\
 71  &  2001 Nov 12  &  52225 &  362  &   $-$43.0  &   471  & $-$34.5 & 513 \\
 72  &  2001 Dec 18  &  52262 &  282  &   $-$47.0  &   214  & $-$34.5 & 493 \\
 73  &  2001 Jan 19  &  52294 &  365  &   $-$49.0  &   180  & $-$34.5 & 419 \\
 74  &  2002 Jan 29  &  52303 &  334  &   $-$50.0  &   112  & $-$35.5 &  -  \\
 75  &  2002 Jan 31  &  52305 &  344  &   $-$49.5  &   084  & $-$35.0 & 412 \\ 
 76  &  2002 Feb 01  &  52306 &  331  &   $-$51.5  &   014  & $-$35.5 & 406 \\
 77  &  2002 Feb 26  &  52332 &   -   &     -      &   178  & $-$35.0 &  -  \\
 78  &  2002 Jul 07  &  52462 &  455  &   $-$48.0  &   462  & $-$32.0 & 392 \\
\enddata
\end{deluxetable}

\clearpage

\begin{deluxetable}{cccrc}
\tabcolsep0.07cm 
\tablecaption{Heliocentric radial velocity values of $\rho$ Cas from Oak Ridge Observatory\label{tbl-3}}
\tablehead{
\colhead{Observation No.}  &  \colhead{Obs Date} & \colhead{Jul. Date}   &  Radial velocity    & $\sigma$-error  \\
\colhead{}                &  \colhead{}         & \colhead{2400000$+$}  &  ($\rm km\,s^{-1}$) & ($\rm km\,s^{-1}$)
}
\startdata
 1   &  1990 Jul 10  &       48082.7792 &    $-$49.25  &  1.76   \\
 2   &  1991 Jan 26  &       48282.5157 &    $-$57.44  &  1.64   \\
 3   &  1991 Jan 26  &       48283.4890 &    $-$56.81  &  1.69   \\ 
 4   &  1991 Feb 22  &       48309.5056 &    $-$53.49  &  1.57   \\
 5   &  1991 Jul 02  &       48439.7185 &    $-$43.87  &  1.31   \\
 6   &  1991 Jul 25  &       48462.7144 &    $-$43.38  &  1.50   \\
 7   &  1991 Aug 15  &       48483.6314 &    $-$44.75  &  1.46   \\
 8   &  1991 Sep 02  &       48501.8755 &    $-$46.22  &  1.58   \\
 9   &  1991 Oct 13  &       48542.8552 &    $-$46.26  &  1.42   \\
 10  &  1991 Dec 19  &       48609.5897 &    $-$44.32  &  1.61   \\
 11  &  1991 Dec 28  &       48618.6044 &    $-$44.04  &  1.86   \\
 12  &  1992 Jan 16  &       48637.5462 &    $-$44.12  &  1.66   \\
 13  &  1992 May 21  &       48763.8432 &    $-$48.21  &  1.80   \\
 14  &  1992 May 26  &       48768.8483 &    $-$48.86  &  1.76   \\
 15  &  1992 Jun 18  &       48791.7918 &    $-$50.42  &  1.63   \\
 16  &  1992 Jul 12  &       48815.7609 &    $-$51.73  &  1.66   \\
 17  &  1992 Aug 13  &       48847.6561 &    $-$53.35  &  1.91   \\
 18  &  1992 Sep 09  &       48874.5931 &    $-$54.74  &  1.78   \\
 19  &  1992 Sep 15  &       48880.8412 &    $-$56.35  &  1.89   \\
 20  &  1992 Nov 10  &       48936.6925 &    $-$57.40  &  1.90   \\
 21  &  1992 Dec 09  &       48965.6710 &    $-$56.62  &  1.99   \\
 22  &  1993 Jan 08  &       48995.5736 &    $-$55.63  &  1.90   \\
 23  &  1993 Oct 06  &       49266.8169 &    $-$45.37  &  1.48   \\
 24  &  1993 Oct 29  &       49289.7625 &    $-$44.76  &  1.59   \\
 25  &  1993 Nov 23  &       49314.6535 &    $-$48.48  &  2.00   \\
 26  &  1993 Dec 02  &       49323.6625 &    $-$50.23  &  2.08   \\
 27  &  1993 Dec 27  &       49348.6110 &    $-$58.16  &  2.29   \\
 28  &  1994 Jan 06  &       49358.6010 &    $-$60.94  &  2.14   \\
 29  &  1994 Jan 23  &       49375.5034 &    $-$64.57  &  2.08   \\
 30  &  1994 Jan 29  &       49382.4878 &    $-$65.05  &  1.97   \\
 31  &  1994 Feb 23  &       49400.4998 &    $-$67.10  &  1.97   \\
 32  &  1994 Aug 25  &       49589.5946 &    $-$59.09  &  1.61   \\
 33  &  1994 Sep 11  &       49606.8947 &    $-$58.50  &  1.59   \\
 34  &  1994 Sep 20  &       49615.5279 &    $-$58.52  &  1.57   \\
 35  &  1994 Oct 11  &       49636.7901 &    $-$59.44  &  1.53   \\
 36  &  1994 Oct 26  &       49651.7738 &    $-$60.46  &  1.57   \\
 37  &  1995 Jul 06  &       49904.7214 &    $-$49.09  &  1.70   \\
 38  &  1995 Jul 12  &       49910.7263 &    $-$50.05  &  1.75   \\
 39  &  1995 Jul 20  &       49918.7322 &    $-$50.58  &  1.81   \\
 40  &  1995 Sep 13  &       49973.5750 &    $-$52.59  &  1.89   \\
 41  &  1995 Oct 18  &       50008.7730 &    $-$54.22  &  1.80   \\
 42  &  1995 Dec 08  &       50059.6458 &    $-$55.76  &  1.76   \\
 43  &  1995 Dec 27  &       50078.5827 &    $-$54.88  &  1.66   \\
 44  &  1996 Jan 16  &       50098.5535 &    $-$53.49  &  1.65   \\
 45  &  1996 Jan 31  &       50113.5055 &    $-$51.85  &  1.67   \\
 46  &  1996 Feb 07  &       50121.4657 &    $-$51.60  &  1.64   \\
 47  &  1996 Aug 08  &       50303.6662 &    $-$50.19  &  1.74   \\
 48  &  1996 Aug 30  &       50325.6188 &    $-$50.79  &  1.71   \\
 49  &  1996 Oct 27  &       50383.7322 &    $-$51.21  &  1.67   \\ 
 50  &  1996 Nov 18  &       50405.6617 &    $-$49.79  &  1.72   \\
 51  &  1996 Dec 21  &       50438.6331 &    $-$49.88  &  1.65   \\
 52  &  1997 Jan 13  &       50461.5697 &    $-$47.96  &  1.92   \\
 53  &  1997 Jul 01  &       50630.7519 &    $-$39.87  &  1.58   \\ 
 54  &  1997 Jul 27  &       50656.7196 &    $-$41.22  &  1.52   \\ 
 55  &  1997 Sep 21  &       50712.8952 &    $-$42.83  &  1.49   \\
 56  &  1997 Oct 18  &       50739.7600 &    $-$42.76  &  1.53   \\
 57  &  1997 Nov 05  &       50757.7209 &    $-$43.67  &  1.43   \\
 58  &  1997 Nov 21  &       50773.6715 &    $-$44.83  &  1.63   \\
 59  &  1997 Dec 12  &       50794.6395 &    $-$45.93  &  1.54   \\
 60  &  1997 Dec 27  &       50809.5757 &    $-$45.69  &  1.52   \\
 61  &  1998 Jan 26  &       50840.4817 &    $-$45.56  &  1.45   \\
 62  &  1998 Jun 02  &       50966.8395 &    $-$43.42  &  1.53   \\
 63  &  1998 Jul 28  &       51022.6960 &    $-$43.58  &  1.54   \\
 64  &  1998 Sep 06  &       51062.6185 &    $-$46.40  &  1.61   \\
 65  &  1998 Oct 27  &       51113.7562 &    $-$51.44  &  1.71   \\
 66  &  1998 Dec 01  &       51148.6347 &    $-$54.76  &  1.73   \\ 
 67  &  1999 Jan 17  &       51196.4950 &    $-$54.74  &  1.85   \\
 68  &  1999 Jul 12  &       51371.7138 &    $-$42.95  &  1.69   \\
 69  &  1999 Sep 26  &       51447.8544 &    $-$38.65  &  1.56   \\ 
 70  &  1999 Oct 29  &       51480.7145 &    $-$37.05  &  1.49   \\
 71  &  1999 Dec 08  &       51520.6480 &    $-$36.80  &  1.47   \\
 72  &  2000 Jan 02  &       51545.6049 &    $-$36.62  &  1.34   \\ 
 73  &  2000 Jan 18  &       51561.5488 &    $-$37.37  &  1.36   \\
 74  &  2000 Feb 07  &       51582.4827 &    $-$39.44  &  1.74   \\
 75  &  2000 Feb 23  &       51598.4778 &    $-$40.74  &  2.02   \\
 76  &  2000 Jul 03  &       51728.7458 &    $-$64.23  &  3.01   \\
 77  &  2000 Aug 26  &       51782.6216 &    $-$55.54  &  2.99   \\
 78  &  2000 Oct 26  &       51843.7270 &    $-$51.68  &  2.21   \\
 79  &  2000 Nov 26  &       51874.6841 &    $-$48.57  &  2.22   \\
 80  &  2001 Jan 04  &       51913.5566 &    $-$43.90  &  2.17   \\
 81  &  2001 Feb 10  &       51951.4604 &    $-$43.70  &  1.86   \\
 82  &  2001 Aug 27  &       52148.6168 &    $-$40.94  &  1.65   \\
 83  &  2001 Sep 17  &       52169.5344 &    $-$40.96  &  1.54   \\
 84  &  2001 Oct 11  &       52193.7849 &    $-$40.65  &  1.49   \\
 85  &  2001 Nov 20  &       52233.6801 &    $-$41.80  &  1.60   \\
 86  &  2001 Dec 22  &       52265.6062 &    $-$42.93  &  1.63   \\
 87  &  2002 Jan 14  &       52288.5766 &    $-$42.77  &  1.59   \\
 88  &  2002 Feb 07  &       52313.4838 &    $-$43.99  &  1.71   \\
\enddata
\end{deluxetable}


\begin{thebibliography}{}
\bibitem[Anders \& Grevesse(1989)]{and89} Anders, E., \& Grevesse, N. 1989, Geochim. Cosmochim. Acta, 53, 197
\bibitem[Arnett \& Livne(1994]{arn94} Arnett, D., \& Livne, E. 1994, \apj, 427, 315
\bibitem[Beardsley(1953)]{bea53} Beardsley, W. R. 1953, \apjs, 5, 381
\bibitem[Beardsley(1961)]{bea61} Beardsley, W. R. 1961, \aj, 58, 34
\bibitem[Bidelmann \& McKellar(1957]{bid57} Bidelmann, W. P., \& McKellar, A. 1957, \pasp, 69, 31
\bibitem[Boyarchuk et al.(1988)]{boy88} Boyarchuk, A. A., Boyarchuk, M. E., \& Petrov P. P. 1988, Proc. of the sixth Soviet-Finnish Astronomical Meeting held in Tallinn, Nov. 10-15, 1986, U. Hanni \& I. Tuominen (eds.), Tartu Astrophysical Obs. Teated, 92, 40
\bibitem[Bruenn, Arnett, \& Schramm(1977)]{bru77} Bruenn, S. W., Arnett, W. D., \& Schramm, D. N. 1977, \apj, 213, 213
\bibitem[Clayton(1996)]{cla96} Clayton, G. C 1996, \pasp, 108, 225
\bibitem[Davidson \& Humphyres(1997)]{dav97} Davidson, K., \& Humphreys, R. M. 1997, ARA\&A, 35, 1
\bibitem[de Jager(1998)]{dej98} de Jager, C. 1998, \aap Rev., 8, 145
\bibitem[de Jager, Lobel, \& Israelian(1997)]{dej97a} de Jager, C., Lobel, A., \& Israelian, G. 1997, \aap, 325, 714
\bibitem[de Jager et al.(2001)]{dej01} de Jager, C., Lobel, A., Nieuwenhuijzen, H., \& Stothers, R. 2001, \mnras, 327, 452 
\bibitem[de Jager \& Nieuwehuijzen(1997)]{dej97b} de Jager, C., \& Nieuwenhuijzen, H. 1997, \mnras, 290, L50
\bibitem[Dupree et al.(1984)]{dup84} Dupree, A. K., Hartmann, L., \& Avrett, E. H. 1984, \apj, 281, L37
\bibitem[El Eid \& Champagne(1995)]{ele95} El Eid, M., \& Champagne, A. E. 1995, \apj, 451, 298
\bibitem[Fill et al.(2000)]{fil00} Fill, E., Pretzler, G., Tommasini, R., \& Witzel, B. 2000, Europhys. Lett., 49 (1), 27 
\bibitem[Galazutdinov(1992)]{gal92} Galazutdinov, H. A. 1992, Preprint of the Special Astrophysical Observatory of the 
Russian Academy of Science No. 92 
\bibitem[Gapschkin(1949)]{gap49} Gaposchkin, S. 1949, Harvard Bulletin 919, 18
\bibitem[Gaposchkin-Payne, Gaposchkin, \& Mayall(1947)]{gap47} Gaposchkin, C. P., Gaposchkin, S., \& Mayall, M. 1947, \aj, 52, 123
\bibitem[Greenstein(1948)]{gre48} Greenstein, J. L. 1948, \apj, 108, 78
\bibitem[Harpaz(1998)]{har98} Harpaz, A. 1998, \apj, 498, 293
\bibitem[Hassforther(2001)]{hass01} Hassforther, B. 2001, BAV Rundbrief, 50, 34
\bibitem[Humphreys \& Davidson(1994)]{hum94} Humphreys, R. M., \& Davidson, K. 1994, \pasp, 106, 1025
\bibitem[Humphreys et al.(1997)]{hum97} Humphreys, R. M., Smith, N., Davidson, K., Jones, T. J., Gehrz, R. T., Mason, C. G., Hayward, T. L., Houck, J. R., \& Krautter, J. 1997, \aj, 114, 2778
\bibitem[Ilyin(2000)]{il00} Ilyin, I. 2000, PhD Thesis, Univ. of Oulu, Finland
\bibitem[Israelian, Lobel, \& Schmidt(1999)]{isr99} Israelian, G., Lobel, A., \& Schmidt M. R. 1999, \apj, 523, L145
\bibitem[Jiang \& Huang(1997)]{jia97} Jiang, S. Y., \& Huang, R. Q. 1997, \aap, 317, 121
\bibitem[Jura \& Kleinmann(1990]{jur} Jura, M., \& Kleinmann, S. G. 1990, \apj, 351, 583
\bibitem[Keenan(1947)]{kee47} Keenan, P. C. 1947, \apj, 106, 295
\bibitem[Kurtz \& Mink(1998)]{Kurtz1998} Kurtz, M.\ J., \& Mink, D.\ J. 1998, \pasp, 110, 934
\bibitem[Kurucz(1996)]{kur96} Kurucz, R. L. 1996, in Model Atmospheres and Spectrum Synthesis, ed. S. Adelman, F. Kupka and W. Weiss (ASP Conf. Ser., 108), 160
\bibitem[Kurucz(1999)]{kur99} Kurucz, R. L. 1999, Kurucz CD-ROM No. 24. Cambridge, Mass., Smithsonian Astrophysical Observatory, 199
\bibitem[Latham(1992)]{Latham1992} Latham, D.\ W. 1992, in IAU Coll.\ 135, Complementary Approaches to Double and Multiple Star Research, eds.\ H.\ A.\ McAlister \& W.\ I.\ Hartkopf (San Francisco: ASP), 110
\bibitem[Leiker \& Hoff(1987)]{lei87} Leiker, P.S., \& Hoff, D. B. 1987, Information Bulletin on Variable Stars, 3020, 1
\bibitem[Leiker, Hoff, \& Milton(1989)]{lei89} Leiker, P. S., Hoff, D. B., \& Milton, R. 1989, Information Bulletin on Variable Stars, 3345, 1
\bibitem[Leiker et al.(1988)]{lei88} Leiker, P. S., Hoff, D. B., Nesbella, J., Gainer, M., Milton, R., \& Pray, D. 1988, Information Bulletin on Variable Stars, 3172, 1
\bibitem[Lobel(1997)]{lob97} Lobel, A. 1997, Pulsation and Atmospherical Instability of Luminous F and G-type Stars, PhD. thesis, Brussels Univ., ISBN 90-423-0014-0, Shaker Publ., Maastricht
\bibitem[Lobel(2001a)]{lob01a} Lobel, A. 2001a, \apj, 558, 780
\bibitem[Lobel(2001b)]{lob01b} Lobel, A. 2001b, http://lanl.arXiv.org/abs/astro-ph/0108358
\bibitem[Lobel et al.(1992)]{lob92} Lobel, A., Achmad, L., de Jager, C., \& Nieuwenhuijzen, H. 1992, \aap, 246, 147  
\bibitem[Lobel et al.(1994)]{lob94} Lobel, A., de Jager, C., Nieuwenhuijzen, H., Smolinski, J., \& Gesicki, K. 1994, 291, 226
\bibitem[Lobel \& Dupree(2000a)]{lob00a} Lobel, A., \& Dupree, A. K. 2000a, BAAS 197, \#44.15
\bibitem[Lobel \& Dupree]{lob00b} Lobel, A., \& Dupree, A. K. 2000b, \apj, 545, 454
\bibitem[Lobel \& Dupree(2001)]{lob01} Lobel, A., \& Dupree, A. K. 2001, \apj, 558, 815
\bibitem[Lobel et al.(1998)]{lob98} Lobel, A., Israelian, G., de Jager, C., Musaev, F., Parker, J. Wm., \& Mavrogiorgou, A. 1998, \aap, 330, 659
\bibitem[Lucy(1967)]{luc67} Lucy, L. B. 1967, \aj, 72, 813 
\bibitem[Massey(2000)]{mas00} Massey, P. 2000, \pasp, 112, 144
\bibitem[Meynet et al.(1994)]{mey94} Meynet, G., Maeder, A., Schaller, G., Schaerer, D., \& Charbonnel, C. 1994, \aaps, 103, 97
\bibitem[Mihalas \& Weibel Mihalas(1984)]{mih84} Mihalas, D., \& Weibel Mihalas, B. 1984, Foundations of Radiation Hydrodynamics, Oxford Univ. Press, Oxford
\bibitem[Moore(1945)]{moo45} Moore, C. E. 1945, Princeton Obs. Contr. No. 20. 
\bibitem[Morrison et al.(1997)]{mor97} Morrison, N. D., Knauth, C. D., Mulliss, C. L., \& Lee, W. 1997, \pasp, 109, 676
\bibitem[Nieuwenhuijzen \& de Jager(1995)]{nie95} Nieuwenhuijzen, H., \& de Jager, C. 1995, \aap, 302, 811
\bibitem[Nordstr\"om et al.(1994)]{Nordstrom1994} Nordstr\"om, B., Latham, D.\ W., Morse, J.\ A., Milone, A.\ A.\ E., Kurucz, R.\ L., Andersen, J., \& Stefanik, R.\ P. 1994, \aap, 287, 338 
\bibitem[Paczynski \& Ziolkowski(1968)]{pac68} Paczynski, B., Ziolkowski, J. 1968, Proc. IAU Symposium No. 34, D. E. Osterbrock and C. R. O'Dell (eds.), Dordrecht, D. Reidel Publ., 396
\bibitem[Payne-Gaposchkin \& Mayall(1946)]{pay46} Payne-Gaposchkin, C., \& Mayall, M. W. 1949,  Harvard Bulletin 918, 11 
\bibitem[Percy \& Kolin(2000)]{per00} Percy, J. R., Kolin, D. L. 2000, JAVSO, 28, 1
\bibitem[Percy, Kolin, \& Henry(2000)]{per00} Percy, J. R., Kolin, D. L., \& Henry, G. W. 2000, \pasp, 112, 363
\bibitem[Popper(1947)]{pop47} Popper, D. M. 1947, \aj, 52, 129
\bibitem[Schuster \& Humphreys]{sch01} Schuster, M. T., \& Humphreys, R. M. 2001, BAAS 199, \#92.17
\bibitem[Schwenke(1998)]{sch98} Schwenke, D. W. 1998, Faraday Discuss., 109, 321
\bibitem[Sedov(1958)]{sed58} Sedov, L. I. 1958, Proc. IAU Symposium No. 8, J. M. Burgers and R. N. Thomas (eds.), 1077
\bibitem[Smith et al.(2001)]{smi01} Smith, N., Humphreys, R. M., Davidson, K., Gehrz, R. D., Schuster, M. T., \& Krautter, J. 2001, \aj, 121, 1111
\bibitem[Soker \& Harpaz(1992)]{sok92} Soker, N., \& Harpaz, A. 1992, \pasp, 104, 923
\bibitem[Stothers \& Chin(1997)]{sto97} Stothers, R. B., \& Chin, C.-W. 1997, \apj, 489, 319
\bibitem[Stothers \& Chin(2001)]{sto01} Stothers, R. B., \& Chin, C.-W. 2001, \apj, 560, 934
\bibitem[Tai \& Thackeray(1948)]{tai48} Tai, W. S., \& Thackeray, A. D. 1948, \mnras, 108, 271
\bibitem[Takeda \& Takeda-Hidai(1994)]{tak94} Takeda, Y., \& Takeda-Hidai, M. 1994, \pasj, 46, 395
\bibitem[Th\'{e}venin et al.(2000)]{the00} Th\'{e}venin, F., Parthasarathy, M., \& Jasniewicz, G. 2000, \aap, 359, 138
\bibitem[Thackeray(1948)]{tha48} Thackeray, A. D. 1948, \mnras, 108, 279
\bibitem[Tuchman(1983)]{tuc83} Tuchman, Y. 1983, in Planetary nebulae, Proc. of the Symposium, London, England, Aug. 
9-13, 1982, Dordrecht, D. Reidel Publ. Co., 281
\bibitem[Tuchman, Sack, \& Barkat(1978)]{tuc78} Tuchman, Y., Sack, N., \& Barkat, Z. 1978, \apj, 219, 183
\bibitem[van Genderen(2001)]{vge00} van Genderen, A. M. 2001, \aap, 366, 508
\bibitem[Wagenhuber \& Weiss(1994)]{wag94} Wagenhuber, J., \& Weiss, A. 1994, \aap, 290, 807 
\bibitem[Zsoldos \& Percy(1991)]{zol91} Zsoldos, E., \& Percy, J. R. 1991, \aap, 246, 441 
\end{thebibliography}
\end{document}